\documentclass[11pt]{article}
\usepackage{epsfig}
\usepackage{latexsym}
\usepackage{amsmath}
\usepackage{amssymb}
\usepackage{amsfonts}
\usepackage[english]{babel}
\usepackage{booktabs}
\usepackage{cellspace}
\usepackage{slashbox}
\usepackage[margin=1in]{geometry}
\usepackage{pdflscape}
\usepackage{authblk}
\usepackage{multirow}
\usepackage{hyperref}
\usepackage{graphicx}
\usepackage{float}
\usepackage{tabularx, rotating}
\usepackage{epsfig} 
\usepackage{eepic}
\usepackage{pict2e}
\usepackage{setspace}
\usepackage{amsthm}
\theoremstyle{plain}
\usepackage{enumitem}
\setlist[1]{itemsep=-5pt}

\newtheorem{thm}{Theorem}[section]

\newtheorem{lem}[thm]{Lemma}
\newtheorem{prop}[thm]{Proposition}
\newtheorem{defn}[thm]{Definition}

\newtheorem{expl}[thm]{Example}
\newtheorem*{example}{Example}

\newcommand{\Real}{\mathbb{R}}

\numberwithin{equation}{section}

\newfont{\sfl}{cmssi12}
\begin{document}

\title{Characterization of production sets through individual returns-to-scale: a non parametric specification and an illustration with the U.S industries}
\date{}
\author[1]{Jean-Philippe Boussemart}
\author[2]{Walter Briec}
\author[1]{Raluca 
Parvulescu}
\author[,3]{Paola Ravelojaona\thanks{Corresponding author: \href{mailto:paola.ravelojaona@icn-artem.com}{paola.ravelojaona@icn-artem.com}}}
\affil[1]{\footnotesize IESEG School of Management, Univ. Lille, CNRS, UMR 9221-LEM, F-59000 Lille, France.}
\affil[2]{\footnotesize University of Perpignan, LAMPS.}
\affil[3]{\footnotesize ICN Business School, CEREFIGE, University of Lorraine, Nancy.}

\maketitle

\begin{abstract} 
This paper proposes to estimate the returns-to-scale of production sets by considering the individual return of each observed firm through the notion of $\Lambda$-returns to scale assumption. Along this line, the global technology is then constructed as the intersection of all the individual technologies. Hence, an axiomatic foundation is proposed to present the notion of $\Lambda$-returns to scale. This new characterization of the returns-to-scale encompasses the definition of $\alpha$-returns to scale, as a special case as well as the standard  non-increasing and non-decreasing returns-to-scale models. A non-parametric procedure based upon the goodness of fit approach is proposed to assess these individual returns-to-scale. To illustrate this notion of $\Lambda$-returns to scale assumption, an empirical illustration is provided based upon a dataset involving 63 industries constituting the whole American economy over the period 1987-2018.
\end{abstract}



\newpage

\section{Introduction}

As an important feature of the production process, returns-to-scale provide information about the production technology such as marginal products and linearity of the process (Podinovski et al., 2016; Podinovski, 2022; Sahoo and Tone, 2013; Tone and Sahoo, 2003). In addition, returns-to-scale (RTS) are also related to the notion of economies of scale and of scope that are involved in performance measurement.

Considering the importance of RTS and based upon the notion of homogeneous multi-output technologies (Lau, 1978; F\"are and Mitchell, 1993), Boussemart et al. (2009, 2010) introduce an approach allowing to characterize either strictly increasing or strictly decreasing RTS alongside the traditional RTS (non increasing, non decreasing, constant), in multi-output technologies. This approach is named the $\alpha$-returns to scale model. It provides more theoretical foundation than the traditional data envelopment analysis (DEA) models (Charnes et al., 1978; Banker et al.,1984) by considering all RTS involved in the production process. Indeed, the traditional DEA models just allow either constant or variable RTS such as variable RTS only encompasses non increasing, non decreasing and constant RTS. In addition, the $\alpha$-returns to scale model allows involving zero data within the production set, and also proposes to model production sets with strictly increasing RTS, as defined in the literature.

The $\alpha$-returns to scale model (Boussemart et al., 2009) is defined from a global standpoint (i.e. by considering all the observation as a whole). This means that $\alpha$ which represents the RTS, is a singleton that is applicable to the whole production set. The estimated $\alpha$-returns to scale is optimal when it characterizes the production frontier that minimizes the inefficiency of the set of observations. As $\alpha$ is a singleton representing only one RTS that characterizes the production frontier, the $\alpha$-returns to scale model does not consider the local structure of RTS that individually applies to each observation. To overcome this issue, this paper extends the $\alpha$-returns to scale model to a more general case through the $\Lambda$-returns to scale model. Indeed, the $\Lambda$-returns to scale model defines $\Lambda$ which represents the RTS of the production set as a subset of the non negative real line that may contain an infinity of elements. Hence, $\Lambda$ can encompasses all kind of RTS. The $\Lambda$-returns to scale model is more general since it encompasses as special cases the $\alpha$-returns to scale model (i.e. if $\Lambda$ is a singleton then it reduces to $\alpha$) as well as variable, non increasing and non decreasing RTS models. Moreover, the $\Lambda$-returns to scale model has as limits in zero and infinity, the input and the output ray disposabilities, respectively. As the $\Lambda$-returns to scale model takes into account the local structure of RTS then, the production set can be either convex or non convex. Hence, the production set is not \emph{a priori} assumed to be convex as in traditional economic literature. Relaxing the convexity property is of particular interest mainly by allowing to take account strictly increasing marginal products as well as  possible non linearity in the production process.
$\Lambda$ is first defined through an individual point of view. This means that each observation is associated to its individual $\Lambda$-returns to scale that corresponds to a local RTS of the global production technology. Once the individual $\Lambda$-returns to scale determined, the global $\Lambda$-returns to scale can be deduced as the union of the individual $\Lambda$. This global $\Lambda$-returns to scale characterize the overall technology. Notice that the optimal $\Lambda$ is the RTS that minimize the inefficiency of firms. The global production set that involves all the observations is then the intersection of each individual production technology with respect to $\Lambda$. This means that $\Lambda$ allows to introduce a new class of production sets as they are defined regarding the RTS.

Boussemart et al. (2009, 2010) provide a non parametric approach to implement the $\alpha$-returns to scale model. Indeed, they defined $\alpha$ through the constant elasticity of substitution (CES) - constant elasticity of transformation (CET) production technology (F\"are et al., 1988). This first approach exogenously assesses $\alpha$ since the efficiency of firms are evaluated with respect to an imposed set of possible values of $\alpha$. Leleu et al. (2012) applied this exogenous procedure by using a DEA approach, to provide an empirical analysis of the optimal productive size of hospitals in intensive care units. More recently, Boussemart et al. (2019) propose to consider $\alpha$ as an endogenous variable. They propose to assess $\alpha$ through a minimum extrapolation principle and the Free Disposal Hull (FDH) model (Deprins et al., 1984; Tulkens, 1993). This approach allows to evaluate the optimal RTS through a non parametric scheme and a linear program. In this line, this paper proposes to assess the $\Lambda$-returns to scale model following the approach provided in Boussemart et al. (2019). 

To summarize, the main objective of this paper is threefold. (i) An axiomatic foundation of a generalized RTS is defined through the $\Lambda$-returns to scale model. It allows to characterize a production technology minimizing the inefficiency of firms. (ii) A new class of production sets is introduced regarding the $\Lambda$-returns to scale model.  (iii) A non parametric procedure is proposed to assess the $\Lambda$-returns to scale through linear programs. 

The remainder of this paper is structured as follows. Section 2 presents the backgrounds on the production technology, efficiency measurement and the $\alpha$-returns to scale model. The notion of $\Lambda$-returns to scale and its connection with standard models are presented in section 3. Section 4 proposes a general procedure to estimate the $\Lambda$-returns to scale based upon the individual $\alpha$-returns to scale and from an input oriented standpoint.  Section 5 provides an empirical illustration by the means of a dataset about 63 industries constituting the whole American economy such that the data is composed with one output and three inputs and covers the period 1987-2018. Section 6 resumes and concludes.

\section{Backgrounds}

This section aims to introduce the notation and the theoretical basis used throughout this paper. Subsection 2.1 defines the production technology as well as the efficiency measure. Subsection 2.2 describes the $\alpha$-returns to scale model.

\subsection{Production technology: assumptions and key concepts}

The production technology $T$ is the process transforming an input vector $x=(x_{1},\cdots
,x_{n})\in \mathbb{R}_{+}^{n}$ composed of $n\in \mathbb{N}$ components into an output vector $y\,=(y_{1},\cdots ,y_{p})\in \mathbb{R}_{+}^{p}$ containing $p\in \mathbb{N}$ elements, and defined by:
\begin{equation}
T=\left\{ {(x,y)\in \mathbb{R}_{+}^{n+p}:x\text{ can produce
}y}\right\}
\end{equation}%
The technology satisfies the following regular axioms: ({$T1$}) no free lunch and inaction; ({$T2$}) infinite outputs cannot be obtained from a finite input vector; ({$T3$}) the production set is closed; ({$T4$}) the inputs and outputs are freely disposable. Remark that the technology is convex neutral meaning that the convexity of the production set is not \emph{a priori} assumed. Also notice that when setting empirical analyses, the production sets may not satisfy all the axioms $T1-T4$.

The efficiency measure of production units can be evaluated through distance functions that assess the distance between the observation and the efficient frontier. One of the most used efficiency measure is the Farrell efficiency measure (Debreu, 1951; Farrell, 1957) that can be either input oriented or output oriented. For any $(x,y)\in \Real_+^{n+p}$ the input Farrell measure provides the maximum radial contraction of the input vector for a given level of outputs and is defined as $D^I(x,y)=\inf_\theta\left\{\theta\geq 0: (\theta x, y)\in T\right\}$. In the same vein, the output Farrell measure gives the maximum radial expansion of the output vector for a given amount of inputs and is defined as $D^O(x,y)=\sup_\mu\left\{\mu\geq 0: (x, \mu y)\in T\right\}$. Remark that the input Farrell measure takes value between 1 and 0. If the observation does not belong to the production set then, both input and output Farrell measures are indeterminate $(\pm \infty)$. Besides, if the production unit is efficient i.e. belongs to the efficient frontier then, the input and/or the output Farrell measure is equal to 1.

Following the proposed definition of F\"are and Mitchell (1993), a production technology $T$ is \textit{homogeneous of degree }$\mathit{\alpha }$ if for any $\eta >0$ and any $\alpha \in \Real_+$, $(x,y)\in T\Rightarrow (\eta x,\eta^{\alpha }y)\in T. \label{alphareturn}$
Obviously, this notion of homogeneity of degree $\alpha$ is connected to the notion of returns-to-scale. Indeed, constant returns-to-scale (CRS) corresponds to $\alpha =1$ while strictly increasing returns (IRS) correspond to $\alpha >1$ and strictly decreasing returns (DRS) correspond to $\alpha <1$. Boussemart et al. (2009, 2010) termed this property of the technology as ``$\mathit{\alpha }$\textit{-returns to scale}". Boussemart et al. (2010) show that under such an assumption, some existing measures (Farrell output measure, hyperbolic efficiency measure of F\"are et al., 1985; proportional distance function of Briec, 1997) can be related in closed form under an $\alpha$-returns to scale assumption.

\subsection{$\alpha$-returns to scale : non-parametric approach and extrapolation principle}\label{DefAlpha}

In the line of Boussemart et al. (2009), Boussemart et al. (2019) propose a non-parametric approach to estimate the best returns-to-scale allowing to maximize the global efficiency of the whole considered production set. To do so, they consider a constant elasticity of substitution - constant elasticity of transformation (CES-CET) production set (F\"are et al., 1988) and apply the minimum extrapolation principle to a Free Disposal Hull (FDH - Deprins et al., 1984) type model by means of input and output-oriented Farrell efficiency measures. For any firm $k\in \mathcal{J}$ belonging to the set of $J$ firms $A=\left\{(x_1,y_1, \cdots, (x_J,y_J) \right\}$, the global technology $T_{\gamma,\delta}$ is the union of each $k$ individual technology $Q_{\gamma,\delta}(x_k,y_k)$ with $\gamma,\delta >0$, where:
\begin{align}
& T_{\gamma,\delta}=\bigcup_{k\in \mathcal J} Q_{\gamma,\delta}
(x_k,y_k),\\
\text{and}\qquad & Q_{\gamma,\delta}
(x_k,y_k)=\Big\{(x,y)\in \Real_+^{n+p}: x\geq \lambda^{1/\gamma}
x_k, y\leq  \lambda^{1/\delta}y_k, \lambda \geq
0\Big\}. \label{individual} 
\end{align}

Remark that the global technology $T_{\gamma,\delta}$ is the production set including all the observations whereas the individual technology $Q_{\gamma,\delta}(x_k,y_k)$ is a production possibility set derived from and related to the observation $k$. Boussemart et al. (2009) prove that $T_{\gamma,\delta}$ satisfies $T1$-$T4$. The efficiency of each production unit is then assessed with respect to each individual technology $Q_{\gamma,\delta}(x_k,y_k)$. In such case, for $k,j \in \mathcal{J}$, the efficiency measures $D^I_k(x_j,y_j;\gamma,\delta)$ and $D^O_k(x_j,y_j;\gamma,\delta)$ are respectively the input and the output Farrell measure of the observation $(x_j,y_j)$ with respect to $Q_{\gamma,\delta}(x_k,y_k)$. Boussemart et al. (2009, 2010) demonstrate that for $\alpha=\gamma/\delta$,
\begin{align}
& D^I_k(x_j,y_j;\gamma,\delta)= \left[\max_{h\in card (y)}
\frac{y_{j,h}}{y_{k,h} }\right]^{1/\alpha}\cdot\left[\max_{i\in
card(x)}\frac{x_{k,i}}{x_{j,i} }\right], \label{Dikx0y0}\\[0.7cm]
\text{and}\qquad & D^O_k(x_j,y_j;\gamma,\delta)= \left[\min_{h\in card (y)}
\frac{y_{k,h}}{y_{j,h} }\right]\cdot\left[\min_{i\in
card(x)} \frac{x_{j,i}}{x_{k,i} }\right]^{\alpha}. \label{Dokx0y0}
\end{align}

By definition, $D^I_j(x_j,y_j;\gamma,\delta)= 1$ since the observation $(x_j,y_j)$ is evaluated with respect to its own individual technology $Q_{\gamma,\delta} (x_j,y_j)$. Remark that $x_{k,i}$ stands for the $i$-th component of the input vector $x$ of firm $k$. Also, note that $card(x)=\{1, \cdots, n\}$ and $card(y)=\{1,\cdots, p\}$ where $n$ and $p$ are the number of elements in input and output vectors, respectively.
Notice that the following convention is adopted: for any $a,b\in \Real_+$, if $a>0$ and $b=0$, then $\frac{a}{b}=+\infty$.\\

\begin{figure}[htpb!]
\begin{minipage}{9cm}
\centering
{\scriptsize \unitlength 0.3mm 
\linethickness{0.4pt}
\ifx\plotpoint\undefined\newsavebox{\plotpoint}\fi 
\begin{picture}(187.75,176.25)(0,0)
\put(89,124.25){\vector(3,4){.07}}\multiput(4.5,14)(.03373253493,.044011976048){2505}{\line(0,1){.044011976048}}
\put(4.5,14){\vector(1,0){175.75}}
\put(4.75,13.75){\vector(0,1){152.25}}
\put(80,134.75){\circle*{1.8}}
\put(131,98){\circle*{1.8}}
\put(4.75,14.75){\line(1,0){.25}}
\qbezier(95.75,171.75)(85,135.25)(5.25,14.75)
\qbezier(5.25,14.75)(148.625,88)(150.5,137.25)
\multiput(80.5,128.5)(-.0336990596,-.046630094){638}{\line(0,-1){.046630094}}
\put(59,98.75){\line(1,0){46.25}}
\multiput(137.25,114.5)(-.03371592539,-.04555236729){697}{\line(0,-1){.04555236729}}
\put(113.75,82.75){\line(1,0){39.25}}
\put(122,153){\circle*{1.8}}
\qbezier(5.25,15.5)(125.75,138.125)(130.25,176.25)
\multiput(117.5,143.75)(-.0337301587,-.0436507937){378}{\line(0,-1){.0436507937}}
\put (125,185){\makebox(0,0)[cc]{\tiny $Q_{\gamma, \delta}(x_1,y_1)$}}
\put (180,170){\makebox(0,0)[cc]{\tiny $Q_{\gamma, \delta}(x_2,y_2)$}}
\put (170,105){\makebox(0,0)[cc]{\tiny $Q_{\gamma, \delta}(x_3,y_3)$}}
\put(104.75,127.25){\line(1,0){23.25}}
\put(93.25,129.75){\makebox(0,0)[cc]{$x'$}}
\put(187.75,15.25){\makebox(0,0)[cc]{$x$}}
\put(4.25,174.25){\makebox(0,0)[cc]{$y$}}
\put(0,5.25){\makebox(0,0)[cc]{$0$}}
\put(145.5,157){\makebox(0,0)[cc]{$(x_2,y_2)$}}
\put(155.25,97.25){\makebox(0,0)[cc]{$(x_3,y_3))$}}
\put(60.5,135.25){\makebox(0,0)[cc]{$(x_1,y_1)$}}
\end{picture}
}
\caption{2-dimensional non-convex technology}\label{NCtechnologies}
\end{minipage}
\hfill
\begin{minipage}{6cm}
\linespread{1.5}\selectfont
Figure \ref{NCtechnologies} describes the individual production sets  for each observation. Notice that the represented production sets are non convex resulting in illustrating efficient frontiers related to strictly increasing RTS for $Q_{\gamma, \delta}(x_1,y_1)$, $Q_{\gamma, \delta}(x_2,y_2)$, and $Q_{\gamma, \delta}(x_3,y_3)$. 
\end{minipage}
\end{figure}
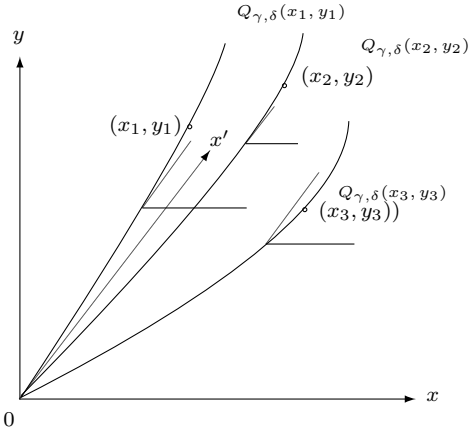

\section{On some extended notions of $\alpha$-returns to scale }

This section aims to propose the axiomatic foundation of the generalized $\Lambda$-returns to scale model. Subsection 3.1 presents its definition while Subsection 3.2. displays the connection with this new RTS model and the traditional ones.

\subsection{$\Lambda$-returns to scale: definition and some basic properties}

Boussemart et al. (2019) show that empirical procedures may provide infinite ($\infty$) and null ($0$) returns-to-scale. Formally, a production set $T$ satisfies a {\bf $0$-returns to scale assumption} if for any scalar $\lambda>0$ and if the production unit $(x,y)$ belongs to $T$ then, $(\lambda x,y)$ also belongs to $T$. This definition is obtained from the standard definition setting $\alpha=0$. Note that, equivalently, a production set satisfies a {\bf $\alpha$-returns to scale assumption} if for all $\lambda>0$, an observation  $(x,y)$ belonging to $ T$ means that $(\lambda^{\frac{1}{\alpha}} x,\lambda y)$ is also part of $ T.$
Along this line, a production set $T$ satisfies an {\bf $\infty$-returns to scale assumption}, if for any $\lambda>0$ and if the production unit $(x,y)\in T$ then, $( x,\lambda y)\in T$.

The $0$-returns to scale assumption and $\infty$-returns to scale assumption are limit cases of the $\alpha$-returns to scale assumption when $\alpha \rightarrow 0$ and $\alpha \rightarrow \infty$. Surprisingly, $\infty$-returns to scale assumption and $0$-returns to scale assumption correspond to the input and output ray disposability (T4) assumption, respectively.\\

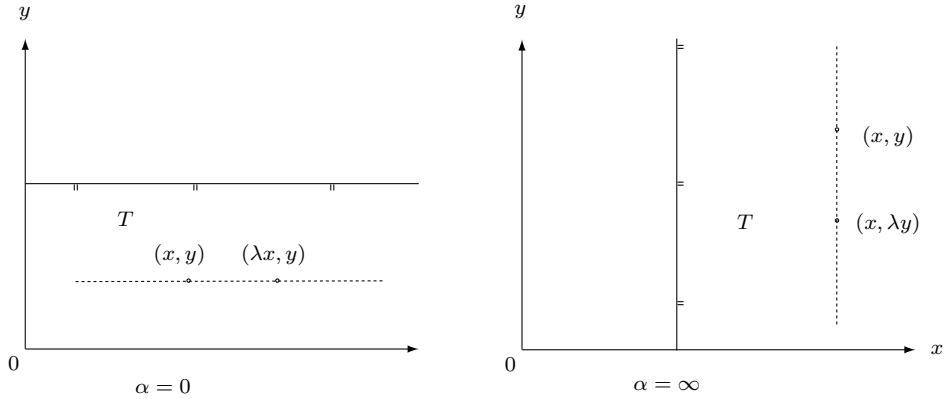
\begin{figure}[htpb!]
\centering{\scriptsize
\unitlength 0.4mm 
\linethickness{0.4pt}
\ifx\plotpoint\undefined\newsavebox{\plotpoint}\fi 
\unitlength 0.4mm 
\linethickness{0.4pt}
\ifx\plotpoint\undefined\newsavebox{\plotpoint}\fi 
\begin{picture}(306.85,138.25)(0,0)
\put(3.8,26.05){\vector(0,1){103.2}}
\put(0,21.25){\makebox(0,0)[cc]{$0$}}
\put(3.85,26){\vector(1,0){130.75}}
\put(168.9,25.8){\vector(0,1){103.2}}
\put(165.1,21){\makebox(0,0)[cc]{$0$}}
\put(168.95,25.75){\vector(1,0){130.75}}
\put(306.85,26){\makebox(0,0)[cc]{$x$}}
\put(168.35,138.25){\makebox(0,0)[cc]{$y$}}
\put(3.6,137.5){\makebox(0,0)[cc]{$y$}}
\put(3.85,81){\line(1,0){130.75}}
\put(220.35,25.5){\line(0,1){103.75}}
\put(49.6,13.75){\makebox(0,0)[cc]{$\alpha=0$}}
\put(217.1,13.75){\makebox(0,0)[cc]{$\alpha=\infty$}}
\put(37.1,69.25){\makebox(0,0)[cc]{$T$}}
\put(243.1,68.25){\makebox(0,0)[cc]{$T$}}
\put(20.1,80.75){\line(0,-1){2}}
\put(220.6,40.75){\line(1,0){2}}
\put(59.85,80.75){\line(0,-1){2}}
\put(220.6,80.5){\line(1,0){2}}
\put(105.35,80.75){\line(0,-1){2}}
\put(220.6,126){\line(1,0){2}}
\put(21.1,80.75){\line(0,-1){2}}
\put(220.6,41.75){\line(1,0){2}}
\put(60.85,80.75){\line(0,-1){2}}
\put(220.6,81.5){\line(1,0){2}}
\put(106.35,80.75){\line(0,-1){2}}
\put(220.6,127){\line(1,0){2}}
\put(273.6,68.75){\circle*{1.118}}
\put(58.1,48.75){\circle*{1.118}}
\put(20.53,48.43){\line(1,0){.9904}}
\put(22.51,48.435){\line(1,0){.9904}}
\put(24.491,48.439){\line(1,0){.9904}}
\put(26.472,48.444){\line(1,0){.9904}}
\put(28.453,48.449){\line(1,0){.9904}}
\put(30.434,48.454){\line(1,0){.9904}}
\put(32.414,48.459){\line(1,0){.9904}}
\put(34.395,48.463){\line(1,0){.9904}}
\put(36.376,48.468){\line(1,0){.9904}}
\put(38.357,48.473){\line(1,0){.9904}}
\put(40.337,48.478){\line(1,0){.9904}}
\put(42.318,48.483){\line(1,0){.9904}}
\put(44.299,48.487){\line(1,0){.9904}}
\put(46.28,48.492){\line(1,0){.9904}}
\put(48.26,48.497){\line(1,0){.9904}}
\put(50.241,48.502){\line(1,0){.9904}}
\put(52.222,48.507){\line(1,0){.9904}}
\put(54.203,48.511){\line(1,0){.9904}}
\put(56.184,48.516){\line(1,0){.9904}}
\put(58.164,48.521){\line(1,0){.9904}}
\put(60.145,48.526){\line(1,0){.9904}}
\put(62.126,48.531){\line(1,0){.9904}}
\put(64.107,48.535){\line(1,0){.9904}}
\put(66.087,48.54){\line(1,0){.9904}}
\put(68.068,48.545){\line(1,0){.9904}}
\put(70.049,48.55){\line(1,0){.9904}}
\put(72.03,48.555){\line(1,0){.9904}}
\put(74.01,48.56){\line(1,0){.9904}}
\put(75.991,48.564){\line(1,0){.9904}}
\put(77.972,48.569){\line(1,0){.9904}}
\put(79.953,48.574){\line(1,0){.9904}}
\put(81.934,48.579){\line(1,0){.9904}}
\put(83.914,48.584){\line(1,0){.9904}}
\put(85.895,48.588){\line(1,0){.9904}}
\put(87.876,48.593){\line(1,0){.9904}}
\put(89.857,48.598){\line(1,0){.9904}}
\put(91.837,48.603){\line(1,0){.9904}}
\put(93.818,48.608){\line(1,0){.9904}}
\put(95.799,48.612){\line(1,0){.9904}}
\put(97.78,48.617){\line(1,0){.9904}}
\put(99.76,48.622){\line(1,0){.9904}}
\put(101.741,48.627){\line(1,0){.9904}}
\put(103.722,48.632){\line(1,0){.9904}}
\put(105.703,48.636){\line(1,0){.9904}}
\put(107.684,48.641){\line(1,0){.9904}}
\put(109.664,48.646){\line(1,0){.9904}}
\put(111.645,48.651){\line(1,0){.9904}}
\put(113.626,48.656){\line(1,0){.9904}}
\put(115.607,48.66){\line(1,0){.9904}}
\put(117.587,48.665){\line(1,0){.9904}}
\put(119.568,48.67){\line(1,0){.9904}}
\put(121.549,48.675){\line(1,0){.9904}}
\put(273.53,34.18){\line(0,1){.9946}}
\put(273.53,36.169){\line(0,1){.9946}}
\put(273.53,38.158){\line(0,1){.9946}}
\put(273.53,40.147){\line(0,1){.9946}}
\put(273.53,42.137){\line(0,1){.9946}}
\put(273.53,44.126){\line(0,1){.9946}}
\put(273.53,46.115){\line(0,1){.9946}}
\put(273.53,48.104){\line(0,1){.9946}}
\put(273.53,50.094){\line(0,1){.9946}}
\put(273.53,52.083){\line(0,1){.9946}}
\put(273.53,54.072){\line(0,1){.9946}}
\put(273.53,56.061){\line(0,1){.9946}}
\put(273.53,58.051){\line(0,1){.9946}}
\put(273.53,60.04){\line(0,1){.9946}}
\put(273.53,62.029){\line(0,1){.9946}}
\put(273.53,64.018){\line(0,1){.9946}}
\put(273.53,66.008){\line(0,1){.9946}}
\put(273.53,67.997){\line(0,1){.9946}}
\put(273.53,69.986){\line(0,1){.9946}}
\put(273.53,71.975){\line(0,1){.9946}}
\put(273.53,73.965){\line(0,1){.9946}}
\put(273.53,75.954){\line(0,1){.9946}}
\put(273.53,77.943){\line(0,1){.9946}}
\put(273.53,79.932){\line(0,1){.9946}}
\put(273.53,81.922){\line(0,1){.9946}}
\put(273.53,83.911){\line(0,1){.9946}}
\put(273.53,85.9){\line(0,1){.9946}}
\put(273.53,87.889){\line(0,1){.9946}}
\put(273.53,89.879){\line(0,1){.9946}}
\put(273.53,91.868){\line(0,1){.9946}}
\put(273.53,93.857){\line(0,1){.9946}}
\put(273.53,95.846){\line(0,1){.9946}}
\put(273.53,97.836){\line(0,1){.9946}}
\put(273.53,99.825){\line(0,1){.9946}}
\put(273.53,101.814){\line(0,1){.9946}}
\put(273.53,103.803){\line(0,1){.9946}}
\put(273.53,105.793){\line(0,1){.9946}}
\put(273.53,107.782){\line(0,1){.9946}}
\put(273.53,109.771){\line(0,1){.9946}}
\put(273.53,111.76){\line(0,1){.9946}}
\put(273.53,113.75){\line(0,1){.9946}}
\put(273.53,115.739){\line(0,1){.9946}}
\put(273.53,117.728){\line(0,1){.9946}}
\put(273.53,119.717){\line(0,1){.9946}}
\put(273.53,121.707){\line(0,1){.9946}}
\put(273.53,123.696){\line(0,1){.9946}}
\put(273.53,125.685){\line(0,1){.9946}}
\put(55.1,57.5){\makebox(0,0)[cc]{$(x,y)$}}
\put(86.35,57.25){\makebox(0,0)[cc]{$(\lambda x,y)$}}
\put(290.6,96.5){\makebox(0,0)[cc]{$( x, y)$}}
\put(290.6,67.75){\makebox(0,0)[cc]{$(x,\lambda y)$}}
\put(87.6,48.75){\circle*{1.118}}
\put(273.6,99){\circle*{1.118}}
\end{picture}
}
\vspace{-0.6cm}\caption{Infinite and null returns to scale}\label{limitcases}
\end{figure}

Figure \ref{limitcases} illustrates the $0$-returns to scale (figure on the left) and $\infty$-returns to scale (figure on the right) assumptions. Obviously, the figure on the left illustrate the input strong disposability since for $\lambda\geq 1$, the same level of output ($y$) is provided by a higher level of input ($\lambda x$). Besides, the figure on the right presents the output strong disposability, since for $\lambda \in [0,1]$, the same level of input ($x$) can produce a lower level of outputs ($\lambda y$). Note that $0$-returns to scale may be not compatible with the no free lunch axiom ($T1$) for any positive output. In addition 
$\infty$-returns to scale do not hold if $T2$ holds.

In the following the notion of $\alpha$-returns to scale is extended. 

\begin{defn}
Let $\Lambda$ be a subset of $[0,+\infty]$. We say that a technology $T$ {\bf satisfies a $\Lambda$-returns to scale assumption} if there exists a family $\{T_\alpha\}_{\alpha\in \Lambda}$ of production sets where for any
$\alpha\in \Lambda$,  $T_\alpha$ satisfies an $\alpha$-returns to scale assumption and such that:
\begin{equation}
T=\bigcap_{\alpha\in \Lambda}T_\alpha.
\end{equation}

\end{defn}

\noindent Notice that there is no specific restriction on $\Lambda$ except that it is a set of positive real numbers and $\Lambda$ may be finite. This definition has the advantage to give information about the local nature of the technology and to propose a more general class of technologies involving some special local returns-to-scale. Indeed, $\alpha$ is a singleton representing an individual returns-to-scale associated to a given observation allowing to minimize its inefficiency whereas $\Lambda$ combines any individual returns-to-scale aiming to provide the global returns-to-scale and the characterization of the overall production set. An expanded explanation of this definition is displayed in Section 4 (Proposition \ref{indextrap}).\\

\begin{figure}[htpb!]
\begin{minipage}{9cm}
\centering {\scriptsize \unitlength 0.3mm 
\linethickness{0.4pt}
\resizebox{0.75\textwidth}{!}{
\ifx\plotpoint\undefined\newsavebox{\plotpoint}\fi 
\begin{picture}(243.25,198)(0,0)
\put(3,8){\vector(1,0){232}}
\put(2.75,8.25){\vector(0,1){184}}
 \qbezier(69.25,186.75)(68,70.5)(2.75,8.25)
\put(66.75,146){\circle*{1.8}}
 \qbezier(192.75,189)(23.75,183)(2.75,9)
\put(243.25,7.25){\makebox(0,0)[cc]{$x$}}
\put(2.25,198){\makebox(0,0)[cc]{$y$}}
\put(0,3){\makebox(0,0)[cc]{$0$}}
\put(89.25,143){\makebox(0,0)[cc]{$(x_0,y_0) $}}
\multiput(54,87.75)(.1413043,-.0326087){23}{\line(1,0){.1413043}}
\multiput(55,90.25)(.1083333,-.0333333){30}{\line(1,0){.1083333}}
\multiput(31,41.5)(.05,-.0333333){30}{\line(1,0){.05}}
\multiput(32.25,43.5)(.0652174,-.0326087){23}{\line(1,0){.0652174}}
\multiput(104.5,170)(.0333333,-.0916667){30}{\line(0,-1){.0916667}}
\multiput(107.5,171.25)(.03289474,-.08552632){38}{\line(0,-1){.08552632}}
\multiput(65.5,138.25)(.34375,-.03125){8}{\line(1,0){.34375}}
\multiput(66.25,141)(.2,-.0333333){15}{\line(1,0){.2}}
\multiput(71.75,150.75)(.03333333,-.04166667){60}{\line(0,-1){.04166667}}
\multiput(74.25,152)(.03289474,-.04605263){38}{\line(0,-1){.04605263}}
\multiput(166.75,187)(.0333333,-.1666667){15}{\line(0,-1){.1666667}}
\multiput(169.25,187.25)(.03125,-.21875){8}{\line(0,-1){.21875}}
\put(134.25,78){\makebox(0,0)[cc]
{$T=Q _{\alpha_1,1} (x_0,y_0)\cap Q _{\alpha_2,1} (x_0,y_0)$}}
\end{picture}}}
\caption{$\Lambda$-returns to scale with $\{\alpha_1,\alpha_2\}$ and a single firm.}\label{lambdartsfig}
\end{minipage}
\hfill
\begin{minipage}{6cm}
\linespread{1.5}\selectfont
In Figure \ref{lambdartsfig}, the production set $Q _{\alpha_1,1} (x_0,y_0)\cap Q _{\alpha_2,1} (x_0,y_0) $  satisfies a $\{\alpha_1,\alpha_2\}$-returns to scale assumption. However it does not satisfy an $\alpha_1$-returns to scale assumption neither an $\alpha_2$-returns to scale assumption, if they are considered separately as singletons. 
\end{minipage}
\end{figure}
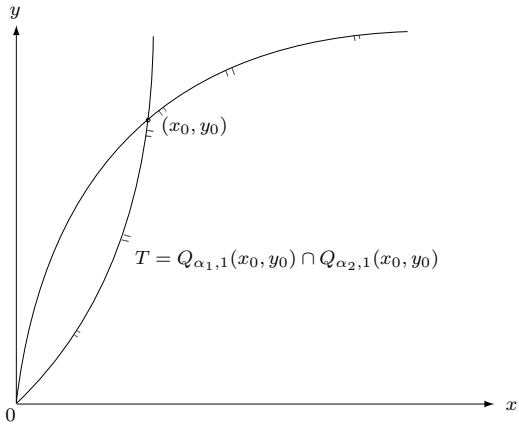

Remark that larger is the collection $\Lambda$, larger is the potential number of technologies satisfying a $\Lambda$-returns to scale assumption. Indeed, if $\Lambda$ is large then there is more class of technologies that is embedded within the technology satisfying a $\Lambda$-returns to scale assumption.\\

We consider the following examples.

\begin{expl} Suppose that $n=2$ and $p=1$ with $T=\big\{(x_1,x_2,y)\in \Real_+^3: x_1\geq 2, x_2\geq 3, y\leq 4\big\}$. By construction, 
$$T= \big\{(x_1, x_2, y): x_1\geq 0, x_2\geq 0, 0\leq y\leq 4\big\}\cap \big\{(x_1, x_2, y): x_1\geq 2, x_2\geq 3, y \in \Real_+\big\}. $$
Let us denote $T_0=\big\{(x_1, x_2, y): x_1\geq 0, x_2\geq 0, 0\leq y\leq 4\big\}$ and $T_\infty=\big\{(x_1, x_2, y): x_1\geq 2, x_2\geq 3, y \in \Real_+\big\}$. We have $T=T_0\cap T_{\infty}$. Clearly, $T_0$ satisfies a $0$-returns to scale assumption and $T_\infty$ satisfies a $\infty$-returns to scale assumption. Therefore $T$ satisfies a $\{0,\infty\}$-returns to scale assumption. Also if $\Lambda\supset \{0,\infty\}$  it also satisfies a $\Lambda$-returns to scale assumption. $T$ only satisfies $T2-T4$. \end{expl}

\begin{expl} Suppose that $n=1$ and $p=1$ with $T=\big\{(x,y)\in \Real_+^2: 2\leq  x\leq 5,  y\leq x^{\frac{1}{3}}\big\}$. By construction 
$$T= \big\{(x , y): 2\leq x\leq 5, y\geq 0 \big\}\cap \big\{(x, y): x \geq 0,y \leq x^{\frac{1}{3}}  \big\}. $$
Let us denote $T_ {\frac{1}{3} }=\big\{(x , y): x \geq 0,   y\leq x^{\frac{1}{3}}  \big\}$ and $T_\infty=\big\{(x, y): 2\leq x\leq 5, y\geq 0\big\}$. We have $T=T_{\frac{1}{3} }\cap T_{\infty}$. Clearly, $T_ {\frac{1}{3} }$ satisfies a $ {\frac{1}{3} }$-returns to scale assumption and $T_\infty$ satisfies a $\infty$-returns to scale assumption. Therefore $T$ satisfies a $\{  {\frac{1}{3} },\infty\}$-returns to scale assumption. This situation might occur when the available input quantity is limited. $T$ only satisfies $T2-T3$. Another symmetrical case is given by the production set 
$S=\big\{(x,y)\in \Real_+^2: 2\leq  y\leq 5,  y\leq x^{\frac{1}{3}}\big\}$. Clearly, $S$ satisfies a $\{  0,{\frac{1}{3} }\}$-returns to scale assumption. This type of situation might arise when a technology requires the production of a minimal amount of output. \end{expl}

\begin{expl} Suppose that $n=2$ and $p=1$ with $T=\big\{(x_1,x_2,y)\in \Real_+^3: y\leq x_1 x_2,y\leq x_1 + x_2, y\leq 4\big\}$.  
Let us denote $T_0=\big\{(x_1, x_2, y)\in \Real_+^3:y\leq 4\big\}$,  $T_1=\big\{(x_1, x_2, y)\in \Real_+^3:y\leq x_1 + x_2\big\}$ and $T_2=\big\{(x_1, x_2, y)\in \Real_+^3:y\leq x_1 x_2\big\}$. We have $T=T_0\cap T_{1}\cap T_{2}$. Clearly, $T_0$ satisfies a $0$-returns to scale assumption, $T_1$ a constant returns to scale assumption and $T_2$ satisfies a $2$-returns to scale assumption. Therefore $T$ satisfies a $\{0, 1,2\}$-returns to scale assumption.  \end{expl}

Notice that although the production process satisfies some standard axioms, there is no guarantee that  the smallest technology satisfying an $\alpha$-returns to scale assumption and containing the production process, obeys these axioms.  For example the conical hull of a closed set may not be closed and  {$ T3$}  might be violated. Also, in the case where $\alpha=+\infty$, {$ T2$} no longer holds true. \\

\noindent Remark that a production set $T$ satisfies {\bf a minimal $\Lambda$-returns to scale assumption} if $T$ satisfies a $\Lambda$-returns to scale assumption and if for all $\beta\in \Lambda$
\begin{equation}T\varsubsetneqq \bigcap_{\alpha\in \Lambda\backslash\{\beta\}}T_\alpha
\end{equation}
This means that the identified $\Lambda$ is the minimal returns-to-scale allowing to characterize the entire production set. \bigskip

\begin{expl} Suppose that $n=1$ and $p=1$ with $T=\big\{(x ,y)\in \Real_+^2: y\leq \min\{x, \sqrt{x}\}\big\}$.  
Let us denote $T_{\frac{1}{2}}=\big\{(x , y)\in \Real_+^2:y \leq   \sqrt{x}\big\}$,  $T_1=\big\{(x , y)\in \Real_+^2:y\leq x \big\}$ and $T_2=\big\{(x ,   y)\in \Real_+^2:y\leq x^2\big\}$. We have $T=T_{\frac{1}{2}}\cap T_{1}\cap T_{2}$. Clearly, $T_{\frac{1}{2}}$ satisfies a $\frac{1}{2}$-returns to scale assumption when $T_1$ and $T_2$ satisfy a constant returns to scale and a $2$-returns to scale assumptions, respectively. Therefore, $T$ satisfies a $\{0, 1,2\}$-returns to scale assumption. However, $T_1$ is not an active constraint. Thus, we also have   $T=T_{\frac{1}{2}}\cap   T_{2}$. Hence, $T$ is $\{\frac{1}{2},2\}$-minimal but not   $\{0, 1,2\}$-minimal. \end{expl}\bigskip

Let us define, for all $\alpha\geq 0$, the $\alpha$-conical hull of any subset $S$ of $\Real_+^{n+m}$ as:
\begin{equation}
K_\alpha(S)=\Big\{(\lambda x,\lambda^{\alpha}y): (x,y)\in S, \lambda\geq 0 \Big \}. 
\end{equation}
Notice that if $\alpha=+\infty$, then $K_\infty(S)=\big\{( x,\lambda y): (x,y)\in S, \lambda\geq 0 \big \}.$ Obviously, when $ \alpha=1$, we retrieve the standard concept of conical hull. \\

We say that the returns-to-scale of a production set $T$ are {\bf $\{\alpha\}$-bounded } if $T$ is contained by at least one production set satisfying $T2$  and an assumption of $\{\alpha\}$-returns to scale. In addition $T$ is {\bf $\Lambda$-bounded } if $T$ is $\{\alpha\}$-bounded for all $\alpha\in \Lambda$. This implies that the RTS of a production set $T$ belong to $\Lambda$ if $T$ is contained by at least one technology satisfying a $\Lambda$-returns to scale assumption. For example in Figure \ref{lambdartsfig}, $Q _{\alpha_1,1} (x_0,y_0) $ is obviously $\{\alpha_1\}$-bounded. Nonetheless, $Q _{\alpha_1,1} (x_0,y_0) $ is not $\{\alpha_2\}$-bounded. Remark that this statement do not postulate any assumption on $T$.\\

The next statement show that if the RTS of a production set are  $\Lambda$-bounded, then there exists a smallest technology containing it and satisfying a $\Lambda$-returns to scale assumption. 
 
\begin{lem}\label{lem32} Let $T$ be a production  set and
suppose that the returns to scale of $T$ are $\{\alpha\}$-bounded, for some $\alpha\in [0,+\infty]$. Let us denote, $T_\alpha:=K_\alpha(T)$. Then,   $T_\alpha$
  is the smallest technology satisfying an assumption of $\alpha$-returns to scale   that contains $T$.
\end{lem}

\noindent See proof in Appendix 2.\\[-0.2cm]

\begin{figure}[htpb!]\centering {\scriptsize \unitlength 0.3mm 
\linethickness{0.4pt}
\ifx\plotpoint\undefined\newsavebox{\plotpoint}\fi 

\unitlength 0.25mm 
\linethickness{0.4pt}
\resizebox{.75\textwidth}{!}{
\ifx\plotpoint\undefined\newsavebox{\plotpoint}\fi 
\begin{picture}(545.552,238)(0,0)
\put(3,48){\vector(1,0){232}}
\put(305.302,46.318){\vector(1,0){232}}
\put(2.75,48.25){\vector(0,1){184}}
\put(305.052,46.568){\vector(0,1){184}}
\put(243.25,47.25){\makebox(0,0)[cc]{$x$}}
\put(545.552,45.568){\makebox(0,0)[cc]{$x$}}
\put(2.25,238){\makebox(0,0)[cc]{$y$}}
\put(304.552,236.318){\makebox(0,0)[cc]{$y$}}
\put(0,43){\makebox(0,0)[cc]{$0$}}
\put(302.302,41.318){\makebox(0,0)[cc]{$0$}}
\multiput(54,127.75)(.464286,-.107143){7}{\line(1,0){.464286}}
\multiput(356.302,126.068)(.464286,-.107143){7}{\line(1,0){.464286}}
\multiput(55,130.25)(.361111,-.111111){9}{\line(1,0){.361111}}
\multiput(357.302,128.568)(.361111,-.111111){9}{\line(1,0){.361111}}
\multiput(14.75,1)(.166667,-.111111){9}{\line(1,0){.166667}}
\multiput(16,3)(.214286,-.107143){7}{\line(1,0){.214286}}
\multiput(104.5,207.5)(.111111,-.305556){9}{\line(0,-1){.305556}}
\multiput(406.802,205.818)(.111111,-.305556){9}{\line(0,-1){.305556}}
\multiput(106.5,209.25)(.113636,-.295455){11}{\line(0,-1){.295455}}
\multiput(408.802,207.568)(.113636,-.295455){11}{\line(0,-1){.295455}}
\multiput(45,99.5)(.91667,-.08333){3}{\line(1,0){.91667}}
\multiput(347.302,97.818)(.91667,-.08333){3}{\line(1,0){.91667}}
\multiput(45.75,102.25)(.6,-.1){5}{\line(1,0){.6}}
\multiput(348.052,100.568)(.6,-.1){5}{\line(1,0){.6}}
\multiput(26.25,67)(.1176471,-.1470588){17}{\line(0,-1){.1470588}}
\multiput(28.75,68.25)(.113636,-.159091){11}{\line(0,-1){.159091}}
\multiput(331.75,68.25)(.113636,-.159091){11}{\line(0,-1){.159091}}
\multiput(331.052,66.568)(.113636,-.159091){11}{\line(0,-1){.159091}}
\multiput(166.75,225.75)(.1,-.5){5}{\line(0,-1){.5}}
\multiput(469.052,224.068)(.1,-.5){5}{\line(0,-1){.5}}
\multiput(169.25,226)(.08333,-.58333){3}{\line(0,-1){.58333}}
\multiput(471.552,224.318)(.08333,-.58333){3}{\line(0,-1){.58333}}
\put(134.25,118){\makebox(0,0)[cc]{$T$}}
\put(436.552,116.318){\makebox(0,0)[cc]{$T$}}
\put(52.25,206.5){\makebox(0,0)[cc]{$T_{\alpha_1}=K_{\alpha_1}(T)$}}
\put(374.552,224.818){\makebox(0,0)[cc]{$T_{\alpha_2}=K_{\alpha_2} (T)$}}
\qbezier(128.25,218.75)(76.375,204.75)(54,129.75)
\qbezier(430.75,216.25)(378.875,202.25)(356.5,127.25)
\qbezier(193.25,228.25)(167.5,226.875)(128.75,219)
\qbezier(495.552,226.568)(469.802,225.193)(431.052,217.318)
\multiput(70.5,168.5)(.25,-.111111){9}{\line(1,0){.25}}
\multiput(372.802,166.818)(.25,-.111111){9}{\line(1,0){.25}}
\multiput(71.75,171)(.227273,-.113636){11}{\line(1,0){.227273}}
\multiput(374.052,169.318)(.227273,-.113636){11}{\line(1,0){.227273}}
\qbezier(192,230.75)(21.25,218.875)(2.5,48.5)
\qbezier(3,49)(33.5,60)(54,129)
\qbezier(305.5,46.5)(336,57.5)(356.5,126.5)
\qbezier(305.5,46)(364.25,115)(380,218)
\end{picture}}}
\vspace{-1cm}\caption{$\alpha$- extrapolation of a production set $T$ with $ \alpha_1<1$ and $\alpha_2>1$.}\label{lambdartsfig2}
\end{figure}
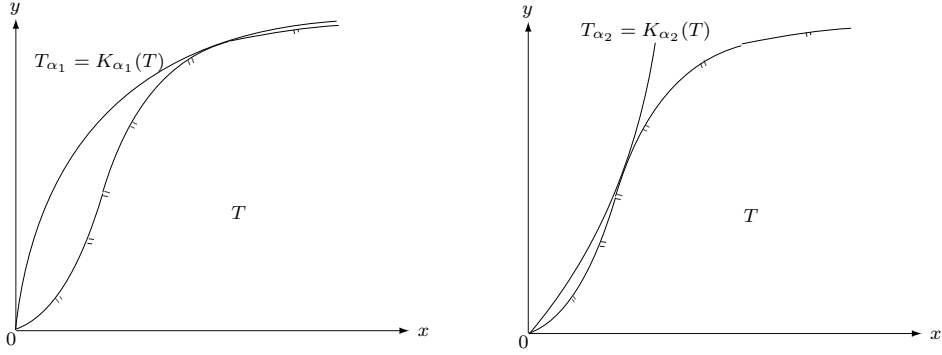

Figure \ref{lambdartsfig2} illustrates Lemma \ref{lem32} where $T$ is contained by a technology $T_\alpha$ satisfying an $\alpha$-RTS assumption. Depending on the value of $\alpha$, $T_\alpha$ can be either convex or non-convex. With $\alpha_1<1$, the production technology $T_{\alpha_1}$ is a convex set demonstrating a strictly decreasing RTS. Besides, with $\alpha_2>1$, the technology $T_{\alpha_2}$ is a non-convex set showing a strictly increasing RTS.\\

The result in Lemma \ref{lem32} extends the construction proposed in Section \ref{DefAlpha} to the case of a general technology. In particular, note that for each production vector $(x_k,y_k)$ we have $Q_{\alpha,1}=K_\alpha\big(S(x_k,y_k)\big)$ where $S(x_k,y_k)=\{(x,y)\in \Real_{+}^{n+p}: x\geq x_k,y\leq y_k\}$. Also notice that we do exclude the situation where $K_\alpha(T)=\Real_+^{n+p}.$ Remark that a specific situation arises when a technology satisfies a $\alpha$-returns to  scale assumption and when we consider its $\beta$-extrapolation with $\alpha\not=\beta. $

\begin{lem} \label{compatibility} Suppose that $T_\alpha$ is a production set satisfying an $\alpha$-returns to scale assumption. Suppose that $\beta>0$, is a positive real number with $\beta\not=\alpha$. Then, either $K_\beta\big(T_\alpha\big)$ fails to satisfy  $T2$ or $y=0$, for all $(x,y)\in T_\alpha.$
\end{lem}

\noindent See proof in Appendix 2.\\[-0.2cm]

The next result is an immediate consequence of Lemma \ref{lem32}. 

\begin{prop}\label{prop32} Let $T$ be a production  set and suppose that the returns-to-scale of $T$ are $\Lambda$-bounded, where $\Lambda$ is a subset of $[0,+\infty]$. For all $\alpha\in \Lambda$, let us denote $T_\alpha:=K_\alpha(T)$. Then,  $$T_\Lambda=\bigcap_{\alpha\in \Lambda}T_\alpha$$
  is the smallest production set satisfying a $\Lambda$-returns to scale assumption that contains $T$, i.e. if $T\subset S$ and $S$ satisfies a $\Lambda$-returns to scale assumption, then $S\supset T_\Lambda$.
\end{prop}

\noindent See proof in Appendix 2.\\[-0.2cm]

\begin{figure}[htpb!]
\begin{minipage}{9cm}
\centering {\scriptsize \unitlength 0.3mm 
\linethickness{0.4pt}
\resizebox{.8\textwidth}{!}{
\ifx\plotpoint\undefined\newsavebox{\plotpoint}\fi 
\unitlength 0.3mm 
\linethickness{0.4pt}
\ifx\plotpoint\undefined\newsavebox{\plotpoint}\fi 

\ifx\plotpoint\undefined\newsavebox{\plotpoint}\fi 
\begin{picture}(243.25,238)(0,0)
\put(3,48){\vector(1,0){232}}
\put(2.75,48.25){\vector(0,1){184}}
\qbezier(69.25,226.75)(68,110.5)(2.75,48.25)
\put(243.25,47.25){\makebox(0,0)[cc]{$x$}}
\put(2.25,238){\makebox(0,0)[cc]{$y$}}
\put(0,43){\makebox(0,0)[cc]{$0$}}
\multiput(54,127.75)(.1413043,-.0326087){23}{\line(1,0){.1413043}}
\multiput(55,130.25)(.1083333,-.0333333){30}{\line(1,0){.1083333}}
\multiput(14.75,1)(.05,-.0333333){30}{\line(1,0){.05}}
\multiput(16,3)(.0652174,-.0326087){23}{\line(1,0){.0652174}}
\multiput(104.5,207.5)(.0333333,-.0916667){30}{\line(0,-1){.0916667}}
\multiput(106.5,209.25)(.03289474,-.08552632){38}{\line(0,-1){.08552632}}
\multiput(45,99.5)(.34375,-.03125){8}{\line(1,0){.34375}}
\multiput(45.75,102.25)(.2,-.0333333){15}{\line(1,0){.2}}
\multiput(26.25,67)(.03333333,-.04166667){60}{\line(0,-1){.04166667}}
\multiput(28.75,68.25)(.03289474,-.04605263){38}{\line(0,-1){.04605263}}
\multiput(166.75,225.75)(.0333333,-.1666667){15}{\line(0,-1){.1666667}}
\multiput(169.25,226)(.03125,-.21875){8}{\line(0,-1){.21875}}
\put(134.25,118){\makebox(0,0)[cc]{$T$}}
\put(52.25,186.5){\makebox(0,0)[cc]{$T_\Lambda$}}
\qbezier(128.25,218.75)(76.375,204.75)(54,129.75)
\qbezier(193.25,228.25)(167.5,226.875)(128.75,219)
\multiput(70.5,168.5)(.075,-.0333333){30}{\line(1,0){.075}}
\multiput(71.75,171)(.06578947,-.03289474){38}{\line(1,0){.06578947}}
\qbezier(192,230.75)(21.25,218.875)(2.5,48.5)
\qbezier(3,49)(33.5,60)(54,129)
\end{picture}}}
\vspace{-1cm}
\caption{$\Lambda$-returns to scale extrapolation of a production set $T$ with $\{\alpha_1,\alpha_2\}$.}\label{lambdartsfig3}
\end{minipage}
\hfill
\begin{minipage}{7cm}
\linespread{1.5}\selectfont
Figure \ref{lambdartsfig3} illustrates Proposition \ref{prop32}. This figure is the combination of the two possibilities in Figure \ref{lambdartsfig2} where $T$ can be contained by either $T_{\alpha_1}$ with $\alpha_1<1$ or $T_{\alpha_2}$ where $\alpha_2>1$. Resulting from Proposition \ref{prop32}, $T \subset T_\Lambda = T_{\alpha_1} \cap T_{\alpha_2}$.
\end{minipage}
\end{figure}

\noindent The existence of a minimal intersecting technology allows to state that it is always possible to associate any $\Lambda$-bounded production set with its minimal extrapolation. For any production set $T \in \Real^{n+p}$, the production process $T_\Lambda$ is a {\bf $\Lambda$-minimal extrapolation} of $T$ if $T_\Lambda$ is the smallest production set satisfying a $\Lambda$-returns to scale assumption that contains $T$. It follows that if $T$ satisfies a $\Lambda$-returns to scale assumption, then:
\begin{equation}
T=T_\Lambda. 
\end{equation}

\begin{prop}\label{interproperty}  Let $\Lambda$ and $\Lambda'$  be two subsets of $[0,+\infty]$. Suppose that $T$ is a  $\Lambda$-bounded and $\Lambda'$-bounded production set. Then the $\Lambda$ and $\Lambda'$ minimum extrapolation of $T$ satisfy the following properties:  
\begin{itemize}
\item[  $(i)$] $T_{\Lambda}\subset T_{\Lambda'} $ if $\Lambda\subset \Lambda'$,
\item[  $(ii)$] $T_{\Lambda }\cap T_{\Lambda' }=T_{\Lambda\cup \Lambda'}$,
\item[  $(ii)$] $T_{\Lambda }\cup T_{\Lambda' }\subset T_{\Lambda\cap \Lambda'} $ if $\Lambda\cap  \Lambda'\not=\emptyset$.
\end{itemize}
\end{prop}

\noindent See proof in Appendix 2.\\[-0.2cm]

In the following, we show that if $\Lambda$ contains some $\alpha>0$, then the $\Lambda$-minimal extrapolation of $T$ satisfies $T1$, $T2$ and $T4$ independently of $T$. This condition avoids the shortcomings of the cases $\alpha=0$ and $\alpha=\infty.$ An additional condition is required for the closedness of $T.$

\noindent Proposition \ref{interproperty} shows that the larger is the set $\Lambda$ and the more vague is the global returns-to-scale structure of the production set. Remark that if $\Lambda$ is a singleton then the technology satisfies an $\alpha$-returns to scale assumption. Doing so, we introduce an alternative formulation of the notion of $\alpha$-returns to scale.\\[0.2cm]
A production set $T$ satisfies a :
\begin{itemize}
\item[(i)]{\bf right}-$\alpha$ returns to scale assumption if for all $\lambda\geq 1$, 
\begin{equation}(x,y)\in T\Rightarrow (\lambda x,\lambda^{\alpha}y)\in T. \end{equation}
\item[(ii)] {\bf left}-$\alpha$ returns to scale assumption if for all $\lambda\in ]0,1]$ 
\begin{equation}(x,y)\in T\Rightarrow (\lambda x,\lambda^{\alpha}y)\in T. \end{equation}
\end{itemize}

\noindent Remark that when $\alpha=\infty$ then, $T$ satisfies a {\bf left}-$\infty$ returns to scale assumption if, for all $\lambda\in ]0,1]$,
$(x,y)\in T\Rightarrow (  x,\lambda y)\in T. $ The next proposition connects the alternative formulation of $\alpha$-returns to scale with the notion of $\Lambda$-returns to scale.

\begin{prop}\label{rightleft}
Let $\Lambda$ be a closed interval of $[0,+\infty]$. Let us denote $\alpha_-=\min \{\alpha: \alpha\in \Lambda\}$ and
$\alpha_+=\max \{\alpha: \alpha\in \Lambda\}$. Suppose that   $T$ satisfies a $\Lambda$-returns to scale assumption and that for each $\alpha\in \Lambda$, $T_\alpha:=K_\alpha(T)$ satisfies $T1-T4$. Then, $T$ satisfies a {  right-}$\alpha_-$ returns to scale  assumption and a left-$\alpha_+$ returns to scale assumption.

\end{prop}

\noindent See proof in Appendix 2.\\

The key intuition is that, when $\Lambda$ is an interval, the returns-to-scale are characterized by its lower and upper bounds. Notice that if a production set $T$ satisfies a {  right-$0$ returns to scale assumption} then,
for all $(x,y)\in T$ and all $\lambda \geq 1$, we have $(\lambda x,y)\in T$.  Besides, if $T$ satisfies a { left-$\infty$ returns to scale assumption} then
for all $(x,y)\in T$ and all $\lambda \in [0, 1]$, we have $( x,\lambda y)\in T$.\\

\begin{figure}[htpb!]
\centering
\begin{minipage}{5.5cm}
\centering {\scriptsize \unitlength 0.3mm 
\resizebox{1.15\textwidth}{!}{%
\linethickness{0.4pt}
\ifx\plotpoint\undefined\newsavebox{\plotpoint}\fi 
\begin{picture}(243.25,208)(0,0)
\put(3,18){\vector(1,0){232}}
\put(2.75,18.25){\vector(0,1){184}}
 \qbezier(69.25,196.75)(68,80.5)(2.75,18.25)
\put(66.75,156){\circle*{1.8}}
 \qbezier(192.75,199)(23.75,193)(2.75,19)
\put(243.25,17.25){\makebox(0,0)[cc]{$x$}}
\put(2.25,208){\makebox(0,0)[cc]{$y$}}
\put(0,13){\makebox(0,0)[cc]{$0$}}
\put(89.25,153){\makebox(0,0)[cc]{$(x_0,y_0) $}}
\multiput(54,97.75)(.1413043,-.0326087){23}{\line(1,0){.1413043}}
\multiput(55,100.25)(.1083333,-.0333333){30}{\line(1,0){.1083333}}
\multiput(31,51.5)(.05,-.0333333){30}{\line(1,0){.05}}
\multiput(32.25,53.5)(.0652174,-.0326087){23}{\line(1,0){.0652174}}
\multiput(104.5,180)(.0333333,-.0916667){30}{\line(0,-1){.0916667}}
\multiput(107.5,181.25)(.03289474,-.08552632){38}{\line(0,-1){.08552632}}
\multiput(65.5,148.25)(.34375,-.03125){8}{\line(1,0){.34375}}
\multiput(66.25,151)(.2,-.0333333){15}{\line(1,0){.2}}
\multiput(71.75,160.75)(.03333333,-.04166667){60}{\line(0,-1){.04166667}}
\multiput(74.25,162)(.03289474,-.04605263){38}{\line(0,-1){.04605263}}
\multiput(166.75,197)(.0333333,-.1666667){15}{\line(0,-1){.1666667}}
\multiput(169.25,197.25)(.03125,-.21875){8}{\line(0,-1){.21875}}

\put(100.75,124.75){\makebox(0,0)[cc]{$(x,y)$}}
\put(110.5,112){\circle*{1.8}}
\qbezier(151,130.75)(122.375,124)(110.25,112.25)
\put(177,120.25){\makebox(0,0)[cc]
{$(\lambda x, \lambda^{\min\{\alpha_1,\alpha_2\}}y)$}}
\put(151.5,130.5){\circle*{1.8}}
\qbezier(100.25,70.5)(109,85.375)(109.75,111.75)
\put(109.5,60.75){\makebox(0,0)[cc]{$(\mu x, \mu^{\max\{\alpha_1,\alpha_2\}}y)$}}
\put(100,70.5){\circle*{1.8}}
\end{picture}}}
\vspace{-1cm}\caption{Left and Right $\alpha$-returns to scale.}\label{leftrightalphafig}
\end{minipage}
\hspace{3cm}
\begin{minipage}{5.5cm}
\centering{\scriptsize
\resizebox{1.15\textwidth}{!}{%
\unitlength 0.4mm 
\linethickness{0.4pt}
\ifx\plotpoint\undefined\newsavebox{\plotpoint}\fi 
\begin{picture}(177.5,156.25)(0,0)
\put(4.75,17.5){\vector(0,1){129}}
\put(0,11.5){\makebox(0,0)[cc]{$0$}}
\put(3.25,156.25){\makebox(0,0)[cc]{$y$}}
\put(177.5,17.75){\makebox(0,0)[cc]{$x$}}
\bezier{2059}(4.75,17.75)(47.25,50.5)(65.75,82.25)
\bezier{2504}(65.75,82.25)(83,111.625)(158.25,121.5)
\put(91.55,82.75){\line(1,0){65}}
\put(91.225,81.775){\line(0,-1){48.1}}
\put(124.7,83.075){\circle*{1.838}}
\put(91.225,57.725){\circle*{1.95}}
\put(84,90){\makebox(0,0)[cc]{$(x,y)$}}
\put(123.25,90){\makebox(0,0)[cc]{$(\lambda x,y)$}}
\put(78,58.5){\makebox(0,0)[cc]{$( x,\lambda y)$}}
\put(91,82.75){\circle*{1.803}}
\put(50.5,30.25){\makebox(0,0)[cc]{$T$}}
\put(4.75,17.5){\vector(1,0){168.5}}
\end{picture}}}
\vspace{-1cm}
\caption{Weak disposability assumption and $\alpha$-returns to scale.}\label{freedispofig}
\end{minipage}
\end{figure}

Figures \ref{leftrightalphafig} and \ref{freedispofig} describe the notions of right- and left-$\alpha$ RTS. Specifically, Figure \ref{leftrightalphafig} illustrates these notions within Proposition \ref{rightleft} framework by considering right-$\alpha_-$ and left-$\alpha_+$ RTS assumptions in general cases. Besides, Figure \ref{freedispofig} illustrates the limit cases of the left-$\alpha_+$ RTS assumption with $\alpha_+ = \infty$ and of the right-$\alpha_-$ RTS assumption where $\alpha_-=0$.

\subsection{From $\Lambda$-returns to scale to Non Increasing, Non Decreasing and Variable returns-to-scale}

In this subsection, we show that traditional convex and non-convex models involving a returns-to-scale assumption, follow a special case of the $\Lambda$-returns to scale assumption.\\

For a given set of production units $A= \left\{ \left( x_{1},y_{1}\right) ,...,\left(
x_{J},y_{J}\right) \right\} \subset \mathbb{R}_{+}^{n+p}$, each individual production possibility set is based upon (i) a single production unit $(x_k,y_k )$ with $k \in \mathcal{J}$, (ii) the strong disposability assumption and (iii) some hypotheses of returns-to-scale. Notice that some of these returns-to-scale assumptions are namely Constant (CRS), Non Increasing (NIRS), Non Decreasing (NDRS) and Variable (VRS) returns-to-scale. Starting from the notation introduced in  Eq.\eqref{individual}, let us denote the individual production set as:
\begin{equation}\label{qindiv}
Q_{1,1}^\Gamma(x_k,y_k)=\big \{(x,y)\in \Real_+^n \times \Real
_+^p: x\geq \mu x_k, y\leq \mu y_k, \tau \in \Gamma\big \},
\end{equation}
where $\Gamma \in \{\Gamma_{CRS},\Gamma_{NDRS},\Gamma_{NIRS},\Gamma_{VRS}\}$, with: $ (\text{i})\;$  $ \Gamma_{CRS}  = \left\{\mu :\; \mu  \ge 0 \right\}$;
 $(\text{ii})$  $ \Gamma_{NDRS}  = \left\{ \mu :\; \mu  \ge 1 \right\}$;
 $(\text{iii})$  $\Gamma_{NIRS}  = \left\{ \mu :\; 0 \le \mu  \le 1 \right\}$;
 $(\text{iv})$ $  \Gamma_{VRS}  = \left\{ \mu :\; \mu  = 1 \right\}.$\\

\noindent Remark that $\Gamma_{\bullet}$ denotes the returns-to-scale assumption whereas $Q_{1,1}^\Gamma(x_k,y_k)$ means that $\gamma=\delta=1$.

Union and convex union of these individual production possibility sets yield non convex (NC) technologies on the one hand and traditional convex (C) possibility sets on the other hand, as follows:
\begin{equation}\label{UnionTechC&NC}
T^{ {NC} ,\Gamma} = \bigcup_{k\in \mathcal J} Q_{1,1}^\Gamma(x_k,y_k) \quad \text{ and } \quad T^{ {C} ,\Gamma} = Co \Big( \bigcup_{k\in \mathcal J} Q_{1,1}^\Gamma(x_k,y_k) \Big),
\end{equation}
where $Co$ is the convex hull operator.\bigskip

\noindent Regarding the convex case, we retrieve  the standard DEA model initiated by Charnes et al. (1978) and Banker et al. (1984). Besides, the non-convex case provides the model proposed by Deprins et al. (1984) and Tulkens (1993).\\

Additionally to the returns-to-scale assumption, convexity constraints can be added to the characterization of the production technology. Consider the following notations:
\begin{equation}
\Theta_{NC} = \left\{ \sum_{k \in \mathcal K} z_k = 1, \; z_k \in \{ 0,1\} \right\} \quad \text{ and } \quad \Theta_{C} = \left\{ \sum_{k \in \mathcal K} z_k = 1, \; z_k \ge 0 \right\} .\end{equation}

\noindent An unified algebraic representation of convex and non convex technologies under different
returns-to-scale assumptions for a sample of $J$ observations is as follows (Briec et al., 2004):
\begin{equation}\label{TechC&NC}
T^{\Theta ,\Gamma} = \left\{ (x,y)\in \mathbb \Real_+^n \times \Real_+^p :  (x,-y) \ge \sum_{k\in \mathcal J} \tau z_k (x_k,-y_k) ,\; z_k \in \Theta,\; \tau \in \Gamma \right\},
\end{equation}
\noindent where $\Theta\in \{\Theta_{NC},\Theta_C\}$. Notice that ($z$) is an activity vector related to either a convexity ($C$) or a non convexity
($NC$) constraint. Moreover, ($\tau $) is a scaling parameter allowing the particular scaling of
all $J$ observations involved in the technology. This scaling parameter is positive and smaller than or equal to 1 under $NIRS$, larger than or equal to 1 under $NDRS$, fixed at unity under $VRS$ and, free under $CRS$ assumptions.\\
However, there is a shortcoming with this formalism. It does not include $(0,0)$ and therefore, does not take into account inaction in the $VRS$ and $NDRS$ cases. To circumvent this problem we slightly modify the above definition (Eq. \ref{TechC&NC}) by introducing the sets
\begin{equation}\label{TechC&NCINAC}
\widetilde T^{\Theta ,\Gamma} =T^{\Theta ,\Gamma}\cup S(0,0)=T^{\Theta ,\Gamma}\cup \Real^n\times \{0\}.
\end{equation}
Clearly if either $\Gamma=\Gamma_{NIRS}$ or  $\Gamma=\Gamma_{CRS}$ then $\widetilde T ^{\Theta ,\Gamma} =T^{\Theta ,\Gamma}$.

The next statement is an immediate consequence of our earlier results and shows that the standard DEA convex models satisfy special cases of the $\Lambda$-returns to scale assumption.

\begin{prop}\label{lambdagammaIntersect}
For any subset $A=\big\{  ( x_{1},y_{1} ) ,..., (
x_{J},y_{J} ) \big\} $ of $ \mathbb{R}_{+}^{n+p}$ and   any $\Theta\in \{\Theta_{NC},\Theta_{C}\}$ we have the following properties:\vspace{-0.4cm}
\begin{itemize} \setlength\itemsep{0em}
\item[$(i)$] If $\Gamma=\Gamma_{CRS}$ then $T^{\Theta, \Gamma}=  T_{\{1\}}^{\Theta, VRS}$ and satisfies a $\{1\}$-returns to scale assumption.
\item[$(ii)$]If $\Gamma=\Gamma_{NIRS}$ then $T^{\Theta, \Gamma}=\bigcap\limits_{\alpha\in  [0,1]}  T_\alpha^{\Theta, \Gamma_{VRS}}$ and satisfies a $[0,1]$-returns to scale assumption.
\item[$(iii)$]If $\Gamma=\Gamma_{NDRS}$ then $\widetilde  T ^{\Theta, \Gamma}=\bigcap\limits_{\alpha\in  [ 1, +\infty]}  T_\alpha^{\Theta, \Gamma_{VRS}}$ satisfies a $[1,+\infty]$-returns to scale assumption.
\item[$(iv)$]If $\Gamma=\Gamma_{VRS}$ then $\widetilde T^{\Theta, \Gamma}=\bigcap\limits_{\alpha\in [0,+\infty]}T_\alpha^{\Theta, \Gamma_{VRS}}$ and satisfies a $[0,+\infty]$-returns to scale assumption.
\end{itemize}
\end{prop}

\noindent See proof in Appendix 2.\\

\noindent Along this line, the next proposition shows that $CRS$, $NIRS$ and $NDRS$ production technologies can be derived from an extrapolation of the $\Lambda$-returns to scale assumption with the $VRS$ model.

\begin{prop} \label{lambdavrsprop}
For any subset $A=\big\{  ( x_{1},y_{1} ,..., (
x_{J},y_{J} ) \big\} $ of $ \mathbb{R}_{+}^{n+p}$ and   any $\Theta\in \{\Theta_{NC},\Theta_{C}\}$ we have: 
\begin{itemize} \setlength\itemsep{0em}
\item[$(i)$] $T^{\Theta, \Gamma_{NIRS}}=T^{\Theta, \Gamma_{VRS}}_{[0,1]}$; 
\item[$(ii)$] $\widetilde T^{\Theta, \Gamma_{NDRS}}=T^{\Theta, \Gamma_{VRS}}_{[1,+\infty]}$; 
\item[$(iii)$] $T^{\Theta, \Gamma_{CRS}}=T^{\Theta, \Gamma_{VRS}}_{\{1\}}$.
\end{itemize} 
\end{prop}\bigskip

\noindent

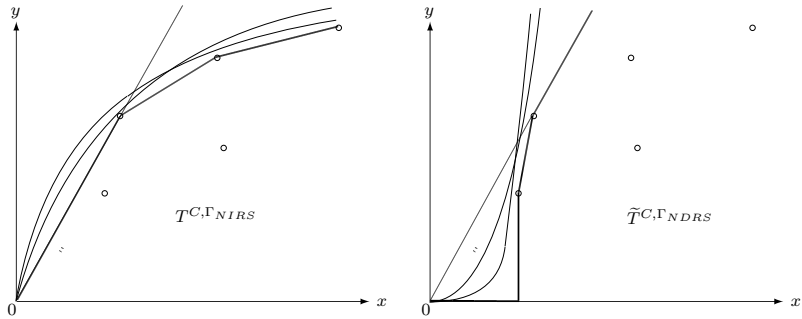
\begin{figure}[htpb!]
\begin{minipage}{10.5cm}
\centering {\scriptsize \unitlength 0.3mm 
\linethickness{0.4pt}
\resizebox{1\textwidth}{!}{
\ifx\plotpoint\undefined\newsavebox{\plotpoint}\fi 
\unitlength 0.25mm 
\linethickness{0.4pt}
\ifx\plotpoint\undefined\newsavebox{\plotpoint}\fi 
\begin{picture}(515.28,199.769)(0,0)
\put(3,5){\vector(1,0){232}}
\put(275.03,5){\vector(1,0){232}}
\put(2.75,5.25){\vector(0,1){184}}
\put(274.78,5.25){\vector(0,1){184}}
\put(243.25,4.25){\makebox(0,0)[cc]{$x$}}
\put(515.28,4.25){\makebox(0,0)[cc]{$x$}}
\put(2.25,195){\makebox(0,0)[cc]{$y$}}
\put(274.28,195){\makebox(0,0)[cc]{$y$}}
\put(0,0){\makebox(0,0)[cc]{$0$}}
\put(272.03,0){\makebox(0,0)[cc]{$0$}}
\multiput(31,38.5)(.166667,-.111111){9}{\line(1,0){.166667}}
\multiput(303.03,38.5)(.166667,-.111111){9}{\line(1,0){.166667}}
\multiput(32.25,40.5)(.214286,-.107143){7}{\line(1,0){.214286}}
\multiput(304.28,40.5)(.214286,-.107143){7}{\line(1,0){.214286}}
\put(70.975,127.032){\circle*{3.568}}
\put(343.005,127.032){\circle*{3.568}}
\put(60.884,76.158){\circle*{3.568}}
\put(332.914,76.158){\circle*{3.568}}
\put(134.883,165.293){\circle*{3.568}}
\put(406.913,165.293){\circle*{3.568}}
\put(214.768,185.054){\circle*{3.568}}
\put(486.798,185.054){\circle*{3.568}}
\put(139.088,106.009){\circle*{3.568}}
\put(411.118,106.009){\circle*{3.568}}
\thicklines
\multiput(214.348,185.895)(-.47518111,-.119417242){169}{\line(-1,0){.47518111}}
\multiput(134.042,165.713)(-.1952545856,-.1197561458){323}{\line(-1,0){.1952545856}}
\multiput(70.975,127.032)(-.1198091353,-.2143169625){565}{\line(0,-1){.2143169625}}
\multiput(381.266,196.406)(-.1199438138,-.2128035407){326}{\line(0,-1){.2128035407}}
\multiput(342.164,127.032)(-.11858796,-.6468434){78}{\line(0,-1){.6468434}}
\put(332.914,76.578){\line(0,-1){71.056}}
\thinlines
\qbezier(2.862,5.943)(47.22,169.707)(210.143,198.088)
\qbezier(274.472,4.681)(324.715,5.522)(347.209,198.088)
\qbezier(324.085,40.84)(319.46,5.102)(274.472,4.681)
\qbezier(324.085,41.26)(335.016,131.236)(340.903,194.304)
\multiput(274.892,5.522)(.1198051344,.2177590179){558}{\line(0,1){.2177590179}}
\qbezier(76.02,140.066)(123.531,175.594)(213.086,190.099)
\multiput(3.283,6.363)(.11992960988,.21300239593){908}{\line(0,1){.21300239593}}
\thicklines
\multiput(274.892,5.522)(14.61058,-.10511){4}{\line(1,0){14.61058}}
\put(134.463,61.862){\makebox(0,0)[cc]{$T^{C, \Gamma_{NIRS}}$}}
\put(432.981,61.442){\makebox(0,0)[cc]{$\widetilde{T}^{C, \Gamma_{NDRS}}$}}
\thinlines
\qbezier(2.442,5.102)(18.84,99.073)(75.601,140.067)
\end{picture}}}
\caption{$\Lambda$-returns Extrapolation of the Convex NIRS and NDRS Models.}\label{ExtrapDEANIRSNDRS}
\end{minipage}
\hfill
\begin{minipage}{5cm}
\linespread{1.5}\selectfont
Figure \ref{ExtrapDEANIRSNDRS} illustrates statements $(i)$ and $(ii)$ of Proposition \ref{lambdavrsprop} with $\Theta=\Theta_C$ for NIRS and NDRS assumptions. Remark that the production frontiers are piecewise linear showing the characteristics of the RTS assumptions.
\end{minipage}
\end{figure}

\begin{figure}[htpb!]
\begin{minipage}{10.5cm}
\centering {\scriptsize \unitlength 0.3mm 
\linethickness{0.4pt}
\resizebox{1\textwidth}{!}{
\ifx\plotpoint\undefined\newsavebox{\plotpoint}\fi 

\unitlength 0.25mm 
\linethickness{0.4pt}
\ifx\plotpoint\undefined\newsavebox{\plotpoint}\fi 

\begin{picture}(515.28,201.861)(0,0)
\put(3,5){\vector(1,0){232}}
\put(275.03,5){\vector(1,0){232}}
\put(2.75,5.25){\vector(0,1){184}}
\put(274.78,5.25){\vector(0,1){184}}
\put(243.25,4.25){\makebox(0,0)[cc]{$x$}}
\put(515.28,4.25){\makebox(0,0)[cc]{$x$}}
\put(2.25,195){\makebox(0,0)[cc]{$y$}}
\put(274.28,195){\makebox(0,0)[cc]{$y$}}
\put(0,0){\makebox(0,0)[cc]{$0$}}
\put(272.03,0){\makebox(0,0)[cc]{$0$}}
\multiput(31,38.5)(.166667,-.111111){9}{\line(1,0){.166667}}
\multiput(303.03,38.5)(.166667,-.111111){9}{\line(1,0){.166667}}
\put(-51.75,-51.25){}
\multiput(32.25,40.5)(.214286,-.107143){7}{\line(1,0){.214286}}
\multiput(304.28,40.5)(.214286,-.107143){7}{\line(1,0){.214286}}
\put(70.975,127.032){\circle*{3.568}}
\put(343.005,127.032){\circle*{3.568}}
\put(60.884,76.158){\circle*{3.568}}
\put(332.914,76.158){\circle*{3.568}}
\put(134.883,165.293){\circle*{3.568}}
\put(406.913,165.293){\circle*{3.568}}
\put(214.768,185.054){\circle*{3.568}}
\put(486.798,185.054){\circle*{3.568}}
\put(139.088,106.009){\circle*{3.568}}
\put(411.118,106.009){\circle*{3.568}}
\thicklines
\multiput(485.538,185.895)(-.475177515,-.119420118){169}{\line(-1,0){.475177515}}
\multiput(405.233,165.713)(-.19525387,-.119755418){323}{\line(-1,0){.19525387}}
\multiput(381.266,196.406)(-.1199447853,-.212803681){326}{\line(0,-1){.212803681}}
\multiput(342.164,127.032)(-.11858974,-.64684615){78}{\line(0,-1){.64684615}}
\put(332.914,76.578){\line(0,-1){71.056}}
\thinlines
\qbezier(274.894,5.102)(319.251,168.866)(482.175,197.247)
\qbezier(274.472,4.681)(324.715,5.522)(347.209,198.088)
\qbezier(324.085,40.84)(319.46,5.102)(274.472,4.681)
\qbezier(324.085,41.26)(335.016,131.236)(340.903,194.304)
\multiput(274.892,5.522)(.1198064516,.2177598566){558}{\line(0,1){.2177598566}}
\qbezier(348.052,139.225)(395.563,174.753)(485.118,189.258)
\thicklines
\multiput(274.892,5.522)(14.61075,-.105){4}{\line(1,0){14.61075}}
\put(134.463,61.862){\makebox(0,0)[cc]{$T^{C, \Gamma_{CRS}}$}}
\put(432.981,61.442){\makebox(0,0)[cc]{$\widetilde{T}^{C, \Gamma_{VRS}}$}}
\thinlines
\qbezier(274.474,4.261)(290.872,98.232)(347.632,139.226)
\thicklines
\multiput(2.488,4.67)(.11993184563,.21550971494){915}{\line(0,1){.21550971494}}
\end{picture}
}}
\caption{$\Lambda$-returns Extrapolation of the Convex CRS and VRS Models.}\label{ExtrapDEACRSVRS}
\end{minipage}
\hfill
\begin{minipage}{5cm}
\linespread{1.5}\selectfont
Figure \ref{ExtrapDEACRSVRS} illustrates statements $(i)$ and $(iv)$ of Proposition \ref{lambdagammaIntersect} with $\Theta=\Theta_C$. The production frontiers are piecewise linear and highlight the specificity of the CRS and the VRS assumptions. 
\end{minipage}
\end{figure}
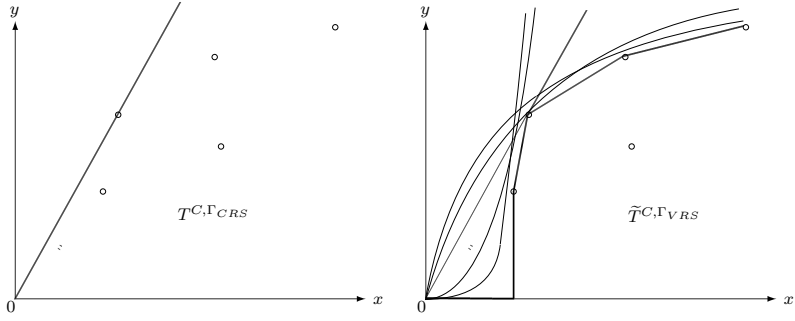

\bigskip

\noindent Since the $VRS$ model satisfies a $[0,\infty]$-returns to scale assumption, it is both the more general and the less informative about the returns-to-scale structure of the production set. Interestingly, union and intersection on $\Lambda$ allow to relate the $NIRS$, $NDRS$ and $CRS$ models. However, these operations are intrinsically derived from  the specific nature of the returns-to-scale. Note that although $\Gamma_{CRS}=\Real_+$
, $T^{\Theta, CRS}$ satisfies a $\{1\}$-returns to scale assumption. Conversely, although $\Gamma_{VRS}=\{1\}$ , $T^{\Theta, VRS}$ satisfies a $[0,\infty]$-returns to scale assumption (see Eq \eqref{qindiv} and Proposition \ref{lambdagammaIntersect}). 

\section{ From individual $\alpha$-returns to scale to global $\Lambda$-returns to scale assumption and technology}

This section presents the minimum extrapolation principle that is used to define the production technology of the data set. Moreover, a procedure is proposed to assess the individual optimal value of $\alpha$ under a generalized FDH technology and through an input-oriented model. This approach fully endogenizes the assessment of $\alpha$ through linear programming. From these individual $\alpha$, the global optimal value of $\Lambda$ is derived allowing to characterize the returns-to-scale of the whole production set. 



\subsection{$\Lambda$-Returns to scale and minimum Extrapolation}

In this section, we introduce the principle of minimum extrapolation allowing to define the notion of rationalized technology.\\

The next proposition presents the minimum extrapolation principle for the individual technologies $Q_{\gamma,\delta}(x_k,y_k)$ and their union $T_{\gamma,\delta}$.
\begin{prop} \label{indextrap}For all $k\in \mathcal J$ and a data set $A=\{(x_k,y_k):k\in \mathcal J\}$, 
\begin{itemize} \setlength\itemsep{0em}
\item[$(i)$] $Q_{\alpha,1} (x_k,y_k)$ is the smallest technology containing $(x_k,y_k)$  that satisfies $T1-T4$ and an $\alpha$-returns to scale assumption.
\item[$(ii)$] $T_{\alpha,1}=\bigcup\limits_{k\in \mathcal J}Q_{\alpha,1}(x_k,y_k)$
 is the smallest technology containing $A$ that satisfies $T1-T4$ and an $\alpha$-returns to scale assumption.
\end{itemize}
\end{prop}

From Proposition \ref{indextrap}, we can now introduce the lower, upper and minimal extrapolation technologies for each observation as follows:
\begin{enumerate}[label=(\roman*)]
\item The {\bf lower individual minimal extrapolation} is denoted
\begin{equation}
  T_{-}^{\star}(x_j,y_j)=\bigcup_{k\in \mathcal J} Q_{\alpha_{-}^\star(j),1}(x_k,y_k).
\end{equation}
\item The {\bf upper individual minimal extrapolation} is defined as:
\begin{equation}
  T_{+}^{\star}(x_j,y_j)=\bigcup_{k\in \mathcal J} Q_{\alpha_{+}^\star(j),1}(x_k,y_k).
\end{equation}
\item The {\bf individual minimal extrapolation} is obtained from the union over $ \Lambda^\star (x_j,y_j)$:
\begin{equation}
T^\star(x_j,y_j)=\bigcap_{\alpha\in \Lambda^\star (j) }\bigcup_{k\in \mathcal J}Q_{\alpha,1}(x_k,y_k).
\end{equation}
\end{enumerate}

\noindent Remark that for the sake of simplicity, we note $\alpha^\star(x_j,y_j)=\alpha^\star(j)$ and $\Lambda^\star(x_j,y_j)=\Lambda^\star(j)$ the $\alpha$-returns to scale and $\Lambda$-returns to scale related to the observation $j\in \mathcal{J}$.

\noindent The two first aforementioned assertions mean that the lower (i) and the upper (ii) individual minimal extrapolations are provided by the union of individual technologies subjected to respectively the lower ($\alpha^\star_-$) and the upper ($\alpha^\star_+$) bounds of $\Lambda^\star(j)=[\alpha^\star_-, \alpha^\star_+]$. The next section introduces these notions of upper and lower bounds.  The third statement means that the individual minimal extrapolation technology is the union of individual technologies and is $\Lambda^\star(j)$-bounded.  \\

\noindent Notice that $Q_{\alpha_-^\star(j),1}(x_k,y_k)$ and $Q_{\alpha_+^\star(j),1}(x_k,y_k)$
 are obtained by replacing $\alpha$ with respectively $\alpha_-^\star(j)$ and $\alpha_+^\star(j)$ in $Q_{\alpha ,1}(x_k,y_k)$, such that :
\begin{align} 
& Q_{\alpha_-^\star(j),1}(x_k,y_k)=\Big\{(x,y)\in \Real_+^{n+p}: x\geq \lambda^{1/\alpha_{-}^\star(j)}
x_k, y\leq  \lambda y_k, \lambda \geq
0\Big\}, \\
& Q_{\alpha_{+}^\star(j),1}(x_k,y_k)=\Big\{(x,y)\in \Real_+^{n+p}: x\geq \lambda^{1/\alpha_{+}^\star(j)}
x_k, y\leq  \lambda y_k, \lambda \geq
0\Big\}.
\end{align}

From the individual scheme, we can deduce the global production possibility set as follows:
\begin{enumerate}[label=(\roman*)]\setcounter{enumi}{3}
\item The {\bf lower minimal extrapolation} of the technology is then defined as:
\begin{equation}T_{-}=\bigcap_{j\in \mathcal J}T_{-}^{\star}(x_j,y_j)=\bigcap_{j\in \mathcal J}\bigcup_{k\in \mathcal J} Q_{\alpha_{-}^\star(j),1}(x_k,y_k).
\end{equation}
\item The {\bf upper minimal extrapolation} of the technology is:
\begin{equation}T_{+}=\bigcap_{j\in \mathcal J} T_{+}^{\star}(x_j,y_j)=\bigcap_{j\in \mathcal J}\bigcup_{k\in \mathcal J} Q_{\alpha_{+}^\star(j),1}(x_k,y_k).
\end{equation}
\item The {\bf global minimal extrapolation} of the technology is similarly defined as:
\begin{equation}T=\bigcap_{j\in \mathcal J}T^\star(x_j,y_j)=\bigcap_{j\in \mathcal J} \bigcap_{\alpha\in \Lambda^\star (j) }\bigcup_{k\in \mathcal J}Q_{\alpha,1}(x_k,y_k).
\end{equation}
\end{enumerate}
 
\noindent Statements (iv)-(vi) mean that the global technology involving all units of the set of observations $A$ is the intersection of all individual minimal extrapolation technologies.

\begin{prop} \label{globaltechlambda}
Let  $A=\{(x_j,y_j)\}_{j\in \mathcal J}$ and let us denote
 $\Lambda^\star=\bigcup_{j\in \mathcal J}\Lambda^\star(x_j,y_j)$. Then,
\begin{itemize} 
\item [$(i)$]  $T$ satisfies a $\Lambda^\star$-returns to scale assumption.
\item[$(ii)$]  $T$ satisfies a right $\alpha_{ -}^\star$-returns to scale assumption and a left $\alpha_{+}^\star$-returns to scale assumption,  with 
 $\quad \alpha_{ -}^\star=\min\{\alpha_{ -}^\star(j):j\in \mathcal J\}\quad \text{and}\quad \alpha_{ +}^\star=\max\{\alpha_{ +}^\star(j):j\in \mathcal J\}.$
\end{itemize}
\end{prop}
\noindent See proofs in Appendix 2. \\

\noindent Proposition \ref{globaltechlambda} means that (i) the returns-to-scale of the global technology is the union of each individual $\Lambda^\star(j)$ which is an interval that reduces to $\alpha^\star(j)$ if it is a singleton. This global returns-to-scale of the global technology is also (ii) upper and lower bounded by $\alpha^\star_+$ and $\alpha^\star_+$, respectively.\\

In the following, we say that a technology $T$ {\bf $\Lambda$-rationalizes} the data set $A$   if $T$ is the smallest technology satisfying a $\Lambda$-returns to scale assumption with $A\subset T$.

\begin{prop}\label{lambdarational}Let  $A=\{(x_j,y_j)\}_{j\in \mathcal J}$ and denote
 $\Lambda^\star=\bigcup_{j\in \mathcal J}\Lambda^\star(x_j,y_j)$ then, $T$ $\Lambda^\star$-rationalizes the data set $A$.
\end{prop}

\noindent See proof in Appendix 2.

\subsection{The non parametric input oriented model}

Through the input oriented model, we look for the individual optimal $\alpha$ related to the observation $(x_j,y_j)$ for any $j\in \mathcal{J}$ and that minimizes the inefficiency of the observed unit. This individual optimal $\alpha$ of $(x_j,y_j)$ is noted $\alpha^\star(x_j,y_j)$ and shortened as $\alpha^\star(j)$. To do so, a goodness of fit index  $L(x_j,y_j; \gamma,\delta)$ minimized by $\alpha^\star(x_j,y_j)$, is defined (Boussemart et al., 2019). Formally, the definition of the goodness of fit index is as follows:
\begin{align}
L(x_j,y_j; \gamma,\delta)& = \min_{k\in \mathcal J} D^I_{k}(x_{j},y_{j}; \gamma,\delta), \notag\\
& = \min\limits_{k\in \mathcal
J}\left(\left[\max\limits_{h\in card (y)}
\frac{y_{j,h}}{y_{k,h} }\right]^{1/\alpha}\cdot\left[\max\limits_{i\in
card(x)} \frac{x_{k,i}}{x_{j,i} }\right]\right).
\end{align}

The goodness of fit index allows to provide the global efficiency measure for each observation given the range of efficiency measures that are related to each of them with respect to the range of individual technologies. Indeed, each individual technology provides an efficiency measure for each observation. This means that for a set of $J$ firms, each firm $k$ has $J$ efficiency measures then, the goodness of fit index allows to identify the optimal efficiency score with respect to $\alpha$.

Basically, the key idea is to consider in a first step process the optimisation of
$\alpha$ only regarding firm $j$. However, for some $\alpha$, all the potential technologies satisfying an $\alpha$-returns to scale assumption should contain the set of all the
observed production units. Notice that we could equivalently consider an approach based upon the output oriented measure $D^O$. This key intuition is depicted in the figure below.\\

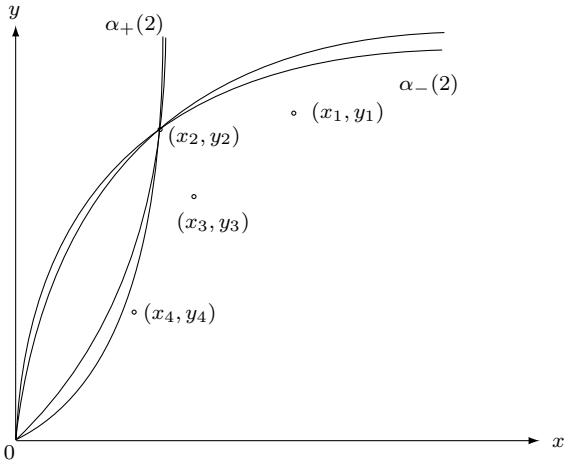
\begin{figure}[htpb!]
\begin{minipage}{8.5cm}
\centering{\scriptsize \unitlength 0.3mm 
\linethickness{0.4pt}
\ifx\plotpoint\undefined\newsavebox{\plotpoint}\fi 
\begin{picture}(243.25,198)(0,0)
\put(3,8){\vector(1,0){232}}
\put(2.75,8.25){\vector(0,1){184}}
\put(55.25,65){\circle*{1.8 }}
\put(81.75,116.25){\circle*{1.8}}
\put(126,153.25){\circle*{1.8}}
 \qbezier(69.25,186.75)(68,70.5)(2.75,8.25)
\put(66.75,146){\circle*{1.8}}
 \qbezier(68,187.25)(67.5,40.125)(3,8.5)
 \qbezier(192.75,189)(23.75,183)(2.75,9)
 \qbezier(191.75,181.25)(10.125,176.5)(3,8.75)
\put(243.25,7.25){\makebox(0,0)[cc]{$x$}}
\put(2.25,198){\makebox(0,0)[cc]{$y$}}
\put(0,3){\makebox(0,0)[cc]{$0$}}
\put(86.25,143){\makebox(0,0)[cc]{$(x_2,y_2) $}}
\put(150,153.5){\makebox(0,0)[cc]{$(x_1,y_1) $}}
\put(90.25,104.5){\makebox(0,0)[cc]{$(x_3,y_3) $}}
\put(75.5,64.25){\makebox(0,0)[cc]{$(x_4,y_4) $}}
\put(55.5,192.5){\makebox(0,0)[cc]{$\alpha_{ +}(2)$}}
\put(186,166){\makebox(0,0)[cc]{$\alpha_{ -}(2)$}}
\end{picture}}
\caption{Individual minimal extrapolation for firm 2} \label{lowsupalpha}
\end{minipage}
\hfill
\begin{minipage}{6cm}
\linespread{1.5}\selectfont
Figure \ref{lowsupalpha} illustrates the lower $\alpha_-(2)$ and the upper $\alpha_+(2)$ bounds of the individual RTS $\Lambda(2)=[\alpha_-(2), \alpha_+(2)]$ of the observation $(x_2,y_2)$. Indeed, $\alpha(2)$ is not a singleton then the individual RTS $\Lambda(2)$ is an interval provided by the union of any $\alpha$-returns to scale related to $(x_2,y_2)$.
\end{minipage}
\end{figure}

\noindent The following program allows to solve the optimization problem :
\begin{align}
\max_\alpha \quad  & \min\limits_{k\in \mathcal
J}\left(\left[\max\limits_{h\in card (y)}
\frac{y_{j,h}}{y_{k,h} }\right]^{1/\alpha}\cdot\left[\max\limits_{i\in
card(x)} \frac{x_{k,i}}{x_{j,i} }\right]\right).
\end{align}
The logarithmic transformation yields:
\begin{align}
\max_\alpha \quad  & \min_{k\in \mathcal
J}\left(\dfrac{1}{\alpha}\ln \left(\left[\max\limits_{h\in card (y)}
\frac{y_{j,h}}{y_{k,h} }\right]\right) + \ln \left(\left[\max\limits_{i\in
card(x)} \frac{x_{k,i}}{x_{j,i} }\right]\right)\right).
\end{align}

\noindent Setting $\beta = \dfrac{1}{\alpha}$,  $f_{j,k}=\ln \left(\left[\max\limits_{h\in card (y)}
\dfrac{y_{j,h}}{y_{k,h} }\right]\right)$, and
$g_{j,k}= \ln \left(\left[\max\limits_{i\in
card(x)} \dfrac{x_{k,i}}{x_{j,i} }\right]\right)$, the program becomes:
\begin{align}
\max_\beta \quad  &  \min_{k\in \mathcal
J}\left(\beta \; f_{j,k}+  g_{j,k}\right),
\end{align}

\noindent and the associated linear program is:
\begin{align}
\begin{array}{lllll}
& \max\limits_{\beta,\lambda} \quad  &  \lambda \nonumber\\
& s.t. & \lambda\leq \left(\beta \; f_{j,k}+ g_{j,k}\right), \quad
k\in \mathcal J \qquad\qquad{(P_j)} .
\end{array}
\end{align}

This linear program has $2$ variables and $ |\mathcal J|$ constraints. Denote $\Lambda^\star(x_j,y_j)$ the set of solutions for the program $(P_j)$ . As $\beta = \dfrac{1}{\alpha}$ then $\alpha^\star(j)=[\beta^\star(j)]^{-1}$.\\

\noindent From $(P_j)$, the derived individual technology related to the observation $(x_j,y_j)$ is then:

\begin{equation}
 T_{\alpha^\star(j),1}(x_j,y_j)=\bigcup_{k\in \mathcal J}Q_{\alpha^\star(j),1}(x_k,y_k)
\end{equation}
where $ Q_{\alpha^\star(j),1}$ is obtained by replacing $\alpha$ with $\alpha^\star(j)$ in $ Q_{\alpha ,1}$ as follows:
\begin{equation} Q_{\alpha^\star(j),1}
(x_k,y_k)=\Big\{(x,y)\in \Real_+^{n+p}: x\geq \lambda^{1/\alpha^\star(j)}
x_k, y\leq  \lambda y_k, \lambda \geq
0\Big\}.
\end{equation}

\noindent A numerical example is proposed in Appendix 2 to illustrate the above notions.\bigskip

Suppose now that there is an infinity of solutions to Program $(P_j)$. Since the optimisation program is linear, the solution set is closed and convex. Therefore, there is an interval
$\Lambda^\star(x_j,y_j)=[\alpha_{-}^\star, \alpha_{+}^\star]$ denoted as $\Lambda^\star(j)$ which contains all the solutions. Let us denote
$\lambda^\star(j)$ the solution in $\lambda$ of $(P_j)$ then, $\alpha_{-}^\star$ and $ \alpha_{+}^\star$ are respectively solutions of the programs below:
\begin{figure}[htpb!]

\begin{minipage}{6cm}
\begin{align}
\begin{array}{lll}
& \max\limits_{\beta}   &  \beta\nonumber\\
& s.t. & \lambda^\star\leq \left(\beta \; f_{j,k}+  g_{j,k}\right), \quad
k\in \mathcal J \qquad (P_j^-).
\end{array}
\end{align}
\end{minipage}
\vrule
\begin{minipage}{6cm}
\begin{align}
\begin{array}{lllll}
&\min\limits_{\beta}   &  \beta\nonumber\\
&s.t.&  \lambda^\star\leq \left(\beta \; f_{j,k}+  g_{j,k}\right), \quad
k\in \mathcal J \qquad (P_j^+) .
\end{array}
\end{align}
\end{minipage}

\end{figure}

\noindent Indeed, as $\beta = 1/\alpha$ then, $\alpha^\star_-$ and $\alpha^\star_+$ are  obtained by maximizing and  minimizing $\beta$, respectively. This means that when the Program $(P_j)$ has an infinity of solutions then, the optimal returns-to-scale of the observation is not a singleton ($\alpha^\star(j))$ but rather an interval $\Lambda^\star(j)$ with a lower ($\alpha^\star_-$) and an upper ($\alpha^\star_+$) bounds.

Note that for all   $j \in \mathcal J$ and all $\alpha\in \Lambda^\star(x_j,y_j)$:
\begin{equation}
D^I(x_j ,y_j ;\alpha_{ -}^\star(j),1)=D^I(x_j ,y_j ;\alpha_{+}^\star(j),1)=D^I(x_j ,y_j ;\alpha,1).
\end{equation}

Remark that $D^I(x_j ,y_j ;\alpha_{ -}^\star(j),1), D^I(x_j ,y_j ;\alpha_{+}^\star(j),1)\quad \text{and} \quad D^I(x_j ,y_j ;\alpha,1)$ are assessed with respect to $ T_{-}^{\star}(x_j,y_j)$, $ T_{+}^{\star}(x_j,y_j)$ and $ T^{\star}(x_j,y_j)$, respectively.\\

Recall that for all $(x,y)\in \Real_+^{n+p}$ we have: $D^I(x  ,y )=\min\{\theta:(\theta x,y)\in T \}.$\\
It follows that
\begin{align}
D^I(x_j ,y_j)&=\min\left\{D^I(x_j ,y_j ;\alpha ,1): \alpha\in \Lambda^\star(x_j,y_j)\right\}\\
&=D^I(x_j ,y_j ;\alpha_{+}^\star(j),1)=D^I(x_j ,y_j ;\alpha_{-}^\star(j),1).
\end{align}
Therefore, it follows that for all $(x_j ,y_j)$, the input measure computed over the global technology $T$ is obtained from the efficiency scores evaluated on $ T_{-}^{\star}(x_j,y_j)$ and $ T_{+}^{\star}(x_j,y_j)$.

It can be useful to compute the efficiency score of any production vectors. This is the case in  super-efficiency models and also for measuring productivity. More importantly, this allows to characterize the production technology.

In the input oriented case, for all $j,k \in \mathcal{J}$ we have:
\begin{equation}\label{genformk}D^I_{(k)}(x_j ,y_j ;\alpha ,1)=\left[\max_{h\in card (y)}
\frac{y_{j,h}}{y_{k, h} }\right]^{1/\alpha } \cdot \left[\max_{i\in
card(x)} \frac{x_{k, i}}{x_{j,i} }\right].
\end{equation}
Since $T=    \bigcap\limits_{\alpha\in \Lambda^\star }\bigcup\limits_{k\in \mathcal J}Q_{\alpha,1}(x_k,y_k),$ we have
\begin{align}\label{genform}D^I(x_j ,y_j ) &=\min\limits_{\alpha\in \bigcup\limits_{j\in \mathcal J}\Lambda_{ }(j)}\max_{k\in \mathcal J}D^I_{(k)}(x_j ,y_j ;\alpha ,1)\\
&=\min_{j\in \mathcal J}\min_{\alpha\in \Lambda(j)}\max_{k\in \mathcal J}\left[\max_{h\in card (y)}
\frac{y_{j,h}}{y_{k, h} }\right]^{1/\alpha } \cdot \left[\max_{i\in
card(x)} \frac{x_{k, i}}{x_{j,i} }\right].
\end{align}

\noindent For a lower minimal extrapolation case:
 \begin{equation}\label{genformlow}D^I(x_j ,y_j ;\alpha_{ -}^\star(j),1)=\min_{j\in \mathcal J}\max_{k\in \mathcal J}\left[\max_{h\in card (y)}
\frac{y_{j,h}}{y_{k, h} }\right]^{1/\alpha_{ -}^\star(j)} \cdot \left[\max_{i\in
card(x)} \frac{x_{k, i}}{x_{j,i} }\right].
\end{equation}

\noindent In the case of the upper minimal extrapolation:
\begin{equation}\label{genformup}D^I(x_j ,y_j ;\alpha_{+}^\star(j),1)=\min_{j\in \mathcal J}\max_{k\in \mathcal J}\left[\max_{h\in card (y)}
\frac{y_{j,h}}{y_{k, h} }\right]^{1/\alpha_{ +}^\star(j)} \cdot \left[\max_{i\in
card(x)} \frac{x_{k, i}}{x_{j,i} }\right].
\end{equation}


\section{Empirical illustration: estimation of individual $\alpha$-returns to scale for the US industries }

This analysis focuses on the evolution of individual $\alpha$-returns to scale for 63 US industries totalling the whole American economy over the period 1987-2018. In this perspective, the theoretical framework developed above is applied to annual underlying technologies retaining one output and three inputs. The production is measured by the gross output while the inputs are intermediate inputs, labour and capital services delivered by equipment, buildings, and intellectual property products.

\subsection{Data description and estimation strategy}

All basic data are estimated by the Bureau of Economic Analysis (BEA) through yearly production accounts established for each specific industry (\url{http://www.bea.gov/}). The decision making units (DMUs) are the 63 industries (Appendix 1, Table 1).

All output and intermediate  quantity indexes are weighted by their respective value levels in 2012 to obtain the gross output expressed in constant US dollar 2012. Volumes of fixed capital consumption are approximated by the cost depreciations of the three types of capital services (also expressed in constant 2012 prices). Finally, full-time employees measure annual changes in labour quantity.

This empirical illustration aims to provide the optimal individual $\Lambda$-returns to scale for each industry (or Decision Making Unit - DMU). To do so, we implement the input-oriented model introduced in Subsection 4.1. Indeed, we first estimate the optimal input oriented individual  $\alpha$-returns to scale which is noted $\alpha^\star$. This $\alpha^\star$   is the maximal $\alpha$ that minimizes the inefficiency score ($D^I$), for each DMU (program $P_j$).  Once $\alpha^\star$ obtained, we apply programs ($P_j^-$) and ($P_j^+$). These two programs allow to determine  if there exists,  for each DMU, an interval ($\Lambda^\star$) having a lower ($\alpha^\star_-$) and an upper ($\alpha^\star_+$) bounds, and which contains $\alpha^\star$. In such case, we established a procedure to characterize the returns to scale as follows. If the lower ($\alpha^\star_-$)  bound is greater than 1, then the industry is characterized by increasing returns (IRS). On the other hand, if the upper ($\alpha^\star_+$) bound is lower than 1, then the industry is characterized by decreasing returns (DRS). In both cases, it is clear that the hypothesis of constant returns to scale (CRS) can be rejected. However, when the interval $\left[\alpha^\star_-, \alpha^\star_+\right]$ includes 1 then, we consider that the hypothesis of CRS cannot be rejected.  

\subsection{Results}

The averages of $\alpha$-returns to scale per industry over the entire period indicates that a majority of sectors are characterized by IRS (26 out of 63). There are 15 industries under DRS and 21 industries for which the CRS hypothesis is not rejected. The existence of strictly IRS as optimal ones, indicates that the production set is locally non-convex and that the production process is non-linear for these industries.  Moreover, these IRS imply that some efficient observations could have increasing marginal products as well as they could also face indivisibilities (Tone and Sahoo, 2003; Sahoo and Tone, 2013) in the production process.  

\begin{figure}[H]
\begin{minipage}{8cm}
\centering
\includegraphics[scale=0.55]{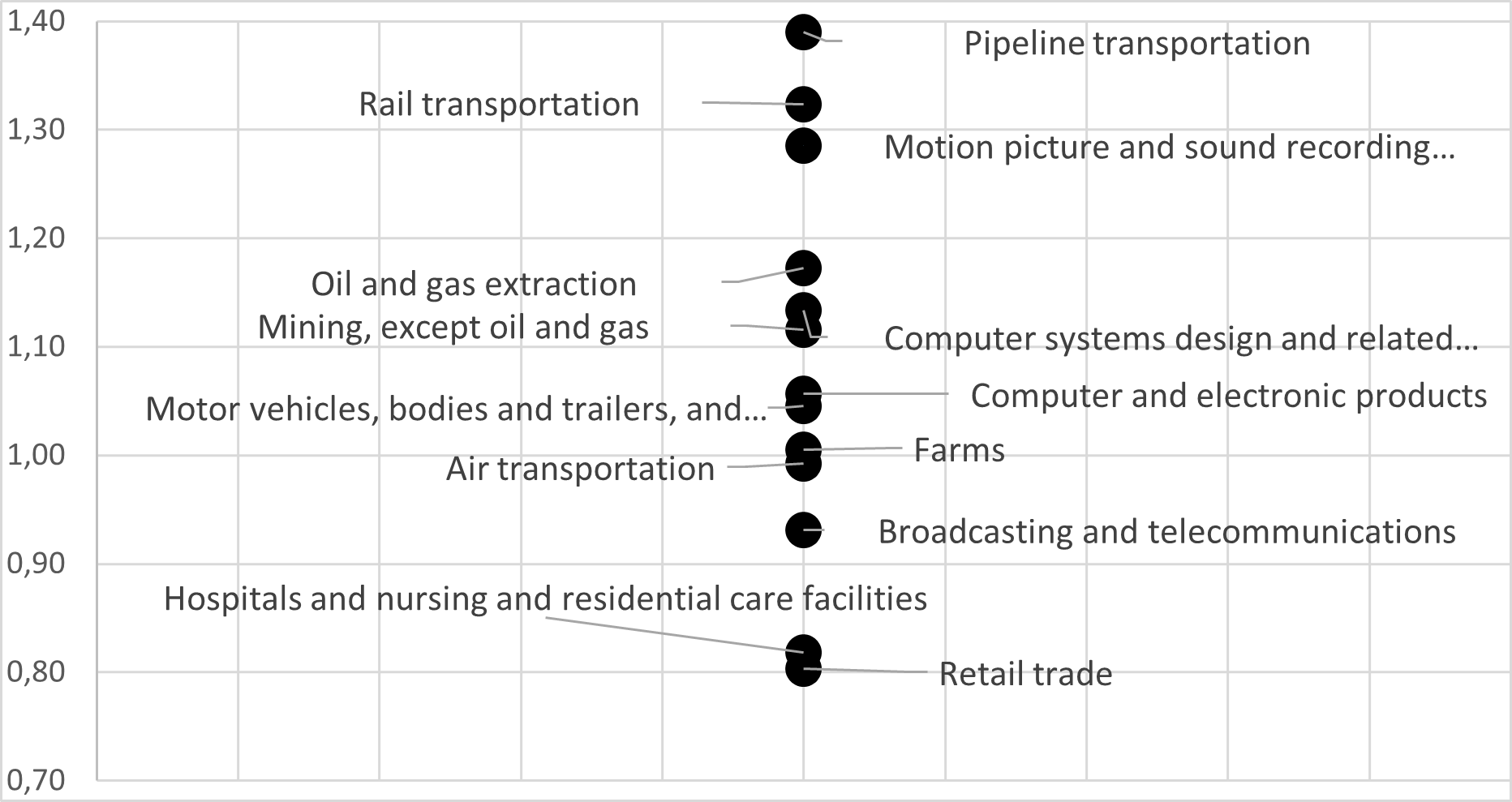}
\caption{\centering Characterization of $\alpha$-returns to scale for several industries \linebreak \footnotesize{(Average by industry over the period 1987-2018).}}\label{fig1}
\end{minipage}
\hfill
\begin{minipage}{8cm}
\linespread{1.5}\selectfont
Figure \ref{fig1} positions several emblematic industries with respect to their RTS. The activities concerning pipeline and rail transportations, motion picture, oil and gas extraction, mining, computer systems design are industries characterized by significantly IRS. The automotive sector and computers and electronics have slightly IRS. In contrast, hospitals and retail trade are clearly activities under DRS while broadcasting and telecommunications have slightly DRS. Finally, farms and air transport operate with CRS technologies.
\end{minipage}
\end{figure}

\begin{figure}[htpb!]

\begin{minipage}{7.2cm}
\centering
\includegraphics[scale=0.75]{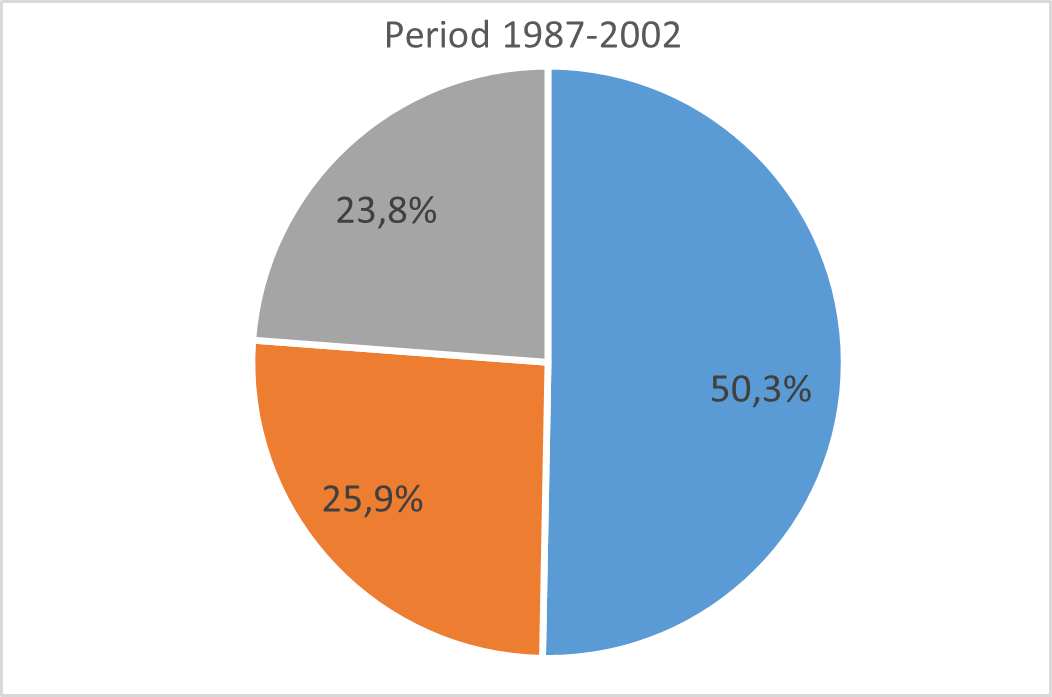}
\end{minipage}
\hfill
\begin{minipage}{7.2cm}
\centering
\includegraphics[scale=0.75]{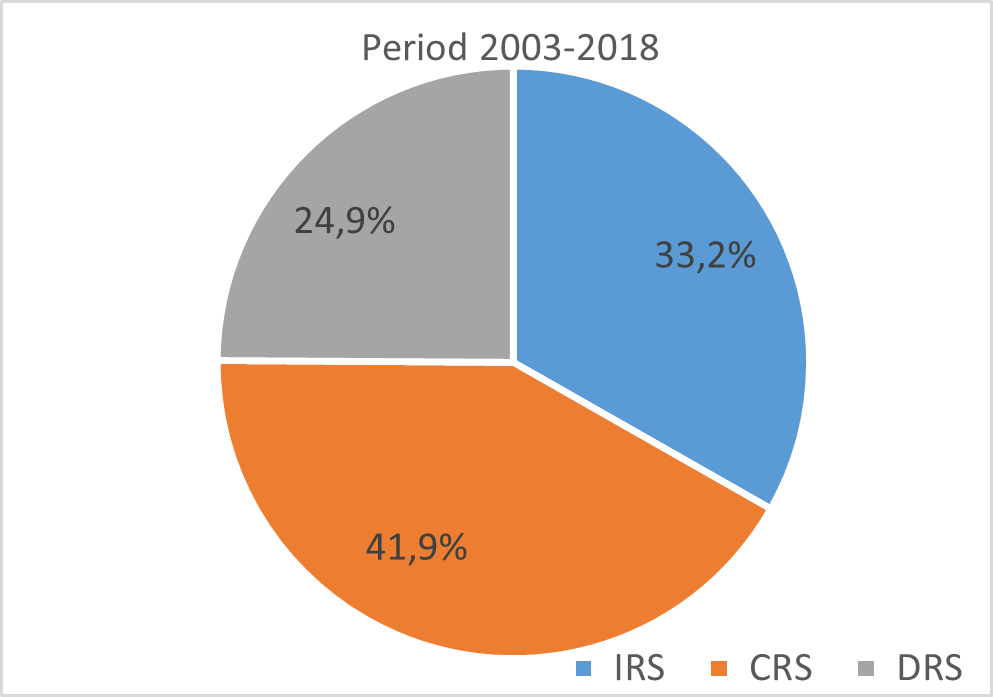}
\end{minipage}
\caption{\centering Distribution of industries according to their respective $\alpha$-returns to scale.}\label{fig2}

\end{figure}

However, the annual changes of individual $\alpha$-RTS strongly this first result established in favour of the IRS which is calculated on a static average over the whole period. According to Figure \ref{fig2},  variations in $\alpha$-RTS per period show a structural evolution of the US economy towards more industrial activities characterized by CRS technologies. At the beginning of the period (1987-2002), more than 50\% of sectors were characterized by IRS while those under CRS weighed only 24\%. Over the more recent period (2003-2018), we observe a substantial decline in the share of IRS industries (33\%) in favour of CRS sectors (42\%). The share of DRS industries remains stable (26\% to 25\%). 

\begin{figure}[htpb!]
\centering
\includegraphics[scale=0.55]{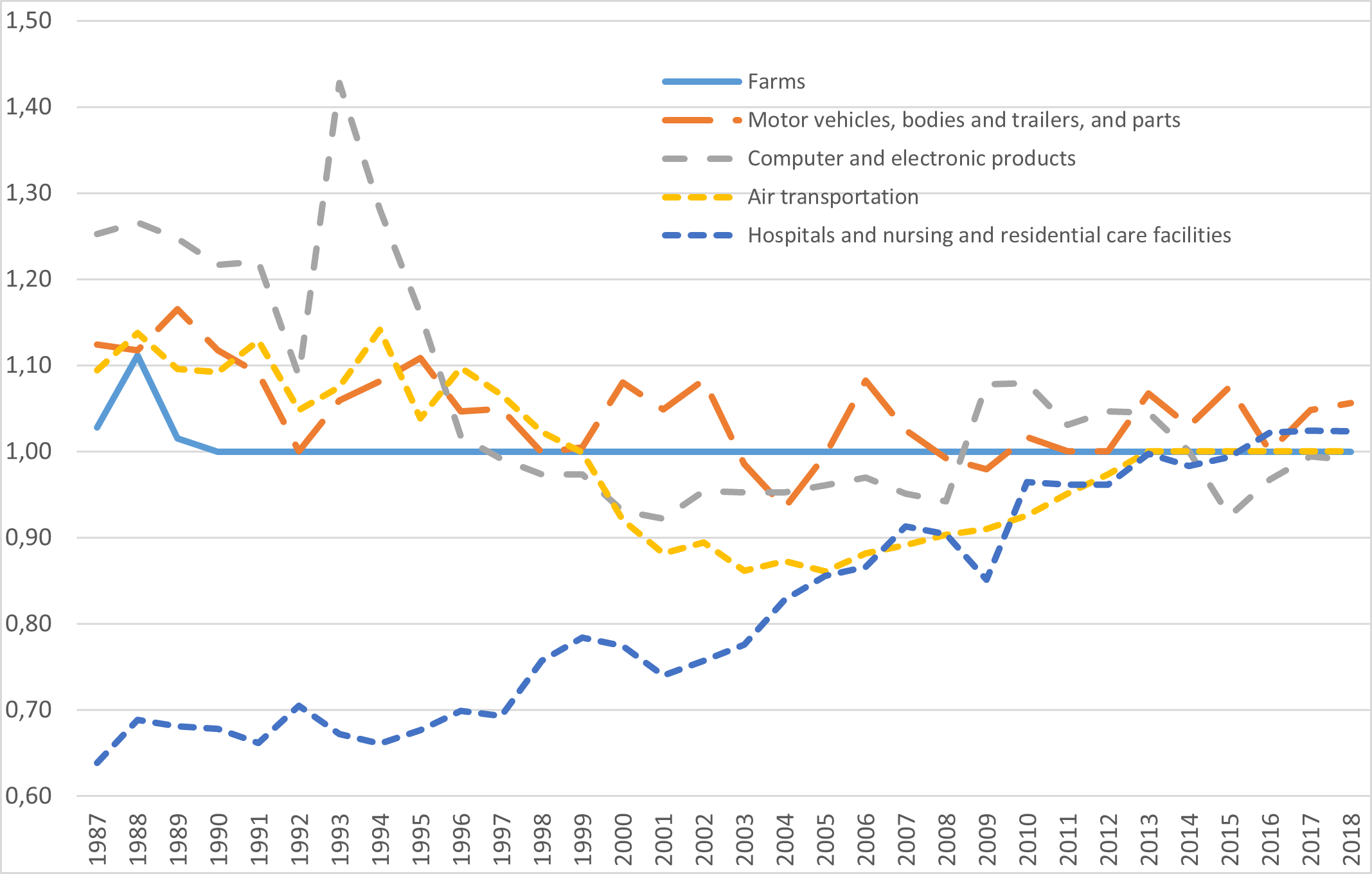}
\caption{\centering Evolution of individual $\alpha$-returns to scale for some industries}\label{fig3}
\end{figure}

These results confirm those previously established by Boussemart et al. (2019) who had shown that estimates of the $\alpha$-RTS for the entire U.S. economy converged clearly towards unity. This indicates that the US economy has nearly converged to a CRS technology implying that industries tend to their most productive scale size (MPSS) improving their total factor productivity levels. Figure \ref{fig3} illustrates this general finding with a few industry examples . The farms sector has maintained CRS throughout the period. The automotive sector's returns to scale fluctuate slightly above unity with some trend convergence toward the CRS area. Air transportation experienced two distinct phases of convergence towards CRS: the first started from a situation characterized by IRS in 1987 to CRS in 1999. Then, after the shock of the early 2000s characterized by DRS, this industry again converged to the CRS zone. The computer and electronic products industry started the period with strongly IRS and finally reached CRS. Conversely, hospitals started from a strongly DRS technology and converged steadily towards a CRS technology. 

The numerical results are displayed in Appendix 2.


\section{Conclusion}

This paper extends the notion of global $\alpha$-returns to scale model proposed by Boussemart et al. (2019). Indeed, the notion of $\Lambda$-returns to scale is introduced as a subset of non-negative real line allowing to characterize the global technology. Indeed, an optimal ``$\alpha$"-returns to scale is estimated for each observation constituting the production set. If $\alpha$ is not a singleton then each observation is associated to an optimal individual $\Lambda$-returns to scale which is an interval containing any optimal individual $\alpha$. Doing so, the local structure of returns-to-scale is considered such that the production possibility set can take into account strictly increasing and decreasing returns-to-scale. These particular returns are not often defined in standard models and hence some features of the production process may be neglected such as non linearity. Thereby, the global production possibility set can locally be non-convex. A non-parametric general procedure is provided to estimate the individual $\alpha$-returns to scale, from an input oriented standpoint. Each optimal individual $\alpha$-returns to scale may have an upper and a lower bounds as well as it can be a singleton. Along this line, the global technology is considered as an intersection of all individual production processes such that each individual returns-to-scale contributes to define the global $\Lambda$-returns to scale of the global technology. Hence, the introduction of $\Lambda$-returns to scale assumption allows to present a new class of production sets allowing to consider any kind of RTS.

These results are illustrated through a dataset composed of 63 industries constituting the whole American economy and which covers 32 years. The empirical results show that the $\Lambda$-returns to scale model allows to identify strictly increasing and decreasing individual returns-to-scale. However, the global technology satisfies a variable returns-to-scale including strictly increasing and decreasing returns-to-scale contrary to standard DEA models. The general procedure proposed in this paper has been presented through an input orientation nonetheless, it is always possible to implement this minimal extrapolation principle from an output oriented standpoint. Also remark that it could be of interest to take account for noise in the efficiency assessment following the approach proposed by Simar and Zelenyuk (2011).

\newpage
\setcounter{page}{1}
\section*{Appendix 1}

\begin{table}[htbp]
  \centering
  \tiny
    \begin{tabular}{cl}
    \toprule
    \textbf{N} & \multicolumn{1}{c}{\textbf{Industry}} \\
    \midrule
    \textbf{1} & Farms \\
    \textbf{2} & Forestry, fishing, and related activities \\
    \textbf{3} & Oil and gas extraction \\
    \textbf{4} & Mining, except oil and gas \\
    \textbf{5} & Support activities for mining \\
    \textbf{6} & Utilities \\
    \textbf{7} & Construction \\
    \textbf{8} & Wood products \\
    \textbf{9} & Nonmetallic mineral products \\
    \textbf{10} & Primary metals \\
    \textbf{11} & Fabricated metal products \\
    \textbf{12} & Machinery \\
    \textbf{13} & Computer and electronic products \\
    \textbf{14} & Electrical equipment, appliances, and components \\
    \textbf{15} & Motor vehicles, bodies and trailers, and parts \\
    \textbf{16} & Other transportation equipment \\
    \textbf{17} & Furniture and related products \\
    \textbf{18} & Miscellaneous manufacturing \\
    \textbf{19} & Food and beverage and tobacco products \\
    \textbf{20} & Textile mills and textile product mills \\
    \textbf{21} & Apparel and leather and allied products \\
    \textbf{22} & Paper products \\
    \textbf{23} & Printing and related support activities \\
    \textbf{24} & Petroleum and coal products \\
    \textbf{25} & Chemical products \\
    \textbf{26} & Plastics and rubber products \\
    \textbf{27} & Wholesale trade \\
    \textbf{28} & Retail trade \\
    \textbf{29} & Air transportation \\
    \textbf{30} & Rail transportation \\
    \textbf{31} & Water transportation \\
    \textbf{32} & Truck transportation \\
    \textbf{33} & Transit and ground passenger transportation \\
    \textbf{34} & Pipeline transportation \\
    \textbf{35} & Other transportation and support activities \\
    \textbf{36} & Warehousing and storage \\
    \textbf{37} & Publishing industries (includes software) \\
    \textbf{38} & Motion picture and sound recording industries \\
    \textbf{39} & Broadcasting and telecommunications \\
    \textbf{40} & Information and data processing services \\
    \textbf{41} & Federal Reserve banks, credit intermediation, and related activities \\
    \textbf{42} & Securities, commodity contracts, and investments \\
    \textbf{43} & Insurance carriers and related activities \\
    \textbf{44} & Funds, trusts, and other financial vehicles \\
    \textbf{45} & Real estate \\
    \textbf{46} & Rental and leasing services and lessors of intangible assets \\
    \textbf{47} & Legal services \\
    \textbf{48} & Computer systems design and related services \\
    \textbf{49} & Miscellaneous professional, scientific, and technical services \\
    \textbf{50} & Management of companies and enterprises \\
    \textbf{51} & Administrative and support services \\
    \textbf{52} & Waste management and remediation services \\
    \textbf{53} & Educational services \\
    \textbf{54} & Ambulatory health care services \\
    \textbf{55} & Hospitals and nursing and residential care facilities \\
    \textbf{56} & Social assistance \\
    \textbf{57} & Performing arts, spectator sports, museums, and related activities \\
    \textbf{58} & Amusements, gambling, and recreation industries \\
    \textbf{59} & Accommodation \\
    \textbf{60} & Food services and drinking places \\
    \textbf{61} & Other services, except government \\
    \textbf{62} & Federal government \\
    \textbf{63} & State and local government \\
    \bottomrule
    \end{tabular}%
     {  \caption{\scriptsize Industries composing the American economy.}}\label{liste}
\end{table}%

\newpage

\section*{Appendix 2}

{\bf Proof of Lemma \ref{lem32}}\\ By construction $T_\alpha$ contains $T$. Let us prove that it satisfies an assumption of $\alpha$ returns to scale assumption. First, suppose that  $\alpha\in \Real_+$. Let $(u,v)\in T_\alpha$, we need to show that for all $\lambda \geq 0$ 
$(\lambda u, \lambda^\alpha v)\in T_\alpha$. By hypothesis if $(u,v)\in T_\alpha$ then there exist $\eta\in \Real_+$ and some $(x,y)\in T$ such that 
$(u,v)=(\eta x,\eta^\alpha y)$. It follows that $ (\lambda u,\lambda^\alpha v)=(\lambda \eta x,\lambda^\alpha \eta^\alpha y)=
\big((\lambda \eta) x,(\lambda \eta)^\alpha y)$. Therefore $(\lambda u,\lambda^\alpha v)\in T_\alpha$. Hence, $T_\alpha$ satisfies an assumption of $\alpha$-returns to scale. Suppose now that $T_\alpha$ is not the smallest set that contains $T$ satisfying an assumption of $\alpha$-returns to scale, and let us show a contradiction. Suppose that this set is $S$ with $T\subset S\subsetneqq T_\alpha$. In such a case there is some $(u,v)\in T_\alpha$ with $(u,v)\notin S$. Since $(u,v)\in T_\alpha$, there is some $(x,y)\in T$ such that $(u,v)=(\lambda x, \lambda ^{\alpha}y) $. However, $T\subset S$ and since $S$ satisfies an assumption of $\alpha$-returns to scale, this is a contradiction.  We deduce that $T_\alpha$ is the smallest set that contains $T$ and satisfies an assumption of $\alpha$-returns to scale. Suppose now that $\alpha= \infty.$ The proof is similar. Let $(u,v)\in T_\infty$, we need to show that for all $\lambda \geq 0$ 
$(  u, \lambda  v)\in T_\infty$. By hypothesis if $(u,v)\in T_\infty$ then there exist $\eta\in \Real_+$ and some $(x,y)\in T$ such that 
$(u,v)=(  x,\eta  y)$. It follows that $ (  u,\lambda  v)=(  x,\lambda \eta  y)=
\big( x,(\lambda \eta)  y)$. Therefore $(  u,\lambda  v)\in T_\infty$. Therefore $T_\infty$ satisfies an assumption of  $\infty$-returns to scale assumption. Suppose now that $T_\infty$ is not the smallest set of this class  that contains $T$. Suppose that this set is  $S$  with  $T\subset S\subsetneqq T_\infty$. In such a case there is some $(u,v)\in T_\infty$ with $(u,v)\notin S$. Since $(u,v)\in T_\infty$, there is some $(x,y)\in T$ such that $(u,v)=(  x, \lambda  y) $. However, $T\subset S$ and since $S$ satisfies an assumption of $\infty$-returns to scale, this is a contradiction.   Therefore $T_\infty$ is the smallest set that contains $T$ and satisfies an assumption of $\alpha$-returns to scale, which complete the proof. $\Box$\\

\noindent {\bf Proof of Lemma \ref{compatibility}}\\  Suppose that $(x,y)\in T_{\alpha}$. By hypothesis,   $(x,y)\in K_\beta(T_\alpha)=T_\alpha$. We consider  two cases.

 $(i)$ $\beta <\alpha$.  For all $\lambda >1$,   $  (\lambda x,  \lambda ^\alpha y)\in T_\alpha$. Moreover   $  (\lambda x, {\lambda }^\beta y)\in  K_\beta(T_\alpha)$. Since $\beta <\alpha $ for all $y\not=0$ we cannot find any $\lambda >1$ such that $\lambda ^\beta y\geq \lambda ^\alpha y$. Consequently, if $K_\beta(T_\alpha)$ satisfies $T2$, then $y=0$ for all $(x,y)\in  K_\beta(T_\alpha)$. 
 
  $(ii)$ $\beta >\alpha$.  For all $\lambda \in ]0,1[$,   $  (\lambda^{\frac{1}{\alpha}} x,  \lambda   y)\in T_\alpha$. Moreover   $  (\lambda^{\frac{1}{\beta}} x, {\lambda }  y)\in  K_\beta(T_\alpha)$. Since $\beta >\alpha $  and  $K_\beta(T_\alpha)\supset   T_\alpha $, we should have $\lambda^{\frac{1}{\alpha}} x\leq \lambda^{\frac{1}{\beta}} x$ for all $x$. However this is true only if $x=0$. Therefore for all $(  x,  y)\in T_\alpha$ and $  \lambda \in ]0,1[$,  $({  \lambda}^{\frac{1}{\beta}}  x,  \lambda    y)\in K_\beta(T_\alpha)$ implies that $(0,  \lambda    y)\in K_\beta(T_\alpha).$ Since $K_\beta(T_\alpha)$ satisfies a $\beta$-returns to scale assumption, it follows that for all $\eta >1$,  $(0,\eta ^\beta   \lambda   y)\in K_\beta(T_\alpha).$ Therefore, if we have not the condition $ y  =0$ for all $(x,y)\in  T_\alpha $ then $ K_\beta(T_\alpha)$ fails to satisfy $T2.$ However,  $y=0$ for all $(x,y)\in T_\alpha$ implies $y=0$ for all $(x,y)\in K_\beta (T_\alpha)$ . $\Box$\\

\noindent {\bf Proof of Proposition \ref{prop32}} \\
Since $T$ is a $\Lambda$-bounded production set there exists at least a production set that contains $T$ and that satisfies a $\Lambda$-returns to scale assumption. For any $\alpha\in \Lambda$, $T$ is $\{\alpha\}$-bounded. Suppose that $T_\alpha$ is not the smallest technology satisfying an assumption of $\Lambda$-returns to scale and let us show a contradiction. For some $\alpha \in \Lambda$, there is a production set $Q_\alpha$ satisfying an $\alpha$-returns to scale assumption with $Q_\alpha\subsetneqq T_\alpha$. However from Lemma \ref{lem32} $T_\alpha=K_\alpha(T)$ is the smallest production set satisfying an assumption of $\alpha$-returns to scale. Therefore this is a contradiction. $\Box$\\

\noindent {\bf Proof of Proposition \ref{interproperty}}\\
$(i)$ and $(ii)$ are immediate.\\
$(ii)$ By definition $T_{\Lambda }$ is the minimal intersection of a collection of production sets satisfying a $\Lambda$-returns to scale assumption.   This implies that there exists a collection
$\{T_\alpha\}_{\alpha\in \Lambda}$ such that $T_{\Lambda}=\bigcap\limits_{\alpha\in \Lambda}T_\alpha.$ Similarly, there exists a collection $\{T_{\alpha}\}_{\alpha\in \Lambda'}$ such that $T_{\Lambda'}=\bigcap\limits_{\alpha\in \Lambda'}T_{\alpha}.$
 Therefore
$$T_{\Lambda}\cap T_{\Lambda'}=\left(\bigcap_{\alpha\in \Lambda}T_\alpha\right)\cap\left(\bigcap_{\alpha'\in \Lambda'}T_\alpha\right)=
\bigcap_{\alpha\in \Lambda\cup \Lambda'}T_\alpha. $$
Let us prove $(ii)$. Since $\Lambda \cap \Lambda'\not=\emptyset$, it follows that $\bigcap_{\alpha\in \Lambda\cap \Lambda'}T_\alpha$ is well defined. Suppose that $(x,y)\in \bigcap_{\alpha\in \Lambda}T_\alpha$, since $\Lambda\cap \Lambda'\subset \Lambda$, it follows from $(i)$ that $(x,y)\in \bigcap_{\alpha\in \Lambda\cap \Lambda'}T_\alpha.$ Similarly, if $(x,y)\in \bigcap_{\alpha\in \Lambda'}T_\alpha$,  we have $(x,y)\in \bigcap_{\alpha\in \Lambda\cap \Lambda'}T_\alpha.$ Consequently  
$$T_{\Lambda}\cup T_{\Lambda'}=\left(\bigcap_{\alpha\in \Lambda}T_\alpha\right)\cup\left(\bigcap_{\alpha'\in \Lambda'}T_\alpha\right)\subset
\bigcap_{\alpha\in \Lambda\cap \Lambda'}T_\alpha. \quad\Box$$\\

\noindent{\bf Proof of Proposition \ref{rightleft}}\\
We first prove that if $(x,y)\in T=\bigcap_{\alpha\in \Lambda}T_\alpha$, then $\lambda \geq 1$ implies $(\lambda x,\lambda^{\alpha_-}y)\in T$. For all $\lambda\geq 1$ and any $\alpha\in \Lambda$, we have ${\alpha_-}\leq  \alpha$ which implies that $\lambda^{\alpha_-}\leq \lambda^\alpha $. By hypothesis, we have $ (x,y)\in T\Rightarrow (\lambda x,\lambda^{\alpha }y)\in T_\alpha.$ Since $\lambda^{\alpha_-}\leq \lambda^\alpha $ we deduce that $\lambda^{\alpha_-}y\leq \lambda^{\alpha}y$ for any $\alpha\in \Lambda$. From the strong disposability assumption we have for all $\alpha \in \Lambda$: $ (x,\lambda ^\alpha y)\in T_\alpha\Rightarrow (\lambda x,\lambda^{\alpha_-}y)\in T_\alpha.$ Therefore $(\lambda x,\lambda^{\alpha_-}y)\in \bigcap_{\alpha\in \Lambda}T_\alpha=T$.
 Thus, $ (x,y)\in T\Rightarrow (\lambda x,\lambda^{\alpha_-}y)\in \bigcap_{\alpha\in \Lambda}T_\alpha=T.$

\noindent Let us prove the second part of the statement. We first assume that $\alpha_+<\infty$. 
 For any $\lambda\leq 1$ and any $\alpha\in \Lambda$, we have $\lambda^{\alpha_+}\leq \lambda^\alpha $. And by hypothesis, for any $\alpha \in \Lambda$, we have $ (x,y)\in T\Rightarrow (\lambda x,\lambda^{\alpha }y)\in T_\alpha.$ Moreover, from the strong disposability assumption we have $ (x,y)\in T\Rightarrow (\lambda x,\lambda^{\alpha_+}y)\in T_\alpha.$ Thus, $ (x,y)\in T\Rightarrow (\lambda x,\lambda^{\alpha_+}y)\in \bigcap_{\alpha\in \Lambda}T_\alpha=T.$ If $\alpha_+=\infty$ we use the fact that equivalently for all $\alpha\in \Real_{++}$ then, $ (x,y)\in T\Rightarrow (\lambda^{\frac{1}{\alpha} } x,\lambda y)\in T_\alpha.$ Hence, for all $\alpha$, if $\lambda \leq 1$ we have $  (\lambda^{\frac{1}{\alpha}} x,\lambda y)\leq ( x,\lambda y).$ Therefore, from the strong disposability assumption, $ (x,y)\in T\Rightarrow (x,\lambda y)\in T_\alpha,$ which ends the proof.  $\Box$\\

\noindent{\bf Proof of Proposition \ref{lambdagammaIntersect}}\\
$(i)$ is obvious since $T^{\Theta, \Gamma_{CRS}}=K_1\Big(T^{\Theta, \Gamma_{VRS}}\Big)$. \\
$(ii)$ In the following we denote $P_{\Theta, \Gamma}(x)$ the output set of $x$ for any $x$. By hypothesis if $(x,y)\in T ^{\Theta, \Gamma_{NIRS}}$, then there is some $(u,v)\in T^{\Theta, \Gamma_{VRS}}$ such that
$(x,y)=(\lambda u,\lambda v)$ with $\lambda \leq 1$. However, since $\lambda \leq 1$, this implies that for any $ \alpha<1$ we have 
$\lambda^\alpha v\geq \lambda v$. Since $(\lambda u, \lambda^\alpha v)\in K_\alpha (T^{\Theta, \Gamma_{VRS}})=T_\alpha^{\Theta, \Gamma_{VRS}}$ which satisfies the strong disposability assumption, we deduce that $(x,y)=(\lambda u,\lambda v)\in T_\alpha^{\Theta, \Gamma_{VRS}}$. Moreover this inclusion is true for any $\alpha\in ]0,1]$ and we have $T ^{\Theta, \Gamma_{NIRS}}\subset T_\alpha^{\Theta, \Gamma_{VRS}}$. In addition, since the strong disposability assumption holds, we have $T ^{\Theta, \Gamma_{NIRS}}\subset T_0^{\Theta, \Gamma_{VRS}}$. We deduce, for all real number $\alpha\leq 1$ the inclusion 
 $T ^{\Theta, \Gamma_{NIRS}}\subset T_\alpha^{\Theta, \Gamma_{VRS}}$. Hence, we have $$T ^{\Theta, \Gamma_{NIRS}}\subset \bigcap_{\alpha\in [0,1]}T_\alpha^{\Theta, \Gamma_{VRS}}.$$
Let us prove the converse inclusion and  note that $\bigcap_{\alpha\in[1,\infty]}T_\alpha^{\Theta,\Gamma_{VRS}}\subset T_1^{\Theta,\Gamma_{VRS}}=T^{\Theta,\Gamma_{CRS}}$. Suppose that $(x,y)\in T^{\Theta,\Gamma_{CRS}}\backslash T ^{\Theta, \Gamma_{NIRS}}$. In such a case there is some $(u,v)\in T^{\Theta,\Gamma_{VRS}}$ and some $\lambda >1$ such that $(x,y)=(\lambda u,\lambda v)$. It follows that for each $\alpha$ there is some $v_\alpha$ with 
 $(x,y)=(\lambda u, \lambda^\alpha v_\alpha)$ and $v_\alpha\in P_{\Theta, \Gamma_{VRS}}(\lambda u)$. Since $\lambda>1$ we deduce that:
$$\|y\|=\lim_{\alpha\longrightarrow \infty}\lambda^\alpha \| v_\alpha\|=+\infty .$$
However, this contradicts $T2$. Therefore, we deduce the converse inclusion
$$\bigcap_{\alpha\in [0,1]}T_\alpha^{\Theta, \Gamma_{VRS}}\subset T ^{\Theta, \Gamma_{NIRS}}.$$
$(iii)$ By hypothesis, if $(x,y)\in T^{\Theta, \Gamma_{NDRS}}$, then there is some $(u,v)\in T^{\Theta, \Gamma_{VRS}}$ such that
$(x,y)=(\lambda u,\lambda v)$ with $\lambda \geq 1$. However, since $\lambda \geq 1$, this implies that for any $ \alpha>1$ we have 
$\lambda^\alpha v\geq \lambda v$. Since $(\lambda u, \lambda^\alpha v)\in K_\alpha (T^{\Theta, \Gamma_{VRS}})=T_\alpha^{\Theta, \Gamma_{VRS}}$ which satisfies the strong disposability assumption, we deduce that $(x,y)=(\lambda u,\lambda v)\in T_\alpha^{\Theta, \Gamma_{VRS}}$. Moreover this inclusion is true for any $\alpha\in [1,+\infty[$ and we have $T ^{\Theta, \Gamma_{NDRS}}\subset T_\alpha^{\Theta, \Gamma_{VRS}}$. In addition, since the strong disposability assumption holds we have $T ^{\Theta, \Gamma_{NDRS}}\subset T_\infty^{\Theta, \Gamma_{VRS}}$. We deduce, for all real number $\alpha\geq 1$ the inclusion 
 $T ^{\Theta, \Gamma_{NDRS}}\subset T_\alpha^{\Theta, \Gamma_{VRS}}$. This inclusion is also true for $S(0,0)$. Hence, we have $$\widetilde T ^{\Theta, \Gamma_{NDRS}}\subset \bigcap_{\alpha\in [1,+\infty]}T_\alpha^{\Theta, \Gamma_{VRS}}.$$
Let us prove the converse inclusion using the fact that $\bigcap_{\alpha\in[1,\infty]}T_\alpha^{\Theta,\Gamma_{VRS}}\subset T_1^{\Theta,\Gamma_{VRS}}=T^{\Theta,\Gamma_{CRS}}$. Suppose that $(x,y)\in T^{\Theta,\Gamma_{CRS}}\backslash \widetilde T ^{\Theta, \Gamma_{NDRS}}$. In such case, there is some $(u,v)\in T^{\Theta,\Gamma_{VRS}}$ and some $\lambda \in ]0,1[$ such that $(x,y)=(\lambda u,\lambda v)$. It follows that for each $\alpha$, there is some $v_\alpha$ with 
 $(x,y)=(\lambda u, \lambda^\alpha v_\alpha)$ and $v_\alpha\in P_{\Theta, \Gamma_{VRS}}(\lambda u)$. Since $P_{\Theta, \Gamma_{VRS}}(\lambda u)$
is bounded and $\lambda\in ]0,1[$, we deduce that:
$$y=\lim_{\alpha\longrightarrow \infty}\lambda^\alpha v_\alpha=0.$$
It follows that $$(x,y)\in S(0,0) \subset T^{\Theta,NDRS}$$
that is a contradiction. Therefore, we deduce the converse inclusion
$$\bigcap_{\alpha\in [1,+\infty]}T_\alpha^{\Theta, \Gamma_{VRS}}\subset T ^{\Theta, \Gamma_{NDRS}}.$$
$(iv)$ We have $\widetilde T^{\Theta, \Gamma_{VRS}}=T^{\Theta, \Gamma_{NIRS}}\cap T^{\Theta, \Gamma_{NDRS}}$. Therefore
$$\widetilde T^{\Theta, \Gamma_{VRS}}=\bigcap_{\alpha\in [0,1]}T_\alpha^{\Theta, \Gamma_{VRS}}\cap \bigcap_{\alpha\in [1,+\infty]}T_\alpha^{\Theta, \Gamma_{VRS}}=\bigcap_{\alpha\in [0,+\infty]}T_\alpha^{\Theta, \Gamma_{VRS}}. \quad\Box$$ 

\noindent {\bf Proof of Proposition \ref{indextrap}}\\
\noindent  Let $T$ satisfies $T1-T4$ and contains $(x_k,y_k)$. We need to prove that $T\supseteq Q_{\alpha,1}$. Suppose that $(x,y)\in Q_{\alpha,1}$. It follows that there is some $\lambda >0$ with $x\geq \lambda^{1/\alpha}
x_k\qquad \text{ and }\qquad y\leq  \lambda y_k. $ Note that $(\lambda x,\lambda^\alpha y)\in Q_{\alpha,1}$. However, one can find some $\bar\lambda=\lambda^{-\frac{1}{\alpha}}>0$ such that
$\bar \lambda x\geq \lambda^{-\frac{1}{\alpha}} \lambda^{\frac{1}{\alpha}} x_k=x_k$
and
$ \bar \lambda ^\alpha y\leq  \lambda^{-1}\lambda y_k =y_k. $
Thus since $T3$ holds and $(x_k,y_k)\in T$, we deduce that $(x,y)\in T$. The second part of the statement is then an immediate consequence. $\Box$\\

\noindent {\bf Proof of Proposition \ref{globaltechlambda}}\\
\noindent {\it Proof of $(i)$} We have  
$$
  T=\bigcap_{j\in \mathcal J}T^\star(x_j,y_j)=\bigcap_{j\in \mathcal J} \bigcap_{\alpha\in \Lambda^\star (j) }\bigcup_{k\in \mathcal J}Q_{\alpha,1}(x_k,y_k).  $$ Thus, $$
  T=   \bigcap_{\alpha\in \bigcup\limits_{j\in \mathcal J}\Lambda^\star (j) }\bigcup_{k\in \mathcal J}Q_{\alpha,1}(x_k,y_k)=\bigcap_{\alpha\in \Lambda^\star }\bigcup_{k\in \mathcal J}Q_{\alpha,1}(x_k,y_k). 
$$
\noindent {\it Proof of $(ii)$}\\
 We have established that
 $T$ satisfies a $\Lambda$-returns to scale assumption. However, by definition
 \begin{align*}\alpha_{-}^\star&=\min\{\alpha_{ -}^\star(j):j\in \mathcal J\}=\min\{\min \{\alpha_{}: \alpha\in \Lambda(x_j,y_j)\}:j\in \mathcal J\}\\&=\min\{\alpha: \alpha\in \Lambda\}. \end{align*}
 Since every technology $\bigcup_{k\in \mathcal J}Q_{\alpha,1}(x_k,y_k)$ satisfies $T1-T4$, from Proposition \ref{rightleft} we deduce the result. $\Box$\\

\noindent {\bf Proof of Proposition \ref{lambdarational}}\\
We have shown that \begin{equation*}T=   \bigcap_{\alpha\in \bigcup\limits _{j\in \mathcal J}\Lambda^\star (j) }\bigcup_{k\in \mathcal J}Q_{\alpha,1}(x_k,y_k)=\bigcap_{\alpha\in \Lambda^\star }\bigcup_{k\in \mathcal J}Q_{\alpha,1}(x_k,y_k).\end{equation*}
From Proposition \ref{indextrap}, $T$ is minimal.
The result immediately follows. $\Box$\\

\begin{example}\label{expl1} 
We consider an example based upon the one proposed in Boussemart et al. (2019). Indeed, we add another observation. Consider that $n=p=1$ such that there is one input and one output. Consider the observations $(x_1,y_1)=(1,1)$, $(x_2,y_2)=(4,2)$, $(x_3,y_3)=(5/2,3/2)$, and $(x_4,y_4)=(3,5)$. The individual technology of each observation is then:
\begin{align*}
& Q_{\alpha,1}(x_1,y_1) = Q_{\alpha,1}(1,1) = \left\{(x,y)\in \Real^2_+: x \geq \lambda^{1/\alpha} \cdot 1 , y \leq \lambda \cdot 1, \lambda \geq 0 \right\};\\
& Q_{\alpha,1}(x_2,y_2) = Q_{\alpha,1}(4,2) = \left\{(x,y)\in \Real^2_+: x \geq \lambda^{1/\alpha} \cdot 4 , y \leq \lambda \cdot 2, \lambda \geq 0 \right\};\\
& Q_{\alpha,1}(x_3,y_3) = Q_{\alpha,1}(5/2,3/2) = \left\{(x,y)\in \Real^2_+: x \geq \lambda^{1/\alpha} \cdot \dfrac{5}{2} , y \leq \lambda \cdot \dfrac{3}{2}, \lambda \geq 0 \right\};\\
& Q_{\alpha,1}(x_4,y_4) = Q_{\alpha,1}(3,5) = \left\{(x,y)\in \Real^2_+: x \geq \lambda^{1/\alpha} \cdot 3 , y \leq \lambda \cdot 5, \lambda \geq 0 \right\}.
\end{align*}
The input efficiency measures of each observation with respect to $Q_{\alpha,1}(x_1,y_1)$, $Q_{\alpha,1}(x_2,y_2)$, $Q_{\alpha,1}(x_3,y_3)$ and $Q_{\alpha,1}(x_4,y_4)$, respectively, are:
\begingroup
\allowdisplaybreaks
\begin{align*}
& D^I(x_1,y_1;\gamma,\delta) = D^I(1,1;\gamma,\delta) = \min \left\{1, 4\left(\dfrac{1}{2}\right)^{1/\alpha}, \dfrac{5}{2}\left( \dfrac{2}{3}\right)^{1/\alpha}, 3 \left(\dfrac{1}{5}\right)^{1/\alpha}\right\};\\
& D^I(x_2,y_2;\gamma,\delta) = D^I(4,2;\gamma,\delta) = \min \left\{\dfrac{1}{4}\left(2\right)^{1/\alpha} , 1, \dfrac{5}{8}\left( \dfrac{4}{3}\right)^{1/\alpha}, \dfrac{3}{4}\left(\dfrac{2}{5}\right)^{1/\alpha} \right\};\\
& D^I(x_3,y_3;\gamma,\delta) = D^I(5/2, 3/2;\gamma,\delta) = \min \left\{\dfrac{2}{5}\left(\dfrac{3}{2}\right)^{1/\alpha}, \dfrac{8}{5}\left(\dfrac{3}{4}\right)^{1/\alpha}, 1, \dfrac{6}{5}\left(\dfrac{3}{10}\right)^{1/\alpha} \right\};\\
& D^I(x_4,y_4;\gamma,\delta) = D^I(3,5;\gamma,\delta) = \min \left\{\dfrac{1}{3}(5)^{1/\alpha}, \dfrac{4}{3}\left(\dfrac{5}{2}\right)^{1/\alpha}, \dfrac{5}{6}\left( \dfrac{10}{3}\right)^{1/\alpha}, 1 \right\}.
\end{align*}\endgroup
As we set $\beta = 1/\alpha$ then, to obtain $\alpha^\star$, we have to solve a maximization-minimization program for each observation, as follows:
\begin{align*}
& \max\limits_\beta\left( D^I(x_1,y_1;\gamma,\delta) \right)  = \max\limits_\beta \left(\min \left\{1, 4\left(\dfrac{1}{2}\right)^{\beta}, \dfrac{5}{2}\left( \dfrac{2}{3}\right)^{\beta}, 3 \left(\dfrac{1}{5}\right)^{\beta} \right\} \right);\\
& \max\limits_\beta\left( D^I(x_2,y_2;\gamma,\delta) \right)  = \max\limits_\beta \left(\min \left\{\dfrac{1}{4}\left(2\right)^{\beta}, 1, \dfrac{5}{8}\left( \dfrac{4}{3}\right)^{\beta}, \dfrac{3}{4}\left(\dfrac{2}{5}\right)^{\beta} \right\}  \right);\\
& \max\limits_\beta\left( D^I(x_3,y_3;\gamma,\delta) \right)  = \max\limits_\beta \left(\min \left\{\dfrac{2}{5}\left(\dfrac{3}{2}\right)^{\beta}, \dfrac{8}{5}\left(\dfrac{3}{4}\right)^{\beta}, 1, \dfrac{6}{5}\left(\dfrac{3}{10}\right)^{\beta}\right\}  \right);\\
& \max\limits_\beta\left( D^I(x_4,y_4;\gamma,\delta) \right)  = \max\limits_\beta \left(\min \left\{ \dfrac{1}{3}(5)^{\beta}, \dfrac{4}{3}\left(\dfrac{5}{2}\right)^{\beta}, \dfrac{5}{6}\left( \dfrac{10}{3}\right)^{\beta}, 1    \right\}  \right).
\end{align*}
Thus, through a maximization process by taking the logarithm, we have to solve:\\
\begin{minipage}{7cm}
\begin{align*}
\begin{array}{lllll}
& \max\limits_{\lambda,\beta} & \lambda_1 \\
& s.t. & \lambda_1 \leq \ln 1 \\
&& \lambda_1 \leq \ln 4 + \beta_1 \ln \dfrac{1}{2} \\
&& \lambda_1 \leq \ln \dfrac{5}{2} + \beta_1 \ln \dfrac{2}{3}\\
&& \lambda_1 \leq \ln 3 + \beta_1 \ln \dfrac{1}{5}.
\end{array}
\end{align*}
\end{minipage}
\hfill
\begin{minipage}{7cm}
\begin{align*}
\begin{array}{lllll}
& \max\limits_{\lambda,\beta} & \lambda_2 \\
& s.t. & \lambda_2 \leq \ln \dfrac{1}{4} + \beta_2 \ln 2   \\
&& \lambda_2 \leq \ln 1 \\
&& \lambda_2 \leq \ln \dfrac{5}{8} + \beta_2 \ln \dfrac{4}{3}\\
&& \lambda_2 \leq \ln \dfrac{3}{4} + \beta_2 \ln \dfrac{2}{5}.
\end{array}
\end{align*}
\end{minipage}
\\
\noindent \begin{minipage}{7cm}
\begin{align*}
\begin{array}{lllll}
& \max\limits_{\lambda,\beta} & \lambda_3 \\
& s.t. & \lambda_3 \leq \ln \dfrac{2}{5} + \beta_3 \ln \dfrac{3}{2}   \\
&& \lambda_3 \leq \ln \dfrac{8}{5} + \beta_3 \ln \dfrac{3}{4}\\
&& \lambda_3 \leq \ln 1 \\
&& \lambda_3 \leq \ln \dfrac{6}{5} + \beta_3 \ln \dfrac{3}{10}.\\
\end{array}
\end{align*}
\end{minipage}
\hfill
\begin{minipage}{7cm}
\begin{align*}
\begin{array}{lllll}
& \max\limits_{\lambda,\beta} & \lambda_4 \\
& s.t. & \lambda_4 \leq \ln \dfrac{1}{3} + \beta_4 \ln {5}   \\
&& \lambda_4 \leq \ln \dfrac{4}{3} + \beta_4 \ln \dfrac{5}{2}\\
&& \lambda_4 \leq \ln \dfrac{5}{6} + \beta_4 \ln \dfrac{10}{3}\\
&& \lambda_4 \leq \ln 1 .
\end{array}
\end{align*}
\end{minipage}\\

\noindent Since $D^I \in [0,1]$, then $\max \lambda\leq 0$. Moreover, since $\alpha\in [0,\infty]$, then $\beta \in \Real_+$. Hence, the solutions are $\beta_1^\star = 0$, $\beta_2^\star = \dfrac{\ln 3}{\ln 5}$, $\beta_3^\star =\dfrac{\ln 6}{\ln 10}$ and, $\beta_4^\star = \dfrac{\ln3}{\ln 5}$. Consequently, $\alpha_1^\star = +\infty$, $\alpha_2^\star = \dfrac{\ln 5}{\ln 3}$, $\alpha_3^\star = \dfrac{\ln 10}{\ln 6}$ and, $\alpha_4^\star = \dfrac{\ln 5}{\ln3}$.

From the results above, we have the efficiency measures of each observation such that $D^I(x_1,y_1;\gamma,\delta) = D^I(1,1;\gamma,\delta) = 1$, $D^I(x_2,y_2;\gamma,\delta) = D^I(4,2;\gamma,\delta) =0.4013$, $D^I(x_3,y_3;\gamma,\delta) = D^I(5/2,3/2;\gamma,\delta) =0.4702$ and, $D^I(x_4,y_4;\gamma,\delta) = D^I(3,5;\gamma,\delta) =1$. We deduce $\lambda^\star$ by taking the logarithmic transformation of these efficiency scores. We can now determine if there exists an infinity of $\beta$ verifying the solutions to the above programs by the means of program $(P_j^-)$ and program $(P_j^+)$. Replacing $\lambda_1$, $\lambda_2$, $\lambda_3$ and, $\lambda_4$ by respectively $\ln(1)$, $\ln (0.4013)$, $\ln (0.4702)$ and $\ln (1)$, we have 
$$\beta_1^\star \in \left[0,\dfrac{\ln 3}{\ln 5}\right],\; \beta_2^\star \in \left[\dfrac{\ln 3}{\ln 5},\dfrac{\ln 3}{\ln 5}\right],\; \beta_3^\star \in \left[0.3989, \dfrac{\ln 6}{\ln 10}\right],\;\beta_4^\star \in \left[\dfrac{\ln 3}{\ln 5}, \infty\right].$$ Consequently, we have 
$$\alpha_1^\star \in \left[\dfrac{\ln 5}{\ln 3},\infty\right],\;\alpha_2^\star= \dfrac{\ln 3}{\ln 5},\; \alpha_3^\star \in \left[\dfrac{\ln 10}{\ln 6}, 2.5059\right],\;\alpha_4^\star \in \left[0, \dfrac{\ln 5}{\ln 3}\right].$$ Moreover, with respect to Proposition \ref{interproperty}, we can establish that the global technology satisfies a $[0,\infty]$-returns to scale assumption.

\end{example}\bigskip


\begin{sidewaystable}[htbp]
\section*{Appendix 3}
{\tiny  \centering
   \begin{tabular}{c|cc|cc|cc|cc|cc|cc|cc|cc}
    \toprule
    \multicolumn{1}{c|}{\multirow{2}[4]{*}{\textbf{DMU}}} & \multicolumn{2}{c|}{\textbf{1987}} & \multicolumn{2}{c|}{\textbf{1988}} & \multicolumn{2}{c|}{\textbf{1989}} & \multicolumn{2}{c|}{\textbf{1990}} & \multicolumn{2}{c|}{\textbf{1991}} & \multicolumn{2}{c|}{\textbf{1992}} & \multicolumn{2}{c|}{\textbf{1993}} & \multicolumn{2}{c}{\textbf{1994}} \\
\cmidrule{2-17}          & $\alpha_-^\star$ & $\alpha_+^\star$ & $\alpha_-^\star$ & $\alpha_+^\star$ & $\alpha_-^\star$ & $\alpha_+^\star$ & $\alpha_-^\star$ & $\alpha_+^\star$ & $\alpha_-^\star$ & $\alpha_+^\star$ & $\alpha_-^\star$ & $\alpha_+^\star$ & $\alpha_-^\star$ & $\alpha_+^\star$ & $\alpha_-^\star$ & $\alpha_+^\star$ \\
    \midrule
    \textbf{1} & 1.028 & 1.156 & \multicolumn{2}{c|}{1.112} & 1.015 & 1.148 & 0.953 & 1.210 & 0.951 & 1.227 & 0.888 & 1.274 & 0.906 & 1.206 & 0.824 & 1.249 \\
    \textbf{2} & 1.550 & 1.617 & \multicolumn{2}{c|}{1.280} & \multicolumn{2}{c|}{1.242} & \multicolumn{2}{c|}{1.207} & \multicolumn{2}{c|}{1.253} & \multicolumn{2}{c|}{1.116} & \multicolumn{2}{c|}{1.059} & 1.018 & 1.018 \\
    \textbf{3} & 1.367 & 1.543 & 1.172 & 1.604 & 1.168 & 1.676 & 1.183 & 1.670 & 1.227 & 1.743 & 1.309 & 1.714 & 1.291 & 1.751 & 1.157 & 1.950 \\
    \textbf{4} & \multicolumn{2}{c|}{1.461} & \multicolumn{2}{c|}{1.676} & \multicolumn{2}{c|}{1.588} & \multicolumn{2}{c|}{1.483} & 1.355 & 1.599 & 1.213 & 1.818 & 1.363 & 1.667 & 1.172 & 1.825 \\
    \textbf{5} & 1.092 & $\infty$ & 1.064 & $\infty$ & 1.050 & $\infty$ & 1.034 & $\infty$ & 1.038 & $\infty$ & 1.064 & $\infty$ & 1.023 & $\infty$ & 1.032 & $\infty$ \\
    \textbf{6} & \multicolumn{2}{c|}{0.854} & \multicolumn{2}{c|}{0.852} & \multicolumn{2}{c|}{0.872} & \multicolumn{2}{c|}{0.910} & \multicolumn{2}{c|}{0.881} & \multicolumn{2}{c|}{0.865} & \multicolumn{2}{c|}{0.854} & \multicolumn{2}{c}{0.889} \\
    \textbf{7} & 0.131 & 1.145 & 0.143 & 1.149 & 0.157 & 1.139 & 0.174 & 1.140 & 0.218 & 1.155 & 0.205 & 1.166 & 0.201 & 1.175 & \multicolumn{2}{c}{0,166} \\
    \textbf{8} & 1.020 & 1.309 & 1.020 & 1.423 & 1.014 & 1.399 & 0.998 & 1.372 & 0.997 & 1.340 & 1.011 & 1.489 & 1.028 & 1.528 & 1.025 & 1.671 \\
    \textbf{9} & \multicolumn{2}{c|}{1.026} & \multicolumn{2}{c|}{1.129} & \multicolumn{2}{c|}{1.072} & \multicolumn{2}{c|}{1.069} & \multicolumn{2}{c|}{1.074} & 0.998 & 1.178 & 1.003 & 1.169 & 0.991 & 1.203 \\
    \textbf{10} & \multicolumn{2}{c|}{1.110} & \multicolumn{2}{c|}{1.140} & \multicolumn{2}{c|}{1.135} & \multicolumn{2}{c|}{1.129} & 1.108 & 1.168 & 1.086 & 1.306 & 1.096 & 1.200 & 1.129 & 1.214 \\
    \textbf{11} & 1.118 & 1.149 & 1.098 & 1.203 & 1.103 & 1.178 & 1.120 & 1.165 & 1.097 & 1.148 & 1.110 & 1.156 & 1.093 & 1.167 & 1.050 & 1.191 \\
    \textbf{12} & \multicolumn{2}{c|}{1.086} & \multicolumn{2}{c|}{1.120} & \multicolumn{2}{c|}{1.132} & \multicolumn{2}{c|}{1.105} & \multicolumn{2}{c|}{1.091} & \multicolumn{2}{c|}{1.096} & \multicolumn{2}{c|}{1.091} & \multicolumn{2}{c}{1.050} \\
    \textbf{13} & \multicolumn{2}{c|}{1.253} & \multicolumn{2}{c|}{1.266} & \multicolumn{2}{c|}{1.247} & \multicolumn{2}{c|}{1.216} & \multicolumn{2}{c|}{1.220} & 1.088 & 1.691 & \multicolumn{2}{c|}{1.428} & \multicolumn{2}{c}{1.281} \\
    \textbf{14} & \multicolumn{2}{c|}{1.023} & \multicolumn{2}{c|}{1.062} & \multicolumn{2}{c|}{1.057} & \multicolumn{2}{c|}{1.032} & \multicolumn{2}{c|}{1.023} & \multicolumn{2}{c|}{1.046} & \multicolumn{2}{c|}{1.051} & \multicolumn{2}{c}{1.051} \\
    \textbf{15} & \multicolumn{2}{c|}{1.124} & \multicolumn{2}{c|}{1.117} & \multicolumn{2}{c|}{1.166} & \multicolumn{2}{c|}{1.118} & \multicolumn{2}{c|}{1.091} & 0.910 & 1.231 & \multicolumn{2}{c|}{1.060} & \multicolumn{2}{c}{1.082} \\
    \textbf{16} & \multicolumn{2}{c|}{0.921} & \multicolumn{2}{c|}{0.922} & \multicolumn{2}{c|}{0.950} & \multicolumn{2}{c|}{0.990} & \multicolumn{2}{c|}{0.969} & \multicolumn{2}{c|}{1.014} & \multicolumn{2}{c|}{0.990} & \multicolumn{2}{c}{1.024} \\
    \textbf{17} & 1.075 & 4.103 & 1.089 & 5.372 & 1.097 & 16.275 & 1.117 & 28.424 & 1.122 & 16.741 & 1.111 & 11.840 & 1.097 & 9.763 & 1.115 & 13.306 \\
    \textbf{18} & \multicolumn{2}{c|}{0.997} & \multicolumn{2}{c|}{1.063} & \multicolumn{2}{c|}{0.998} & \multicolumn{2}{c|}{0.964} & \multicolumn{2}{c|}{0.973} & \multicolumn{2}{c|}{0.962} & \multicolumn{2}{c|}{0.997} & \multicolumn{2}{c}{1.001} \\
    \textbf{19} & 0.453 & 1.255 & 0.420 & 1.369 & 0.397 & 1.394 & 0.390 & 1.378 & 0.375 & 1.404 & 0.373 & 1.438 & 0.378 & 1.443 & 0.379 & 1.462 \\
    \textbf{20} & \multicolumn{2}{c|}{1.040} & \multicolumn{2}{c|}{1.033} & \multicolumn{2}{c|}{1.018} & \multicolumn{2}{c|}{1.022} & \multicolumn{2}{c|}{1.009} & 1.008 & 1.247 & 1.047 & 1.145 & 1.051 & 1.203 \\
    \textbf{21} & \multicolumn{2}{c|}{1.175} & \multicolumn{2}{c|}{1.114} & \multicolumn{2}{c|}{1.085} & \multicolumn{2}{c|}{1.066} & \multicolumn{2}{c|}{1.088} & 1.060 & 6.132 & 1.133 & 1.926 & 1.134 & 1.793 \\
    \textbf{22} & 1.037 & 1.135 & 0.996 & 1.093 & 1.011 & 1.158 & \multicolumn{2}{c|}{1.029} & 1.024 & 1.051 & 1.001 & 1.080 & 0.970 & 1.167 & 0.987 & 1.179 \\
    \textbf{23} & \multicolumn{2}{c|}{1.145} & \multicolumn{2}{c|}{1.149} & \multicolumn{2}{c|}{1.139} & \multicolumn{2}{c|}{1.130} & \multicolumn{2}{c|}{1.137} & \multicolumn{2}{c|}{1.133} & \multicolumn{2}{c|}{1.129} & \multicolumn{2}{c}{1.119} \\
    \textbf{24} & 0.381 & 2.168 & 0.359 & 2.241 & 0.359 & 2.275 & 0.375 & 2.177 & 0.371 & 2.133 & 0.390 & 2.009 & 0.396 & 1.932 & 0.400 & 1.951 \\
    \textbf{25} & 0.645 & 1.126 & 0.605 & 1.140 & 0.574 & 1.168 & 0.555 & 1.129 & 0.521 & 1.124 & 0.559 & 1.083 & 0.566 & 1.076 & 0.547 & 1.140 \\
    \textbf{26} & 1.011 & 1.060 & 0.987 & 1.128 & 0.969 & 1.115 & 0.973 & 1.094 & 0.951 & 1.131 & 0.955 & 1.178 & 0.943 & 1.194 & 0.946 & 1.220 \\
    \textbf{27} & \multicolumn{2}{c|}{0.638} & \multicolumn{2}{c|}{0.688} & \multicolumn{2}{c|}{0.678} & \multicolumn{2}{c|}{0.678} & \multicolumn{2}{c|}{0.662} & 0.534 & 0.993 & \multicolumn{2}{c|}{0.710} & \multicolumn{2}{c}{0.689} \\
    \textbf{28} & \multicolumn{2}{c|}{0.784} & \multicolumn{2}{c|}{0.740} & \multicolumn{2}{c|}{0.758} & \multicolumn{2}{c|}{0.780} & \multicolumn{2}{c|}{0.781} & 0.700 & 0.880 & \multicolumn{2}{c|}{0.807} & \multicolumn{2}{c}{0.795} \\
    \textbf{29} & \multicolumn{2}{c|}{1.094} & \multicolumn{2}{c|}{1.138} & \multicolumn{2}{c|}{1.096} & \multicolumn{2}{c|}{1.092} & \multicolumn{2}{c|}{1.129} & 1.048 & 1.182 & \multicolumn{2}{c|}{1.075} & \multicolumn{2}{c}{1.141} \\
    \textbf{30} & \multicolumn{2}{c|}{2.207} & \multicolumn{2}{c|}{2.106} & \multicolumn{2}{c|}{2.071} & \multicolumn{2}{c|}{1.607} & \multicolumn{2}{c|}{2.298} & 1.232 & 2.896 & \multicolumn{2}{c|}{1.405} & \multicolumn{2}{c}{1.281} \\
    \textbf{31} & 1.112 & $\infty$ & 1.096 & $\infty$ & 1.101 & $\infty$ & 1.094 & $\infty$ & 1.079 & $\infty$ & 1.068 & $\infty$ & 1.066 & $\infty$ & 1.060 & $\infty$ \\
    \textbf{32} & \multicolumn{2}{c|}{0.968} & \multicolumn{2}{c|}{1.022} & \multicolumn{2}{c|}{1.019} & \multicolumn{2}{c|}{1.010} & \multicolumn{2}{c|}{1.023} & 0.983 & 1.166 & 1.070 & 1.107 & 1.035 & 1.137 \\
    \textbf{33} & 0.997 & $\infty$ & 1.016 & $\infty$ & 1.013 & $\infty$ & 1.006 & $\infty$ & 1.018 & $\infty$ & 1.022 & $\infty$ & 1.007 & $\infty$ & 0.994 & $\infty$ \\
    \textbf{34} & 1.155 & $\infty$ & 1.236 & $\infty$ & 1.220 & $\infty$ & 1.235 & $\infty$ & 1.244 & $\infty$ & \multicolumn{2}{c|}{1.180} & $\infty$ & $\infty$ & $\infty$ & $\infty$ \\
    \textbf{35} & 0.996 & 1.872 & \multicolumn{2}{c|}{1.424} & \multicolumn{2}{c|}{1.224} & \multicolumn{2}{c|}{1.136} & \multicolumn{2}{c|}{1.185} & 1.148 & 1.225 & 1.307 & 1.428 & 1.147 & 1.304 \\
    \textbf{36} & 0.889 & $\infty$ & 1.002 & $\infty$ & 1.003 & $\infty$ & 1.042 & $\infty$ & 0.971 & $\infty$ & 0.962 & $\infty$ & 0.938 & $\infty$ & 0.924 & $\infty$ \\
    \textbf{37} & \multicolumn{2}{c|}{1.088} & \multicolumn{2}{c|}{1.063} & \multicolumn{2}{c|}{1.059} & \multicolumn{2}{c|}{1.002} & \multicolumn{2}{c|}{0.953} & \multicolumn{2}{c|}{0.911} & \multicolumn{2}{c|}{0.953} & \multicolumn{2}{c}{0.979} \\
    \textbf{38} & \multicolumn{2}{c|}{1.461} & \multicolumn{2}{c|}{1.676} & \multicolumn{2}{c|}{1.588} & \multicolumn{2}{c|}{1.537} & \multicolumn{2}{c|}{1.582} & \multicolumn{2}{c|}{1.729} & \multicolumn{2}{c|}{1.667} & \multicolumn{2}{c}{1.825} \\
    \textbf{39} & \multicolumn{2}{c|}{0.798} & \multicolumn{2}{c|}{0.837} & \multicolumn{2}{c|}{0.847} & \multicolumn{2}{c|}{0.843} & \multicolumn{2}{c|}{0.844} & 0.807 & 0.870 & \multicolumn{2}{c|}{0.835} & \multicolumn{2}{c}{0.849} \\
    \textbf{40} & 0.899 & $\infty$ & 1.036 & $\infty$ & 1.055 & $\infty$ & 1.060 & 9.469 & 1.103 & 4.488 & 1.119 & 2.674 & 1.134 & 2.094 & 1.170 & 2.106 \\
    \textbf{41} & 0.838 & 0.848 & \multicolumn{2}{c|}{1.058} & 0.899 & 1.011 & 0.839 & 0.953 & 0.809 & 0.996 & 0.846 & 0.957 & 0.841 & 0.978 & 0.935 & 0.989 \\
    \textbf{42} & \multicolumn{2}{c|}{1.021} & \multicolumn{2}{c|}{1.036} & \multicolumn{2}{c|}{1.055} & 1.060 & 1.060 & \multicolumn{2}{c|}{1.151} & \multicolumn{2}{c|}{1.177} & \multicolumn{2}{c|}{1.210} & \multicolumn{2}{c}{1.207} \\
    \textbf{43} & 0.911 & 1.175 & 0.976 & 1.171 & \multicolumn{2}{c|}{1.122} & \multicolumn{2}{c|}{1.130} & \multicolumn{2}{c|}{1.122} & 0.977 & 1.175 & \multicolumn{2}{c|}{1.097} & \multicolumn{2}{c}{1.115} \\
    \textbf{44} & 1.026 & 4.489 & 1.057 & 2.485 & 1.067 & 5.516 & 1.095 & 3.085 & 1.120 & 3.385 & \multicolumn{2}{c|}{1.178} & 1.068 & $\infty$ & 1.095 & $\infty$ \\
    \textbf{45} & 0.075 & 1.681 & 0.063 & 1.614 & 0.063 & 1.581 & 0.058 & 1.582 & 0.062 & 1.598 & 0.058 & 1.593 & 0.058 & 1.542 & 0.054 & 1.531 \\
    \textbf{46} & 0.798 & 1.968 & 0.828 & 1.676 & 0.831 & 1.588 & 0.823 & 1.537 & 0.823 & 1.582 & 0.823 & 1.729 & 0.811 & 1.773 & 0.795 & 1.815 \\
    \textbf{47} & 0.569 & $\infty$ & 0.553 & $\infty$ & 0.548 & $\infty$ & 0.547 & $\infty$ & 0.546 & $\infty$ & 0.570 & $\infty$ & 0.598 & 350.595 & 0.613 & 317.364 \\
    \textbf{48} & \multicolumn{2}{c|}{0.906} & \multicolumn{2}{c|}{1.036} & \multicolumn{2}{c|}{1.055} & \multicolumn{2}{c|}{1.060} & \multicolumn{2}{c|}{1.103} & 1.061 & 1.182 & \multicolumn{2}{c|}{1.134} & \multicolumn{2}{c}{1.183} \\
    \textbf{49} & 0.718 & 0.826 & 0.729 & 0.814 & 0.728 & 0.856 & 0.688 & 0.800 & \multicolumn{2}{c|}{0.671} & \multicolumn{2}{c|}{0.681} & 0.685 & 0.711 & \multicolumn{2}{c}{0.670} \\
    \textbf{50} & 0.768 & 1.166 & \multicolumn{2}{c|}{0.736} & \multicolumn{2}{c|}{0.754} & \multicolumn{2}{c|}{0.767} & \multicolumn{2}{c|}{0.808} & \multicolumn{2}{c|}{0.770} & \multicolumn{2}{c|}{0.783} & \multicolumn{2}{c}{0.829} \\
    \textbf{51} & \multicolumn{2}{c|}{1.765} & \multicolumn{2}{c|}{0.689} & \multicolumn{2}{c|}{0.708} & \multicolumn{2}{c|}{0.715} & \multicolumn{2}{c|}{0.733} & \multicolumn{2}{c|}{0.704} & \multicolumn{2}{c|}{0.701} & \multicolumn{2}{c}{0.719} \\
    \textbf{52} & \multicolumn{2}{c|}{1.861} & \multicolumn{2}{c|}{1.551} & \multicolumn{2}{c|}{1.471} & \multicolumn{2}{c|}{1.487} & \multicolumn{2}{c|}{1.471} & 1.391 & 1.893 & \multicolumn{2}{c|}{1.444} & \multicolumn{2}{c}{1.331} \\
    \textbf{53} & \multicolumn{2}{c|}{1.168} & \multicolumn{2}{c|}{1.254} & \multicolumn{2}{c|}{1.145} & \multicolumn{2}{c|}{1.112} & \multicolumn{2}{c|}{1.091} & \multicolumn{2}{c|}{1.146} & \multicolumn{2}{c|}{1.130} & \multicolumn{2}{c}{1.137} \\
    \textbf{54} & 0.638 & 1.171 & 0.688 & 1.196 & 0.681 & 1.123 & 0.678 & 1.099 & 0.662 & 1.068 & 0.681 & 1.007 & \multicolumn{2}{c|}{0.711} & \multicolumn{2}{c}{0.717} \\
    \textbf{55} & \multicolumn{2}{c|}{0.638} & \multicolumn{2}{c|}{0.688} & \multicolumn{2}{c|}{0.681} & \multicolumn{2}{c|}{0.678} & \multicolumn{2}{c|}{0.662} & 0.668 & 0.705 & \multicolumn{2}{c|}{0.672} & \multicolumn{2}{c}{0.661} \\
    \textbf{56} & 1.053 & 12.020 & 1.015 & $\infty$ & 1.034 & $\infty$ & 1.041 & $\infty$ & 1.088 & $\infty$ & 1.146 & $\infty$ & 1.130 & 120.333 & 1.137 & $\infty$ \\
    \textbf{57} & 1.463 & 2.392 & 1.303 & 2.106 & 1.067 & 1.893 & 1.047 & 1.896 & 1.111 & 2.153 & 1.104 & 2.385 & 1.273 & 1.533 & \multicolumn{2}{c}{1.281} \\
    \textbf{58} & 1.128 & 3.714 & 1.032 & 1.822 & 0.939 & 2.086 & 0.943 & 2.212 & 0.930 & 2.325 & 0.911 & 2.229 & 0.953 & 1.279 & 0.975 & 1.114 \\
    \textbf{59} & \multicolumn{2}{c|}{0.952} & \multicolumn{2}{c|}{0.983} & \multicolumn{2}{c|}{1.008} & \multicolumn{2}{c|}{1.020} & \multicolumn{2}{c|}{1.044} & 0.989 & 1.059 & \multicolumn{2}{c|}{1.004} & \multicolumn{2}{c}{0.994} \\
    \textbf{60} & 1.080 & 1.122 & 1.106 & 1.173 & 1.109 & 1.160 & 1.085 & 1.159 & 1.107 & 1.154 & \multicolumn{2}{c|}{1.133} & \multicolumn{2}{c|}{1.129} & \multicolumn{2}{c}{1.119} \\
    \textbf{61} & \multicolumn{2}{c|}{0.638} & 0.661 & 1.014 & 0.662 & 1.085 & 0.657 & 1.104 & 0.643 & 1.062 & 0.664 & 1.073 & 0.652 & 1.116 & 0.661 & 1.174 \\
    \textbf{62} & \multicolumn{2}{c|}{0.887} & \multicolumn{2}{c|}{0.900} & \multicolumn{2}{c|}{0.911} & \multicolumn{2}{c|}{0.885} & \multicolumn{2}{c|}{0.911} & \multicolumn{2}{c|}{0.907} & 0.925 & 0.940 & 0.892 & 0.965 \\
    \textbf{63} & 0     & 0.888 & 0     & 0.876 & 0     & 0.883 & 0     & 0.881 & 0     & 0.893 & 0     & 0.881 & 0     & 0.893 & 0     & 0.902 \\
    \bottomrule
    \end{tabular}%
    
      \caption{Individual input oriented $\alpha$-returns to scale for 1987-1994}
  \label{IARTS8794}%
}\end{sidewaystable}%


\begin{landscape}

\begin{table}[htbp]
{\tiny  \centering
    \begin{tabular}{c|cc|cc|cc|cc|cc|cc|cc|cc}
    \toprule
    \multicolumn{1}{c|}{\multirow{2}[4]{*}{\textbf{DMU}}} & \multicolumn{2}{c|}{\textbf{1995}} & \multicolumn{2}{c|}{\textbf{1996}} & \multicolumn{2}{c|}{\textbf{1997}} & \multicolumn{2}{c|}{\textbf{1998}} & \multicolumn{2}{c|}{\textbf{1999}} & \multicolumn{2}{c|}{\textbf{2000}} & \multicolumn{2}{c|}{\textbf{2001}} & \multicolumn{2}{c}{\textbf{2002}} \\
\cmidrule{2-17}          & $\alpha_-^\star$ & $\alpha_+^\star$ & $\alpha_-^\star$ & $\alpha_+^\star$ & $\alpha_-^\star$ & $\alpha_+^\star$ & $\alpha_-^\star$ & $\alpha_+^\star$ & $\alpha_-^\star$ & $\alpha_+^\star$ & $\alpha_-^\star$ & $\alpha_+^\star$ & $\alpha_-^\star$ & $\alpha_+^\star$ & $\alpha_-^\star$ & $\alpha_+^\star$ \\
    \midrule
    \textbf{1} & 0.902 & 1.161 & 0.835 & 1.149 & 0.844 & 1.140 & 0.913 & 1.156 & 0.840 & 1.157 & 0.821 & 1.195 & 0.852 & 1.221 & 0.834 & 1.270 \\
    \textbf{2} & \multicolumn{2}{c|}{1.004} & \multicolumn{2}{c|}{1.000} & \multicolumn{2}{c|}{0.974} & \multicolumn{2}{c|}{1.046} & 1.054 & 1.572 & 1.026 & 2.696 & 1.011 & $\infty$ & 0.993 & 15.450 \\
    \textbf{3} & 1.124 & 2.328 & 1.233 & 1.874 & 1.259 & 1.839 & 1.142 & 1.895 & 1.235 & 1.559 & \multicolumn{2}{c|}{1.406} & 1.308 & 1.513 & 1.129 & 1.461 \\
    \textbf{4} & 1.171 & 2.011 & 1.213 & 2.258 & 0.866 & 2.260 & 0.810 & 2.465 & 0.805 & 2.489 & 0.822 & 1.953 & 0.810 & 1.974 & 0.833 & 2.226 \\
    \textbf{5} & 1.052 & $\infty$ & 1.045 & $\infty$ & 1.653 & $\infty$ & 3.577 & $\infty$ & 1.071 & $\infty$ & 0.896 & $\infty$ & 0.978 & 3.727 & \multicolumn{2}{c}{$\infty$} \\
    \textbf{6} & \multicolumn{2}{c|}{0.913} & \multicolumn{2}{c|}{0.897} & \multicolumn{2}{c|}{0.847} & \multicolumn{2}{c|}{0.816} & 0.795 & 0.897 & 0.753 & 1.004 & 0.717 & 1.069 & 0.871 & 0.909 \\
    \textbf{7} & 0.208 & 1.141 & 0.196 & 1.146 & 0.200 & 1.105 & 0.200 & 1.121 & 0.205 & 1.111 & 0.200 & 1.099 & 0.216 & 1.101 & 0.238 & 1.118 \\
    \textbf{8} & 1.011 & 1.697 & 1.013 & 1.810 & 1.016 & 1.936 & 0.996 & 2.177 & 1.018 & 3.086 & 1.010 & 2.137 & 1.009 & 2.042 & 1.008 & 1.589 \\
    \textbf{9} & 0.975 & 1.215 & 0.960 & 1.292 & 0.950 & 1.300 & 0.947 & 1.324 & 0.940 & 1.413 & 0.948 & 1.540 & 0.944 & 1.373 & 0.926 & 1.326 \\
    \textbf{10} & \multicolumn{2}{c|}{1.179} & \multicolumn{2}{c|}{1.152} & \multicolumn{2}{c|}{1.121} & 1.095 & 1.122 & 1.090 & 1.111 & 1.067 & 1.099 & 1.059 & 1.150 & 1.045 & 1.509 \\
    \textbf{11} & 1.022 & 1.172 & 1.023 & 1.168 & 1.003 & 1.116 & 1.018 & 1.105 & 0.996 & 1.120 & 0.978 & 1.138 & 0.978 & 1.124 & 0.941 & 1.077 \\
    \textbf{12} & \multicolumn{2}{c|}{1.022} & \multicolumn{2}{c|}{1.023} & \multicolumn{2}{c|}{1.003} & \multicolumn{2}{c|}{1.003} & \multicolumn{2}{c|}{0.983} & \multicolumn{2}{c|}{0.955} & \multicolumn{2}{c|}{0.972} & \multicolumn{2}{c}{0.961} \\
    \textbf{13} & \multicolumn{2}{c|}{1.160} & \multicolumn{2}{c|}{1.017} & \multicolumn{2}{c|}{0.992} & \multicolumn{2}{c|}{0.973} & \multicolumn{2}{c|}{0.973} & \multicolumn{2}{c|}{0.931} & \multicolumn{2}{c|}{0.922} & \multicolumn{2}{c}{0.954} \\
    \textbf{14} & \multicolumn{2}{c|}{1.054} & \multicolumn{2}{c|}{1.087} & \multicolumn{2}{c|}{0.991} & \multicolumn{2}{c|}{0.964} & \multicolumn{2}{c|}{0.952} & \multicolumn{2}{c|}{0.960} & \multicolumn{2}{c|}{0.962} & \multicolumn{2}{c}{1.020} \\
    \textbf{15} & \multicolumn{2}{c|}{1.108} & \multicolumn{2}{c|}{1.047} & \multicolumn{2}{c|}{1.050} & \multicolumn{2}{c|}{0.999} & \multicolumn{2}{c|}{1.005} & \multicolumn{2}{c|}{1.080} & \multicolumn{2}{c|}{1.049} & \multicolumn{2}{c}{1.083} \\
    \textbf{16} & \multicolumn{2}{c|}{1.028} & \multicolumn{2}{c|}{1.025} & \multicolumn{2}{c|}{1.045} & \multicolumn{2}{c|}{1.008} & \multicolumn{2}{c|}{1.001} & \multicolumn{2}{c|}{1.037} & \multicolumn{2}{c|}{0.992} & \multicolumn{2}{c}{0.996} \\
    \textbf{17} & 1.097 & 22.830 & 1.096 & 11.518 & 1.060 & 7.508 & 1.061 & 6.154 & 1.053 & 4.195 & 1.053 & 4.222 & 1.044 & 3.022 & 1.001 & 2.390 \\
    \textbf{18} & \multicolumn{2}{c|}{1.006} & \multicolumn{2}{c|}{0.995} & \multicolumn{2}{c|}{0.998} & \multicolumn{2}{c|}{1.010} & \multicolumn{2}{c|}{1.024} & \multicolumn{2}{c|}{1.152} & 1.054 & 1.054 & \multicolumn{2}{c}{1.036} \\
    \textbf{19} & 0.378 & 1.472 & 0.400 & 1.458 & 0.392 & 1.435 & 0.403 & 1.190 & 0.424 & 1.433 & 0.431 & 1.400 & 0.427 & 1.222 & 0.428 & 1.388 \\
    \textbf{20} & 1.014 & 1.198 & 1.045 & 1.259 & 1.067 & 1.336 & 1.074 & 1.331 & 1.092 & 1.649 & 1.066 & 2.111 & 1.064 & 1.885 & 1.031 & 2.041 \\
    \textbf{21} & 1.120 & 1.587 & 1.117 & 2.246 & 1.082 & 7.745 & 3.681 & 6.870 & 1.511 & 7.028 & 1.045 & $\infty$ & 1.087 & $\infty$ & 1.209 & $\infty$ \\
    \textbf{22} & 0.962 & 1.123 & 0.957 & 1.215 & 0.960 & 1.313 & 0.926 & 1.168 & 0.892 & 1.187 & 0.909 & 1.320 & 0.956 & 1.254 & 0.938 & 1.093 \\
    \textbf{23} & \multicolumn{2}{c|}{1.101} & \multicolumn{2}{c|}{1.100} & \multicolumn{2}{c|}{1.077} & \multicolumn{2}{c|}{1.105} & \multicolumn{2}{c|}{1.120} & \multicolumn{2}{c|}{1.138} & \multicolumn{2}{c|}{1.114} & \multicolumn{2}{c}{1.028} \\
    \textbf{24} & 0.397 & 1.962 & 0.402 & 1.943 & 0.390 & 2.064 & 0.399 & 2.210 & 0.395 & 2.343 & 0.404 & 2.364 & 0.404 & 2.352 & 0.400 & 2.361 \\
    \textbf{25} & 0.530 & 1.144 & 0.553 & 1.140 & 0.516 & 1.150 & 0.534 & 1.109 & 0.527 & 1.123 & 0.539 & 1.099 & 0.555 & 1.035 & 0.542 & 1.041 \\
    \textbf{26} & 0.989 & 1.211 & 0.957 & 1.242 & 0.934 & 1.245 & 0.909 & 1.219 & 0.916 & 1.200 & 0.915 & 1.244 & 0.926 & 1.204 & 0.931 & 1.172 \\
    \textbf{27} & \multicolumn{2}{c|}{0.731} & \multicolumn{2}{c|}{0.699} & \multicolumn{2}{c|}{0.690} & 0.642 & 0.847 & 0.680 & 0.816 & 0.618 & 0.842 & 0.624 & 0.868 & 0.608 & 0.915 \\
    \textbf{28} & \multicolumn{2}{c|}{0.779} & \multicolumn{2}{c|}{0.747} & \multicolumn{2}{c|}{0.720} & 0.577 & 0.813 & 0.626 & 0.792 & 0.618 & 0.793 & 0.597 & 0.818 & 0.593 & 0.859 \\
    \textbf{29} & 1.038 & 1.151 & 1.097 & 1.246 & 1.066 & 1.212 & \multicolumn{2}{c|}{1.023} & \multicolumn{2}{c|}{0.999} & \multicolumn{2}{c|}{0.920} & \multicolumn{2}{c|}{0.882} & \multicolumn{2}{c}{0.895} \\
    \textbf{30} & \multicolumn{2}{c|}{1.249} & \multicolumn{2}{c|}{1.211} & \multicolumn{2}{c|}{1.302} & \multicolumn{2}{c|}{1.083} & \multicolumn{2}{c|}{1.134} & 1.058 & 1.134 & \multicolumn{2}{c|}{1.037} & 1.252 & 1.290 \\
    \textbf{31} & 1.062 & $\infty$ & 1.024 & $\infty$ & 1.028 & $\infty$ & 1.061 & $\infty$ & 1.062 & $\infty$ & 1.065 & $\infty$ & 1.075 & $\infty$ & 1.105 & $\infty$ \\
    \textbf{32} & 1.045 & 1.100 & 1.016 & 1.138 & 0.998 & 1.087 & 1.008 & 1.140 & 0.988 & 1.154 & 0.983 & 1.190 & 0.980 & 1.131 & 0.957 & 1.086 \\
    \textbf{33} & 0.992 & $\infty$ & 0.983 & $\infty$ & 0.945 & $\infty$ & 0.957 & $\infty$ & 0.982 & $\infty$ & 0.990 & $\infty$ & 0.982 & $\infty$ & 0.983 & $\infty$ \\
    \textbf{34} & 7.426 & $\infty$ & 2.920 & $\infty$ & 23.236 & $\infty$ & 1.124 & $\infty$ & 1.075 & $\infty$ & 1.189 & $\infty$ & 1.037 & $\infty$ & 1.169 & $\infty$ \\
    \textbf{35} & 1.085 & 1.197 & 1.089 & 1.169 & \multicolumn{2}{c|}{1.070} & \multicolumn{2}{c|}{1.131} & \multicolumn{2}{c|}{1.079} & \multicolumn{2}{c|}{1.093} & \multicolumn{2}{c|}{1.145} & 1.081 & 1.090 \\
    \textbf{36} & 0.977 & $\infty$ & 0.967 & $\infty$ & 0.924 & $\infty$ & 0.842 & $\infty$ & 0.931 & $\infty$ & 0.955 & $\infty$ & 1.136 & $\infty$ & 1.220 & $\infty$ \\
    \textbf{37} & \multicolumn{2}{c|}{0.990} & \multicolumn{2}{c|}{0.995} & \multicolumn{2}{c|}{1.018} & \multicolumn{2}{c|}{1.039} & 0.982 & 1.030 & \multicolumn{2}{c|}{1.084} & \multicolumn{2}{c|}{1.032} & \multicolumn{2}{c}{1.001} \\
    \textbf{38} & \multicolumn{2}{c|}{1.898} & \multicolumn{2}{c|}{1.595} & \multicolumn{2}{c|}{1.869} & \multicolumn{2}{c|}{1.389} & \multicolumn{2}{c|}{1.225} & \multicolumn{2}{c|}{0.984} & \multicolumn{2}{c|}{1.029} & \multicolumn{2}{c}{1.003} \\
    \textbf{39} & \multicolumn{2}{c|}{0.853} & \multicolumn{2}{c|}{0.847} & \multicolumn{2}{c|}{0.913} & \multicolumn{2}{c|}{0.897} & \multicolumn{2}{c|}{0.864} & \multicolumn{2}{c|}{0.835} & \multicolumn{2}{c|}{0.814} & \multicolumn{2}{c}{0.833} \\
    \textbf{40} & 1.164 & 2.160 & 1.387 & 1.771 & \multicolumn{2}{c|}{1.275} & \multicolumn{2}{c|}{1.109} & \multicolumn{2}{c|}{1.357} & \multicolumn{2}{c|}{1.238} & \multicolumn{2}{c|}{1.089} & \multicolumn{2}{c}{1.048} \\
    \textbf{41} & \multicolumn{2}{c|}{0.992} & \multicolumn{2}{c|}{1.028} & \multicolumn{2}{c|}{1.045} & \multicolumn{2}{c|}{1.030} & \multicolumn{2}{c|}{1.031} & \multicolumn{2}{c|}{0.967} & \multicolumn{2}{c|}{0.944} & \multicolumn{2}{c}{0.924} \\
    \textbf{42} & \multicolumn{2}{c|}{1.285} & \multicolumn{2}{c|}{1.183} & \multicolumn{2}{c|}{1.160} & \multicolumn{2}{c|}{1.597} & \multicolumn{2}{c|}{0.975} & \multicolumn{2}{c|}{1.025} & \multicolumn{2}{c|}{0.868} & \multicolumn{2}{c}{0.875} \\
    \textbf{43} & 1.086 & 1.135 & 1.055 & 1.133 & 0.986 & 1.193 & \multicolumn{2}{c|}{0.930} & \multicolumn{2}{c|}{0.985} & 0.892 & 1.088 & 0.993 & 1.105 & \multicolumn{2}{c}{1.055} \\
    \textbf{44} & 1.035 & $\infty$ & 1.041 & $\infty$ & 1.050 & 23.862 & 0.999 & 10.261 & 0.999 & 5.878 & 1.035 & 4.282 & 1.049 & 2.808 & 0.998 & 2.147 \\
    \textbf{45} & 0.057 & 1.509 & 0.051 & 1.475 & 0.052 & 1.473 & 0.054 & 1.471 & 0.053 & 1.497 & 0.045 & 1.508 & 0.048 & 1.512 & 0.044 & 1.548 \\
    \textbf{46} & 0.779 & 1.813 & 0.762 & 1.877 & 0.795 & 1.869 & 0.796 & 1.389 & 0.823 & 1.348 & 0.873 & 1.261 & 0.920 & 1.174 & 0.951 & 1.821 \\
    \textbf{47} & 0.608 & 96.449 & 0.601 & 33.229 & 0.601 & 11.707 & 0.615 & 5.756 & 0.634 & 3.597 & 0.672 & 2.635 & 0.694 & 2.173 & 0.713 & 2.296 \\
    \textbf{48} & \multicolumn{2}{c|}{1.281} & \multicolumn{2}{c|}{1.285} & \multicolumn{2}{c|}{1.126} & \multicolumn{2}{c|}{1.432} & \multicolumn{2}{c|}{1.418} & \multicolumn{2}{c|}{1.399} & \multicolumn{2}{c|}{1.355} & \multicolumn{2}{c}{1.309} \\
    \textbf{49} & \multicolumn{2}{c|}{0.693} & \multicolumn{2}{c|}{0.724} & \multicolumn{2}{c|}{0.757} & 0.829 & 0.888 & 0.816 & 0.871 & 0.834 & 0.889 & 0.776 & 0.890 & \multicolumn{2}{c}{0.915} \\
    \textbf{50} & \multicolumn{2}{c|}{0.802} & \multicolumn{2}{c|}{0.783} & 0.829 & 0.847 & \multicolumn{2}{c|}{0.847} & \multicolumn{2}{c|}{0.815} & \multicolumn{2}{c|}{0.832} & \multicolumn{2}{c|}{0.862} & \multicolumn{2}{c}{0.871} \\
    \textbf{51} & \multicolumn{2}{c|}{0.746} & \multicolumn{2}{c|}{0.750} & \multicolumn{2}{c|}{0.751} & \multicolumn{2}{c|}{0.757} & \multicolumn{2}{c|}{0.761} & \multicolumn{2}{c|}{0.744} & \multicolumn{2}{c|}{0.728} & \multicolumn{2}{c}{0.725} \\
    \textbf{52} & 1.078 & 1.356 & 1.000 & 1.473 & 0.992 & 1.511 & 0.984 & 1.624 & 0.971 & 1.854 & 0.974 & 1.868 & 0.992 & 1.186 & 0.993 & 1.306 \\
    \textbf{53} & \multicolumn{2}{c|}{1.145} & \multicolumn{2}{c|}{1.183} & \multicolumn{2}{c|}{1.185} & \multicolumn{2}{c|}{1.339} & \multicolumn{2}{c|}{1.357} & \multicolumn{2}{c|}{1.399} & \multicolumn{2}{c|}{1.355} & \multicolumn{2}{c}{1.309} \\
    \textbf{54} & \multicolumn{2}{c|}{0.718} & \multicolumn{2}{c|}{0.727} & \multicolumn{2}{c|}{0.749} & \multicolumn{2}{c|}{0.777} & \multicolumn{2}{c|}{0.761} & \multicolumn{2}{c|}{0.769} & \multicolumn{2}{c|}{0.779} & \multicolumn{2}{c}{0.782} \\
    \textbf{55} & \multicolumn{2}{c|}{0.677} & \multicolumn{2}{c|}{0.699} & \multicolumn{2}{c|}{0.693} & \multicolumn{2}{c|}{0.758} & \multicolumn{2}{c|}{0.784} & \multicolumn{2}{c|}{0.775} & \multicolumn{2}{c|}{0.740} & \multicolumn{2}{c}{0.757} \\
    \textbf{56} & 1.145 & $\infty$ & 1.183 & $\infty$ & 1.126 & $\infty$ & 1.296 & $\infty$ & 1.308 & $\infty$ & 1.608 & 11.164 & 1.287 & 5.505 & 1.304 & 3.798 \\
    \textbf{57} & \multicolumn{2}{c|}{1.249} & \multicolumn{2}{c|}{1.190} & \multicolumn{2}{c|}{1.010} & 0.964 & 1.039 & \multicolumn{2}{c|}{0.982} & \multicolumn{2}{c|}{0.990} & 0.981 & 0.983 & 0.929 & 1.062 \\
    \textbf{58} & 0.990 & 1.174 & 0.995 & 1.256 & 0.959 & 1.279 & 0.987 & 1.154 & 1.016 & 1.259 & 1.074 & 1.291 & 1.128 & 1.148 & \multicolumn{2}{c}{1.082} \\
    \textbf{59} & \multicolumn{2}{c|}{0.992} & \multicolumn{2}{c|}{1.005} & \multicolumn{2}{c|}{0.989} & \multicolumn{2}{c|}{0.987} & \multicolumn{2}{c|}{1.016} & 1.012 & 1.190 & \multicolumn{2}{c|}{1.079} & 0.987 & 1.095 \\
    \textbf{60} & \multicolumn{2}{c|}{1.116} & \multicolumn{2}{c|}{1.104} & \multicolumn{2}{c|}{1.105} & \multicolumn{2}{c|}{1.111} & \multicolumn{2}{c|}{1.064} & 1.074 & 1.078 & 1.055 & 1.133 & 1.047 & 1.167 \\
    \textbf{61} & 0.677 & 1.194 & 0.699 & 1.211 & 0.693 & 1.157 & 0.708 & 0.882 & 0.723 & 0.866 & 0.722 & 0.886 & 0.829 & 0.896 & 0.787 & 0.880 \\
    \textbf{62} & \multicolumn{2}{c|}{0.956} & 0.928 & 0.958 & 0.996 & 1.002 & 0.871 & 1.049 & 0.963 & 1.020 & 0.891 & 0.971 & \multicolumn{2}{c|}{0.938} & \multicolumn{2}{c}{0.930} \\
    \textbf{63} & 0     & 0.912 & 0     & 0.902 & 0     & 0.916 & 0     & 0.913 & 0     & 0.909 & 0     & 0.886 & 0     & 0.869 & 0     & 0.871 \\
    \bottomrule
    \end{tabular}%
  \caption{Individual $\alpha$-returns to scale for 1995-2002}  \label{IARTS95-02}%
}
\end{table}%

\end{landscape}

\begin{landscape}

\begin{table}[htbp]
{\tiny  \centering
    \begin{tabular}{c|cc|cc|cc|cc|cc|cc|cc|cc}
    \toprule
    \multicolumn{1}{c|}{\multirow{2}[4]{*}{\textbf{DMU}}} & \multicolumn{2}{c|}{\textbf{2003}} & \multicolumn{2}{c|}{\textbf{2004}} & \multicolumn{2}{c|}{\textbf{2005}} & \multicolumn{2}{c|}{\textbf{2006}} & \multicolumn{2}{c|}{\textbf{2007}} & \multicolumn{2}{c|}{\textbf{2008}} & \multicolumn{2}{c|}{\textbf{2009}} & \multicolumn{2}{c}{\textbf{2010}} \\
\cmidrule{2-17}           & $\alpha_-^\star$ & $\alpha_+^\star$ & $\alpha_-^\star$ & $\alpha_+^\star$ & $\alpha_-^\star$ & $\alpha_+^\star$ & $\alpha_-^\star$ & $\alpha_+^\star$ & $\alpha_-^\star$ & $\alpha_+^\star$ & $\alpha_-^\star$ & $\alpha_+^\star$ & $\alpha_-^\star$ & $\alpha_+^\star$ & $\alpha_-^\star$ & $\alpha_+^\star$  \\
    \midrule
    \textbf{1} & 0.869 & 1.239 & 0.856 & 1.241 & 0.799 & 1.346 & 0.796 & 1.353 & 0.835 & 1.285 & 0.793 & 1.316 & 0.777 & 1.247 & 0.793 & 1.232 \\
    \textbf{2} & 0.995 & 17.538 & 0.975 & $\infty$ & 1.003 & $\infty$ & \multicolumn{2}{c|}{0.969} & 1.006 & 46.069 & 1.032 & 1.758 & 1.012 & 1.654 & 1.017 & 1.501 \\
    \textbf{3} & 1.248 & 1.427 & 1.200 & 1.366 & 1.173 & 1.360 & 1.090 & 1.333 & 1.101 & 1.291 & \multicolumn{2}{c|}{1.231} & 0.869 & 1.616 & 1.114 & 1.366 \\
    \textbf{4} & 0.827 & 2.499 & 0.860 & 2.821 & 0.850 & 3.068 & 0.882 & 2.111 & 0.861 & 2.691 & 0.876 & 2.176 & 0.862 & 2.806 & 0.829 & 2.568 \\
    \textbf{5} & \multicolumn{2}{c|}{1.029} & \multicolumn{2}{c|}{0.963} & \multicolumn{2}{c|}{1.038} & \multicolumn{2}{c|}{0.963} & 0.982 & 1.307 & 0.961 & 1.399 & 1.027 & 2.311 & 0.960 & 2.018 \\
    \textbf{6} & \multicolumn{2}{c|}{0.896} & \multicolumn{2}{c|}{0.943} & \multicolumn{2}{c|}{0.924} & 0.961 & 1.007 & \multicolumn{2}{c|}{0.934} & 0.915 & 0.958 & \multicolumn{2}{c|}{0.978} & \multicolumn{2}{c}{0.972} \\
    \textbf{7} & 0.221 & 1.108 & 0.204 & 1.092 & 0.199 & 1.080 & 0.227 & 1.070 & 0.278 & 1.075 & 0.329 & 1.059 & 0.412 & 1.101 & 0.424 & 1.093 \\
    \textbf{8} & 1.019 & 1.544 & 1.042 & 1.582 & 1.018 & 1.478 & 1.046 & 1.440 & 1.018 & 1.634 & 0.988 & 1.896 & 0.984 & 1.196 & 0.989 & 1.492 \\
    \textbf{9} & 0.920 & 1.355 & 0.912 & 1.394 & 0.916 & 1.480 & 0.938 & 1.363 & 0.974 & 1.178 & 0.997 & 1.207 & 0.983 & 1.266 & 0.995 & 1.285 \\
    \textbf{10} & 1.099 & 1.574 & 1.123 & 1.609 & 1.121 & 1.408 & 1.150 & 1.386 & 1.180 & 1.373 & 1.152 & 1.297 & \multicolumn{2}{c|}{1.020} & 1.092 & 1.117 \\
    \textbf{11} & 0.942 & 1.073 & 0.950 & 1.070 & 0.924 & 1.044 & 0.863 & 1.054 & 0.807 & 1.075 & 0.798 & 1.066 & 0.873 & 1.044 & 0.835 & 1.064 \\
    \textbf{12} & \multicolumn{2}{c|}{0.925} & \multicolumn{2}{c|}{0.913} & 0.870 & 0.940 & 0.845 & 1.008 & 0.835 & 1.013 & 0.839 & 1.057 & 0.889 & 1.063 & 0.846 & 1.076 \\
    \textbf{13} & \multicolumn{2}{c|}{0.952} & \multicolumn{2}{c|}{0.953} & \multicolumn{2}{c|}{0.961} & \multicolumn{2}{c|}{0.970} & \multicolumn{2}{c|}{0.951} & \multicolumn{2}{c|}{0.942} & \multicolumn{2}{c|}{1.078} & \multicolumn{2}{c}{1.079} \\
    \textbf{14} & 0.994 & 1.068 & \multicolumn{2}{c|}{0.940} & \multicolumn{2}{c|}{0.928} & 0.952 & 1.215 & 0.975 & 1.263 & 0.959 & 1.323 & 0.984 & 1.276 & 0.966 & 1.306 \\
    \textbf{15} & \multicolumn{2}{c|}{0.985} & \multicolumn{2}{c|}{0.935} & \multicolumn{2}{c|}{0.996} & \multicolumn{2}{c|}{1.082} & \multicolumn{2}{c|}{1.025} & \multicolumn{2}{c|}{0.992} & \multicolumn{2}{c|}{0.979} & \multicolumn{2}{c}{1.017} \\
    \textbf{16} & \multicolumn{2}{c|}{1.000} & \multicolumn{2}{c|}{0.975} & \multicolumn{2}{c|}{0.990} & \multicolumn{2}{c|}{0.997} & 0.979 & 1.028 & \multicolumn{2}{c|}{0.974} & 0.923 & 1.057 & 0.942 & 0.994 \\
    \textbf{17} & 0.999 & 2.204 & 0.990 & 2.418 & 0.966 & 2.451 & 0.946 & 2.387 & 0.927 & 2.620 & 0.924 & 2.698 & 0.956 & 2.471 & 0.950 & 2.250 \\
    \textbf{18} & \multicolumn{2}{c|}{1.017} & \multicolumn{2}{c|}{1.007} & \multicolumn{2}{c|}{1.017} & \multicolumn{2}{c|}{1.020} & \multicolumn{2}{c|}{1.041} & 0.979 & 1.080 & 0.990 & 1.137 & 0.949 & 1.023 \\
    \textbf{19} & 0.432 & 0.985 & 0.444 & 0.935 & 0.420 & 0.996 & 0.401 & 1.325 & 0.404 & 1.348 & 0.406 & 1.104 & 0.404 & 1.271 & 0.399 & 1.298 \\
    \textbf{20} & 1.025 & 2.375 & 0.996 & 2.855 & 0.997 & 3.145 & 1.012 & 3.074 & 1.013 & 1.982 & 0.989 & 1.355 & 1.388 & 4.036 & 1.056 & 3.951 \\
    \textbf{21} & 1.248 & $\infty$ & 1.178 & $\infty$ & 1.165 & $\infty$ & 1.174 & $\infty$ & 1.074 & $\infty$ & 1.072 & $\infty$ & 1.117 & $\infty$ & 1.049 & $\infty$ \\
    \textbf{22} & 0.941 & 1.068 & 0.914 & 1.045 & 0.925 & 1.150 & 0.934 & 1.165 & 0.959 & 1.100 & 0.960 & 1.029 & 0.936 & 1.033 & 0.977 & 1.060 \\
    \textbf{23} & \multicolumn{2}{c|}{1.065} & \multicolumn{2}{c|}{1.070} & \multicolumn{2}{c|}{1.052} & 1.056 & 1.095 & 1.072 & 1.240 & 1.132 & 1.235 & 1.062 & 1.071 & 1.032 & 1.244 \\
    \textbf{24} & 0.423 & 2.231 & 0.417 & 2.309 & 0.424 & 2.517 & 0.433 & 2.486 & 0.432 & 2.383 & 0.446 & 2.372 & 0.453 & 2.216 & 0.472 & 2.261 \\
    \textbf{25} & 0.562 & 1.016 & 0.571 & 1.048 & 0.599 & 1.055 & 0.593 & 1.066 & 0.564 & 1.113 & 0.620 & 1.093 & 0.694 & 1.094 & 0.688 & 1.100 \\
    \textbf{26} & 0.929 & 1.156 & 0.905 & 1.113 & 0.920 & 1.109 & 0.982 & 1.046 & 0.971 & 1.133 & 0.957 & 1.064 & 0.930 & 1.039 & 0.938 & 1.096 \\
    \textbf{27} & 0.621 & 0.911 & 0.534 & 0.971 & 0.509 & 1.008 & 0.507 & 1.020 & 0.483 & 1.044 & 0.470 & 1.067 & 0.648 & 0.958 & 0.536 & 1.040 \\
    \textbf{28} & 0.560 & 0.768 & 0.543 & 0.568 & 0.555 & 0.767 & \multicolumn{2}{c|}{0.557} & \multicolumn{2}{c|}{0.986} & \multicolumn{2}{c|}{0.946} & 0.691 & 0.769 & \multicolumn{2}{c}{0.577} \\
    \textbf{29} & \multicolumn{2}{c|}{0.862} & \multicolumn{2}{c|}{0.873} & \multicolumn{2}{c|}{0.861} & \multicolumn{2}{c|}{0.882} & \multicolumn{2}{c|}{0.891} & \multicolumn{2}{c|}{0.903} & \multicolumn{2}{c|}{0.910} & \multicolumn{2}{c}{0.926} \\
    \textbf{30} & 1.093 & 1.506 & 1.009 & 1.388 & 1.026 & 1.194 & 0.962 & 1.066 & 1.049 & 1.050 & 1.004 & 1.336 & 1.038 & 1.261 & \multicolumn{2}{c}{1.703} \\
    \textbf{31} & 1.149 & $\infty$ & 1.148 & $\infty$ & 1.095 & $\infty$ & 1.043 & $\infty$ & 1.025 & $\infty$ & 0.992 & $\infty$ & 0.976 & $\infty$ & 1.011 & $\infty$ \\
    \textbf{32} & 0.970 & 1.103 & 0.947 & 1.074 & 0.949 & 1.054 & 0.912 & 0.980 & 0.981 & 1.037 & \multicolumn{2}{c|}{0.977} & 1.017 & 1.045 & 0.964 & 1.101 \\
    \textbf{33} & 0.936 & $\infty$ & 0.958 & $\infty$ & 0.961 & $\infty$ & 0.945 & $\infty$ & 0.948 & 3.019 & 0.965 & 1.987 & 0.955 & 1.651 & 0.977 & 1.405 \\
    \textbf{34} & 1.179 & $\infty$ & 1.144 & $\infty$ & 1.189 & $\infty$ & 1.160 & $\infty$ & 1.090 & $\infty$ & 1.072 & $\infty$ & 1.120 & $\infty$ & 1.089 & $\infty$ \\
    \textbf{35} & \multicolumn{2}{c|}{1.081} & 0.995 & 1.130 & 0.996 & 1.217 & 0.984 & 1.195 & 1.056 & 1.086 & 1.009 & 1.140 & 0.974 & 1.182 & 0.997 & 1.156 \\
    \textbf{36} & 1.235 & $\infty$ & 1.168 & $\infty$ & 1.029 & $\infty$ & 1.154 & $\infty$ & 1.153 & 3.060 & 1.131 & 2.941 & 1.090 & 2.815 & 1.005 & 2.580 \\
    \textbf{37} & \multicolumn{2}{c|}{0.949} & 0.939 & 0.985 & 0.895 & 1.058 & 0.950 & 0.973 & 0.869 & 1.023 & 0.879 & 1.057 & 0.907 & 0.980 & 0.897 & 0.944 \\
    \textbf{38} & \multicolumn{2}{c|}{0.918} & \multicolumn{2}{c|}{1.076} & \multicolumn{2}{c|}{0.977} & \multicolumn{2}{c|}{1.051} & \multicolumn{2}{c|}{1.105} & 1.077 & 1.156 & 1.043 & 1.333 & 0.956 & 1.376 \\
    \textbf{39} & \multicolumn{2}{c|}{0.827} & \multicolumn{2}{c|}{0.860} & 0.848 & 0.905 & 0.832 & 0.927 & 0.816 & 0.946 & 0.797 & 1.001 & 0.814 & 0.942 & 0.819 & 0.985 \\
    \textbf{40} & \multicolumn{2}{c|}{0.936} & 0.873 & 1.655 & 1.055 & 1.239 & 0.984 & 1.449 & \multicolumn{2}{c|}{0.957} & \multicolumn{2}{c|}{1.001} & \multicolumn{2}{c|}{0.916} & \multicolumn{2}{c}{0.907} \\
    \textbf{41} & 0.903 & 0.952 & \multicolumn{2}{c|}{0.951} & 0.943 & 0.966 & 0.938 & 0.978 & \multicolumn{2}{c|}{0.946} & 0.828 & 0.944 & 0.688 & 0.946 & 0.757 & 0.923 \\
    \textbf{42} & \multicolumn{2}{c|}{1.007} & \multicolumn{2}{c|}{1.092} & \multicolumn{2}{c|}{1.044} & \multicolumn{2}{c|}{1.062} & \multicolumn{2}{c|}{0.971} & \multicolumn{2}{c|}{0.977} & \multicolumn{2}{c|}{1.099} & \multicolumn{2}{c}{1.088} \\
    \textbf{43} & 1.044 & 1.099 & 0.998 & 1.144 & 0.955 & 1.114 & 0.871 & 1.113 & 0.745 & 1.137 & 0.694 & 1.064 & 0.769 & 1.123 & 0.693 & 1.084 \\
    \textbf{44} & 0.956 & 2.060 & 0.962 & 1.972 & 0.945 & 1.710 & 1.002 & 1.157 & 0.999 & 1.092 & 0.955 & 1.091 & 0.945 & 1.085 & 0.934 & 1.336 \\
    \textbf{45} & 0.029 & 1.530 & 0.004 & 1.507 & 0     & 1.565 & 0     & 1.542 & 0     & 1.506 & 0     & 1.507 & 0.010 & 1.478 & 0     & 1.499 \\
    \textbf{46} & 0.931 & 1.186 & 0.943 & 1.190 & 0.955 & 1.209 & 0.970 & 1.233 & 0.980 & 1.258 & 0.942 & 1.296 & 0.978 & 1.439 & 0.939 & 1.534 \\
    \textbf{47} & 0.731 & 1.744 & 0.730 & 1.803 & 0.737 & 1.687 & 0.730 & 1.649 & 0.717 & 1.689 & 0.599 & 2.117 & 0.665 & 1.674 & 0.685 & 1.586 \\
    \textbf{48} & \multicolumn{2}{c|}{1.270} & \multicolumn{2}{c|}{1.217} & \multicolumn{2}{c|}{1.147} & \multicolumn{2}{c|}{1.154} & \multicolumn{2}{c|}{1.304} & \multicolumn{2}{c|}{1.324} & \multicolumn{2}{c|}{1.298} & \multicolumn{2}{c}{0.849} \\
    \textbf{49} & \multicolumn{2}{c|}{0.911} & \multicolumn{2}{c|}{0.971} & \multicolumn{2}{c|}{1.008} & \multicolumn{2}{c|}{1.020} & \multicolumn{2}{c|}{1.044} & \multicolumn{2}{c|}{1.047} & 0.703 & 0.997 & \multicolumn{2}{c}{1.034} \\
    \textbf{50} & 0.905 & 0.918 & \multicolumn{2}{c|}{0.848} & \multicolumn{2}{c|}{0.840} & 0.872 & 0.872 & \multicolumn{2}{c|}{0.950} & \multicolumn{2}{c|}{0.950} & \multicolumn{2}{c|}{0.958} & \multicolumn{2}{c}{0.986} \\
    \textbf{51} & \multicolumn{2}{c|}{0.742} & \multicolumn{2}{c|}{0.732} & \multicolumn{2}{c|}{0.775} & \multicolumn{2}{c|}{0.767} & \multicolumn{2}{c|}{0.763} & \multicolumn{2}{c|}{0.790} & 0.776 & 0.815 & 0.807 & 0.839 \\
    \textbf{52} & 0.993 & 1.123 & \multicolumn{2}{c|}{0.983} & 0.990 & 1.099 & 1.151 & 1.293 & 1.014 & 1.185 & 1.104 & 1.340 & 1.018 & 1.358 & 0.971 & 1.471 \\
    \textbf{53} & \multicolumn{2}{c|}{1.235} & \multicolumn{2}{c|}{1.164} & \multicolumn{2}{c|}{1.124} & \multicolumn{2}{c|}{1.154} & \multicolumn{2}{c|}{1.268} & \multicolumn{2}{c|}{1.214} & \multicolumn{2}{c|}{1.205} & \multicolumn{2}{c}{0.914} \\
    \textbf{54} & \multicolumn{2}{c|}{0.774} & 0.789 & 0.795 & 0.766 & 0.803 & 0.716 & 0.799 & 0.735 & 0.805 & 0.646 & 0.834 & 0.561 & 0.849 & 0.518 & 0.852 \\
    \textbf{55} & \multicolumn{2}{c|}{0.776} & \multicolumn{2}{c|}{0.828} & \multicolumn{2}{c|}{0.856} & \multicolumn{2}{c|}{0.866} & \multicolumn{2}{c|}{0.913} & \multicolumn{2}{c|}{0.904} & \multicolumn{2}{c|}{0.851} & 0.964 & 0.964 \\
    \textbf{56} & 1.393 & 3.216 & 1.401 & 2.905 & 1.321 & 2.701 & 1.283 & 2.465 & 1.247 & 2.822 & 1.079 & 3.194 & 0.928 & 3.259 & 0.909 & 2.897 \\
    \textbf{57} & 0.925 & 1.158 & 0.930 & 1.040 & 0.959 & 0.981 & 0.945 & 0.963 & \multicolumn{2}{c|}{0.951} & 0.954 & 1.004 & \multicolumn{2}{c|}{0.926} & 0.928 & 0.928 \\
    \textbf{58} & \multicolumn{2}{c|}{1.070} & \multicolumn{2}{c|}{0.996} & \multicolumn{2}{c|}{0.979} & \multicolumn{2}{c|}{0.959} & \multicolumn{2}{c|}{0.963} & \multicolumn{2}{c|}{0.970} & \multicolumn{2}{c|}{0.986} & \multicolumn{2}{c}{0.977} \\
    \textbf{59} & 1.011 & 1.051 & 0.962 & 1.087 & 0.967 & 1.037 & \multicolumn{2}{c|}{0.966} & \multicolumn{2}{c|}{0.973} & \multicolumn{2}{c|}{0.972} & \multicolumn{2}{c|}{0.957} & \multicolumn{2}{c}{0.977} \\
    \textbf{60} & 1.023 & 1.145 & 1.005 & 1.154 & 0.999 & 1.138 & 1.000 & 1.116 & 1.043 & 1.121 & 0.972 & 1.127 & 0.899 & 1.108 & 0.851 & 1.075 \\
    \textbf{61} & \multicolumn{2}{c|}{0.816} & 0.825 & 0.843 & \multicolumn{2}{c|}{0.865} & \multicolumn{2}{c|}{0.877} & \multicolumn{2}{c|}{0.919} & \multicolumn{2}{c|}{0.959} & \multicolumn{2}{c|}{0.860} & \multicolumn{2}{c}{0.980} \\
    \textbf{62} & \multicolumn{2}{c|}{0.938} & \multicolumn{2}{c|}{0.925} & \multicolumn{2}{c|}{0.925} & \multicolumn{2}{c|}{0.938} & \multicolumn{2}{c|}{0.999} & \multicolumn{2}{c|}{1.015} & \multicolumn{2}{c|}{1.023} & \multicolumn{2}{c}{1.009} \\
    \textbf{63} & 0     & 0.906 & 0     & 0.888 & 0.087 & 0.904 & 0.083 & 0.895 & 0.056 & 0.876 & 0.016 & 0.850 & 0     & 0.819 & 0.014 & 0.918 \\
    \bottomrule
    \end{tabular}%
   \caption{Individual $\alpha$-returns to scale for 2003-2010} \label{IARTS0310}%
}\end{table}%

\end{landscape}

\begin{landscape}
\begin{table}[htbp]
{\tiny  \centering
    \begin{tabular}{c|cc|cc|cc|cc|cc|cc|cc|cc}
    \toprule
    \multicolumn{1}{c|}{\multirow{2}[4]{*}{\textbf{DMU}}} & \multicolumn{2}{c|}{\textbf{2011}} & \multicolumn{2}{c|}{\textbf{2012}} & \multicolumn{2}{c|}{\textbf{2013}} & \multicolumn{2}{c|}{\textbf{2014}} & \multicolumn{2}{c|}{\textbf{2015}} & \multicolumn{2}{c|}{\textbf{2016}} & \multicolumn{2}{c|}{\textbf{2017}} & \multicolumn{2}{c}{\textbf{2018}} \\
\cmidrule{2-17}          & $\alpha_-^\star$ & $\alpha_+^\star$ & $\alpha_-^\star$ & $\alpha_+^\star$ & $\alpha_-^\star$ & $\alpha_+^\star$ & $\alpha_-^\star$ & $\alpha_+^\star$ & $\alpha_-^\star$ & $\alpha_+^\star$ & $\alpha_-^\star$ & $\alpha_+^\star$ & $\alpha_-^\star$ & $\alpha_+^\star$ & $\alpha_-^\star$ & $\alpha_+^\star$  \\
    \midrule
    \textbf{1} & 0.834 & 1.183 & 0.858 & 1.162 & 0.814 & 1.548 & 0.842 & 1.367 & 0.867 & 1.545 & 0.856 & 1.535 & 0.853 & 1.436 & 0.868 & 1.439 \\
    \textbf{2} & 1.040 & 1.851 & 0.995 & 2.161 & \multicolumn{2}{c|}{1.575} & \multicolumn{2}{c|}{1.351} & 1.100 & 1.346 & 1.121 & 1.426 & 1.186 & 1.359 & 1.100 & 1.642 \\
    \textbf{3} & 1.104 & 1.334 & 1.057 & 1.350 & 1.258 & 1.294 & 1.217 & 1.330 & 0.965 & 1.490 & 0.881 & 1.596 & 0.911 & 1.891 & 1.020 & 1.690 \\
    \textbf{4} & 0.823 & 2.353 & 0.846 & 2.251 & 0.920 & 1.563 & 0.917 & 1.636 & 0.941 & 1.902 & 0.974 & 2.379 & 0.957 & 2.150 & 0.958 & 1.914 \\
    \textbf{5} & 0.950 & 2.100 & 0.946 & 1.530 & \multicolumn{2}{c|}{1.025} & \multicolumn{2}{c|}{1.008} & \multicolumn{2}{c|}{1.417} & \multicolumn{2}{c|}{1.151} & \multicolumn{2}{c|}{1.212} & \multicolumn{2}{c}{1.130} \\
    \textbf{6} & \multicolumn{2}{c|}{1.029} & \multicolumn{2}{c|}{1.050} & 1.026 & 1.035 & \multicolumn{2}{c|}{1.028} & \multicolumn{2}{c|}{1.044} & \multicolumn{2}{c|}{1.100} & \multicolumn{2}{c|}{1.096} & \multicolumn{2}{c}{1.128} \\
    \textbf{7} & 0.422 & 1.056 & 0.406 & 1.065 & 0.753 & 1.060 & 0.749 & 1.038 & 0.659 & 1.038 & 0.571 & 1.047 & 0.659 & 1.055 & 0.795 & 0.977 \\
    \textbf{8} & 0.966 & 1.554 & 0.970 & 1.604 & 0.962 & 1.928 & 0.982 & 1.743 & 0.986 & 1.683 & 0.968 & 1.569 & 0.964 & 1.331 & 0.973 & 1.021 \\
    \textbf{9} & 0.982 & 1.430 & 0.995 & 1.326 & 1.074 & 1.136 & 1.049 & 1.259 & 1.019 & 1.271 & 1.005 & 1.190 & 1.022 & 1.229 & 1.035 & 1.246 \\
    \textbf{10} & 1.092 & 1.137 & 1.001 & 1.167 & 0.996 & 1.280 & 1.004 & 1.106 & 0.966 & 1.455 & 0.929 & 1.633 & 0.965 & 1.359 & 0.994 & 1.292 \\
    \textbf{11} & 0.815 & 1.068 & 0.829 & 1.030 & 0.976 & 1.031 & 0.989 & 1.022 & \multicolumn{2}{c|}{1.003} & 1.001 & 1.003 & 0.981 & 1.047 & 0.999 & 1.050 \\
    \textbf{12} & 0.801 & 1.063 & 0.796 & 1.061 & 0.890 & 1.061 & 0.925 & 1.045 & \multicolumn{2}{c|}{1.039} & \multicolumn{2}{c|}{1.057} & \multicolumn{2}{c|}{1.040} & \multicolumn{2}{c}{1.051} \\
    \textbf{13} & \multicolumn{2}{c|}{1.031} & \multicolumn{2}{c|}{1.046} & \multicolumn{2}{c|}{1.045} & 0.999 & 1.027 & 0.882 & 0.925 & 0.808 & 0.967 & 0.816 & 0.994 & 0.823 & 0.990 \\
    \textbf{14} & 0.954 & 1.221 & 0.971 & 1.203 & 0.989 & 1.228 & 1.025 & 1.123 & 1.043 & 1.232 & 1.057 & 1.103 & 1.014 & 1.032 & \multicolumn{2}{c}{1.029} \\
    \textbf{15} & 0.919 & 1.027 & 0.846 & 1.025 & \multicolumn{2}{c|}{1.067} & 1.029 & 1.029 & \multicolumn{2}{c|}{1.075} & 0.965 & 1.061 & \multicolumn{2}{c|}{1.048} & \multicolumn{2}{c}{1.056} \\
    \textbf{16} & 0.967 & 0.967 & 0.869 & 1.098 & \multicolumn{2}{c|}{0.922} & 0.937 & 0.956 & \multicolumn{2}{c|}{0.957} & \multicolumn{2}{c|}{0.958} & \multicolumn{2}{c|}{0.950} & \multicolumn{2}{c}{0.949} \\
    \textbf{17} & 0.951 & 2.122 & 0.932 & 2.128 & 0.961 & 1.970 & 0.962 & 2.052 & 0.944 & 2.045 & 0.945 & 1.894 & 0.957 & 1.727 & 0.965 & 1.569 \\
    \textbf{18} & 0.944 & 1.098 & 0.960 & 1.071 & 1.012 & 1.075 & 1.035 & 1.035 & \multicolumn{2}{c|}{1.057} & \multicolumn{2}{c|}{1.005} & \multicolumn{2}{c|}{1.018} & \multicolumn{2}{c}{1.008} \\
    \textbf{19} & 0.399 & 1.265 & 0.411 & 1.215 & 0.396 & 1.088 & 0.415 & 1.040 & 0.410 & 1.081 & 0.419 & 1.084 & 0.424 & 1.067 & 0.439 & 1.081 \\
    \textbf{20} & 1.124 & 3.356 & 1.161 & 3.385 & 0.995 & 3.123 & 0.968 & 3.027 & 0.981 & 2.857 & 0.978 & 2.722 & 0.977 & 2.347 & 0.966 & 2.096 \\
    \textbf{21} & 1.041 & $\infty$ & 1.029 & $\infty$ & 1.042 & $\infty$ & 1.155 & $\infty$ & 1.206 & $\infty$ & 1.088 & $\infty$ & 1.152 & $\infty$ & 1.078 & $\infty$ \\
    \textbf{22} & 0.972 & 1.025 & 0.966 & 1.004 & 0.958 & 1.261 & 0.955 & 1.251 & 0.953 & 1.285 & 1.013 & 1.271 & 0.979 & 1.238 & 0.975 & 1.228 \\
    \textbf{23} & 0.990 & 1.288 & 1.013 & 1.279 & 1.039 & 1.123 & \multicolumn{2}{c|}{1.038} & \multicolumn{2}{c|}{1.040} & \multicolumn{2}{c|}{1.023} & \multicolumn{2}{c|}{1.001} & 1.013 & 1.013 \\
    \textbf{24} & 0.468 & 2.252 & 0.474 & 2.285 & 0.393 & 1.056 & 0.418 & 1.029 & 0.427 & 1.075 & 0.424 & 1.047 & 0.428 & 1.048 & 0.431 & 1.056 \\
    \textbf{25} & 0.723 & 1.119 & 0.737 & 1.122 & 0.810 & 0.912 & 0.850 & 0.930 & \multicolumn{2}{c|}{0.882} & \multicolumn{2}{c|}{0.912} & \multicolumn{2}{c|}{0.957} & \multicolumn{2}{c}{0.992} \\
    \textbf{26} & 0.928 & 1.078 & 0.934 & 1.037 & 0.945 & 1.105 & 0.952 & 1.080 & 0.934 & 1.112 & 0.921 & 1.099 & 0.923 & 1.047 & 0.937 & 1.014 \\
    \textbf{27} & 0.492 & 1.033 & 0.498 & 1.031 & 0.267 & 1.120 & 0.265 & 1.125 & 0.283 & 1.099 & 0.313 & 1.070 & 0.304 & 1.073 & 0.301 & 1.086 \\
    \textbf{28} & \multicolumn{2}{c|}{0.556} & \multicolumn{2}{c|}{0.519} & 0.577 & 1.034 & 0.656 & 1.012 & 0.987 & 1.007 & 1.037 & 1.052 & 0.696 & 1.051 & 0.544 & 1.024 \\
    \textbf{29} & \multicolumn{2}{c|}{0.951} & \multicolumn{2}{c|}{0.974} & 0.869 & 1.108 & 0.869 & 1.198 & 0.873 & 1.232 & 0.894 & 1.335 & 0.912 & 1.370 & 0.900 & 1.330 \\
    \textbf{30} & \multicolumn{2}{c|}{1.836} & \multicolumn{2}{c|}{1.807} & 1.008 & 1.393 & 1.031 & 1.411 & 1.029 & 1.096 & \multicolumn{2}{c|}{1.031} & \multicolumn{2}{c|}{1.056} & 1.112 & 1.167 \\
    \textbf{31} & 0.994 & $\infty$ & 0.982 & $\infty$ & 0.942 & $\infty$ & 0.966 & $\infty$ & 1.009 & $\infty$ & 1.072 & $\infty$ & 1.056 & $\infty$ & 1.053 & $\infty$ \\
    \textbf{32} & 0.977 & 1.084 & 1.002 & 1.038 & \multicolumn{2}{c|}{1.004} & \multicolumn{2}{c|}{1.025} & \multicolumn{2}{c|}{1.043} & \multicolumn{2}{c|}{1.057} & \multicolumn{2}{c|}{1.063} & \multicolumn{2}{c}{1.065} \\
    \textbf{33} & 0.995 & 1.259 & 0.990 & 1.215 & 0.971 & 1.575 & 0.961 & 1.400 & 1.006 & 1.435 & 1.005 & 1.462 & 1.001 & 1.395 & 1.005 & 1.309 \\
    \textbf{34} & 0.993 & $\infty$ & 0.966 & $\infty$ & 1.045 & $\infty$ & 1.014 & $\infty$ & 0.888 & $\infty$ & 0.833 & $\infty$ & 0.834 & $\infty$ & 0.832 & $\infty$ \\
    \textbf{35} & 1.004 & 1.087 & 1.002 & 1.072 & 1.078 & 1.163 & 1.064 & 1.072 & \multicolumn{2}{c|}{1.063} & \multicolumn{2}{c|}{1.070} & \multicolumn{2}{c|}{1.073} & \multicolumn{2}{c}{1.086} \\
    \textbf{36} & 0.954 & 2.457 & 0.946 & 2.465 & 0.876 & 2.979 & 0.914 & 2.557 & 0.988 & 2.189 & 1.015 & 2.043 & 1.125 & 1.825 & 1.157 & 1.644 \\
    \textbf{37} & \multicolumn{2}{c|}{0.844} & 0.831 & 0.880 & \multicolumn{2}{c|}{1.059} & \multicolumn{2}{c|}{1.068} & \multicolumn{2}{c|}{1.122} & \multicolumn{2}{c|}{1.184} & 1.076 & 1.177 & 0.910 & 1.160 \\
    \textbf{38} & 0.911 & 1.320 & 0.900 & 1.326 & \multicolumn{2}{c|}{1.045} & \multicolumn{2}{c|}{1.065} & \multicolumn{2}{c|}{1.134} & \multicolumn{2}{c|}{1.184} & 1.270 & 1.270 & \multicolumn{2}{c}{1.137} \\
    \textbf{39} & 0.824 & 1.014 & 0.829 & 1.052 & \multicolumn{2}{c|}{1.105} & \multicolumn{2}{c|}{1.111} & \multicolumn{2}{c|}{1.164} & \multicolumn{2}{c|}{1.203} & \multicolumn{2}{c|}{1.183} & \multicolumn{2}{c}{1.058} \\
    \textbf{40} & \multicolumn{2}{c|}{0.992} & 1.077 & 0.992 & \multicolumn{2}{c|}{1.048} & 1.151 & 1.158 & \multicolumn{2}{c|}{1.174} & 1.118 & 1.121 & \multicolumn{2}{c|}{1.258} & \multicolumn{2}{c}{1.264} \\
    \textbf{41} & 0.776 & 0.903 & 0.798 & 0.905 & 1.020 & 1.035 & \multicolumn{2}{c|}{1.038} & \multicolumn{2}{c|}{1.071} & \multicolumn{2}{c|}{1.086} & \multicolumn{2}{c|}{1.096} & \multicolumn{2}{c}{1.085} \\
    \textbf{42} & \multicolumn{2}{c|}{1.056} & 1.065 & 1.056 & \multicolumn{2}{c|}{1.067} & \multicolumn{2}{c|}{1.038} & \multicolumn{2}{c|}{1.035} & \multicolumn{2}{c|}{1.051} & \multicolumn{2}{c|}{1.073} & \multicolumn{2}{c}{1.086} \\
    \textbf{43} & 0.619 & 1.063 & 0.593 & 1.086 & 0.924 & 1.064 & 0.884 & 1.084 & 0.836 & 1.085 & 0.779 & 1.068 & 0.767 & 1.063 & 0.812 & 1.065 \\
    \textbf{44} & 0.942 & 1.164 & 0.967 & 1.026 & 0.543 & $\infty$ & 0.641 & $\infty$ & 0.630 & $\infty$ & 0.692 & $\infty$ & 0.713 & $\infty$ & 0.721 & $\infty$ \\
    \textbf{45} & 0     & 1.497 & 0     & 1.501 & 0     & 1.602 & 0     & 1.564 & 0     & 1.588 & 0     & 1.524 & 0     & 1.482 & 0     & 1.360 \\
    \textbf{46} & 0.931 & 1.467 & 0.927 & 1.432 & 1.230 & 1.277 & 1.247 & 1.314 & \multicolumn{2}{c|}{1.449} & \multicolumn{2}{c|}{1.499} & \multicolumn{2}{c|}{1.430} & 0.976 & 1.324 \\
    \textbf{47} & 0.664 & 1.640 & 0.661 & 1.622 & 0.761 & 1.597 & 0.774 & 1.609 & 0.761 & 1.613 & 0.748 & 1.600 & 0.751 & 1.559 & 0.781 & 1.458 \\
    \textbf{48} & \multicolumn{2}{c|}{0.786} & \multicolumn{2}{c|}{0.792} & 0.888 & 1.073 & 0.878 & 1.121 & 0.810 & 1.277 & 0.796 & 1.291 & 0.764 & 1.357 & 0.730 & 1.381 \\
    \textbf{49} & \multicolumn{2}{c|}{1.020} & \multicolumn{2}{c|}{1.013} & \multicolumn{2}{c|}{0.958} & 0.922 & 0.960 & 0.868 & 0.956 & \multicolumn{2}{c|}{0.970} & \multicolumn{2}{c|}{0.986} & \multicolumn{2}{c}{1.024} \\
    \textbf{50} & \multicolumn{2}{c|}{0.983} & 1.010 & 0.983 & \multicolumn{2}{c|}{0.901} & \multicolumn{2}{c|}{0.890} & \multicolumn{2}{c|}{0.880} & \multicolumn{2}{c|}{0.851} & \multicolumn{2}{c|}{0.890} & \multicolumn{2}{c}{0.875} \\
    \textbf{51} & 0.796 & 0.863 & 0.764 & 0.872 & \multicolumn{2}{c|}{0.896} & \multicolumn{2}{c|}{0.912} & \multicolumn{2}{c|}{0.938} & \multicolumn{2}{c|}{0.979} & \multicolumn{2}{c|}{0.968} & \multicolumn{2}{c}{1.000} \\
    \textbf{52} & 0.987 & 1.369 & 0.992 & 1.336 & 1.025 & 1.249 & 1.035 & 1.060 & \multicolumn{2}{c|}{1.030} & \multicolumn{2}{c|}{1.012} & \multicolumn{2}{c|}{1.001} & \multicolumn{2}{c}{1.013} \\
    \textbf{53} & 0.890 & 0.890 & 0.943 & 0.890 & \multicolumn{2}{c|}{0.942} & \multicolumn{2}{c|}{0.946} & \multicolumn{2}{c|}{0.944} & \multicolumn{2}{c|}{0.912} & \multicolumn{2}{c|}{0.962} & \multicolumn{2}{c}{0.985} \\
    \textbf{54} & 0.520 & 0.856 & 0.491 & 0.865 & 0.745 & 0.942 & 0.757 & 0.946 & 0.785 & 0.944 & 0.751 & 0.912 & 0.708 & 1.007 & 0.691 & 0.985 \\
    \textbf{55} & \multicolumn{2}{c|}{0.961} & 0.997 & 0.961 & \multicolumn{2}{c|}{0.998} & \multicolumn{2}{c|}{0.983} & \multicolumn{2}{c|}{0.993} & \multicolumn{2}{c|}{1.022} & \multicolumn{2}{c|}{1.024} & \multicolumn{2}{c}{1.024} \\
    \textbf{56} & 1.032 & 2.600 & 1.226 & 2.367 & 0.960 & 1.889 & 0.945 & 1.812 & 0.979 & 1.795 & 0.938 & 1.899 & 0.904 & 1.880 & 0.888 & 1.840 \\
    \textbf{57} & \multicolumn{2}{c|}{0.930} & 0.955 & 1.137 & 0.931 & 1.074 & 0.910 & 1.223 & 0.955 & 1.171 & 0.933 & 1.249 & 0.946 & 1.264 & 0.948 & 1.283 \\
    \textbf{58} & 1.002 & 1.018 & 1.012 & 1.018 & \multicolumn{2}{c|}{0.981} & 0.969 & 0.969 & \multicolumn{2}{c|}{1.006} & 1.041 & 1.090 & \multicolumn{2}{c|}{1.060} & \multicolumn{2}{c}{1.017} \\
    \textbf{59} & \multicolumn{2}{c|}{0.995} & \multicolumn{2}{c|}{1.001} & 0.996 & 1.017 & \multicolumn{2}{c|}{1.015} & \multicolumn{2}{c|}{1.018} & \multicolumn{2}{c|}{1.049} & 1.027 & 1.060 & \multicolumn{2}{c}{1.005} \\
    \textbf{60} & 0.792 & 1.047 & 0.786 & 1.045 & 0.693 & 1.110 & 0.696 & 1.103 & 0.657 & 1.070 & 0.648 & 1.051 & 0.631 & 1.089 & 0.639 & 1.152 \\
    \textbf{61} & \multicolumn{2}{c|}{0.991} & \multicolumn{2}{c|}{0.976} & \multicolumn{2}{c|}{0.931} & \multicolumn{2}{c|}{0.937} & \multicolumn{2}{c|}{0.963} & \multicolumn{2}{c|}{1.005} & \multicolumn{2}{c|}{0.983} & \multicolumn{2}{c}{1.001} \\
    \textbf{62} & \multicolumn{2}{c|}{1.003} & 1.013 & 1.003 & \multicolumn{2}{c|}{1.045} & \multicolumn{2}{c|}{1.014} & \multicolumn{2}{c|}{0.965} & \multicolumn{2}{c|}{0.881} & \multicolumn{2}{c|}{0.911} & \multicolumn{2}{c}{0.917} \\
    \textbf{63} & 0.063 & 0.895 & 0.077 & 0.888 & 0.666 & 0.917 & 0.761 & 0.919 & 0.765 & 0.892 & 0.792 & 0.873 & 0.840 & 0.896 & \multicolumn{2}{c}{0.883} \\
    \bottomrule
    \end{tabular}%
  \caption{Individual $\alpha$-returns to scale for 2011-2018}  \label{IARTS1118}%
}\end{table}%

\end{landscape}

\end{document}